\documentclass{ajour}
\usepackage{amsmath}
\usepackage{amsfonts}
\usepackage{epsfig}
\newcommand{\hs}[1]{\hspace*{#1cm}}
\newcommand{\vs}[1]{\vspace*{#1cm}}
\newcommand{\bea}{\begin{eqnarray}}
\newcommand{\eea}{\end{eqnarray}}
\newcommand{\la}{\langle}
\newcommand{\ra}{\rangle}

\newcommand{\half}{{\textstyle \frac{1}{2}}}
\newcommand{\balpha}{\mbox{\boldmath $\alpha$}}

\newcommand{\bbeta}{\mbox{\boldmath $\beta$}}
\newcommand{\bgamma}{\mbox{\boldmath $\gamma$}}
\newcommand{\bepsilon}{\mbox{\boldmath $\epsilon$}}

\newcommand{\bbo}{\mbox{\boldmath $0$}}

\newcommand{\bp}{\mbox{\boldmath $p$}}
\newcommand{\bq}{\mbox{\boldmath $q$}}
\newcommand{\bx}{\mbox{\boldmath $x$}}

\newcommand{\bA}{\mbox{\boldmath $A$}}
\newcommand{\bB}{\mbox{\boldmath $B$}}
\newcommand{\bE}{\mbox{\boldmath $E$}}

\newcommand{\bS}{\mbox{\boldmath $S$}}

\newcommand{\bM}{\mbox{\boldmath $M$}}
\newcommand{\clf}{\mathcal{F}}
\newcommand{\clh}{\mathcal{H}}

\newcommand{\clp}{\mathcal{P}}
\newcommand{\grad}{\nabla}

\newcommand{\sig}{\sigma}
\newcommand{\sech}{\mbox{sech}}

\def\tightmaths{                                                                
  \thinmuskip=1.5mu                                                             
  \medmuskip=2mu plus 1mu minus 2mu                                             
  \thickmuskip=2.5mu plus 2.5mu
}                         
                                  
\tightmaths

 \let\hat=\widehat

\begin{document}

%


\authorrunninghead{C. E. Dolby and S. F. Gull}
\titlerunninghead{Particle Creation in Electric Fields}




\title{State-Space Based Approach to Particle Creation in Spatially Uniform Electric Fields}


\authors{Carl E. Dolby and Stephen F. Gull}
\affil{Astrophysics Group, Cavendish Laboratory, Madingley Road, Cambridge
CB3 0HE, U.K.}

\email{c.dolby@mrao.cam.ac.uk}

\abstract{
Our formalism described recently in \cite{Me1} is applied to the study of
particle creation in spatially uniform electric fields, concentrating on
the cases of a time-invariant electric field and a so-called `adiabatic'
electric field. Several problems are resolved by incorporating the
`Bogoliubov coefficient' approach and the `tunnelling' approaches into a single
consistent, gauge invariant formulation. The value of a time-dependent
particle interpretation is demonstrated by presenting a coherent account
of the time-development of the particle creation process, in which the
particles are created with small momentum (in
the frame of the electric field) and are then accelerated by the electric
field to make up the `bulge' of created particles predicted
by asymptotic calculations~\cite{GaGi}. An initial state comprising one
particle is also considered, and its evolution is described as being the sum of
two contributions: the `sea of current' produced by the evolved vacuum,
and the extra current arising from the initial particle state.
}

\begin{article}

\section{INTRODUCTION}

It has long been known that a strong classical electromagnetic background can cause 
vacuum instability and generate particle/antiparticle pairs. 
This phenomenon was first
studied in 1951 by Schwinger~\cite{Schw}, who predicted particle creation in a constant electric field.
Schwinger's predictions were based on the 
imaginary component of the effective action,
 but the same result has since been 
derived from a variety of alternative approaches:  Bogoliubov coefficients within 
a Canonical formalism~\cite{GaGi}, tunnelling amplitudes~\cite{Dam,Nik,BMPS}, a 
wave-functional method~\cite{Hal,Keif2}, and particle detectors~\cite{Sr1,Sr2}. 
(In ~\cite{CoMo1,Mott2} back-reaction is also considered.) A brief 
review  is presented by Sriramkumar et 
al.~\cite{Sr1}, who point out that the predictions of these approaches 
do not always agree. 
The same authors point out in \cite{SPad} that 
approaches based on Bogoliubov coefficients or on tunnelling amplitudes are liable 
to suffer from a gauge-dependent particle definition (i.e., choice of `in' 
and `out' modes) which can in turn  lead to gauge-dependent predictions of the number of particles created. 

We have recently introduced an alternative approach to the study 
of particle creation in electromagnetic backgrounds~\cite{Me1} which uses 
methods analogous to those of conventional 
multiparticle quantum mechanics. 
Our approach is to emphasise the 
states of the system, described in terms of Slater
determinants of Dirac states. The vacuum state `at time $\tau$' is described in terms of the Slater determinant of all 
`negative energy' states, and  
provides a concrete realisation of the Dirac Sea. 
This approach has provided 
straightforward derivations of the general S-Matrix element and 
expectation value of the theory, using only  
methods  familiar from multiparticle quantum mechanics. We argued 
in \cite{Me1} that this approach is often quicker and clearer than conventional methods.
More importantly, the `Bogoliubov coefficient' and 
`tunnelling amplitude' methods were combined consistently into a single gauge invariant formulation,
 which could be linked to the motion of an observer (detector).

	As with the Canonical, Tunnelling, and Wave-functional approaches, 
our analysis depends explicitly on a choice of `in' and `out' particle 
interpretation, and on the corresponding categorisation of in and out modes. 
Various categorisation schemes have been proposed, based on 
asymptotic or adiabatic properties of solutions, or on the 
diagonalisation of a suitable Hamiltonian. Unfortunately, most   
schemes depend on a choice of coordinates (as in the Unruh 
effect~\cite{Unruh}) or of gauge~\cite{SPad}, and the relation of these to the 
behaviour of a particle detector is often ambiguous~\cite{Sr1}. Also, 
those schemes based on asymptotic or adiabatic approximations generate 
accurate predictions only at asymptotically late times 
or in sufficiently weak electromagnetic backgrounds. 
Particle detectors 
provide a more operational particle concept, but their predictions are not 
always proportional to the number of particles present~\cite{Sr1,Sr2} 
even when the detector is inertial.

	The categorisation scheme suggested in \cite{Me1} provides  
a consistent particle interpretation at all times without requiring any 
asymptotic conditions on the in and out states. It consistently combines the conventional
`Bogoliubov coefficient' and `tunnelling amplitude' methods, resolving
gauge inconsistencies that trouble each of these~\cite{SPad}. Also, 
by utilising the concept 
of radar time, this definition naturally incorporates the motion of the 
observer (detector), providing a definition which depends only on the observer's  
motion and on the background, and not on the choice of coordinates or gauge.

	The present paper 
presents a simple application of this formulation. We shall examine
the problem of pair creation in time-varying spatially uniform electric 
fields, as viewed by an observer stationary with respect to this electric 
field. We present definite results for a constant electric 
field $\bE(t) = (0,0,E)$ and for the so-called `adiabatic' \cite{GaGi} electric field 
$\bE(t) = \left(0,0,E \sech^2\left(m t/\rho\right)\right)$. Such backgrounds have been 
studied by various authors~\cite{Schw,Keif2,BMPS,Nik}, 
who derived pair-creation rates as time-averages over an 
infinite period of time. The problem of pair creation 
after a finite evolution time has only recently been considered only recently \cite{Hal,GaGi}. 
The electric field was  
taken as constant within a certain time interval (say $0 < t < T$), but was 
zero before and after. 
The `in' and `out' states were defined in the `free regions' 
before the field was turned on ($t < 0$) , and after it was turned off ($t > T$), where the solutions were simple plane wave states. 
Dependence of the results on $T$ was then studied.

	We use the particle definition proposed in \cite{Me1} to consider the background $\bE = (0,0,E)$ for all 
time, and derive the magnitude of pair creation in a state which 
has evolved from the vacuum for a finite length of time. 
That is, we answer the question, ``Suppose the system is in the vacuum state at some 
time $t_0$. What will be the properties of the system at time $t_0 + T$?'' 
We demonstrate the consistency of this particle definition, by obtaining the same answers to this question
as in \cite{GaGi,Hal}. (Although the difference between these two questions is 
purely formal in this case, it 
becomes vital when considering  
gravitational fields, which cannot simply be switched off whilst making 
measurements.) The finite time behaviour of particle creation 
in the `adiabatic' background is also studied here, providing a consistent 
account of the time-dependence of this creation process. We also consider the time-development of initial states containing particles.

	Section 2 is a short review of the approach to pair creation described 
in \cite{Me1,mythesis}. It includes the particle definition appropriate for inertial 
observers, and a brief derivation of the general S-Matrix element and 
expectation value in terms of `Bogoliubov coefficients'. Section 3 is 
devoted to calculating the finite time 
Bogoliubov coefficients in an spatially uniform, time-varying electric field in 
terms of the solution of a simple linear, second order equation. In Section 
4 we solve this equation in the case of a constant electric field, and use the result to find
the number density of created pairs, and also their current density, after a finite 
evolution time.  Section 5 presents a similar treatment for the `adiabatic' electric field 
$\bE(t) = (0,0,E \sech^2(\frac{m t}{\rho}))$. Conclusions are set out in Section 6.

\section{BACKGROUND}

	Let $\clh \equiv L^2(\mathbb{R}^3)^4$ (the space of finite-norm 
spinor-valued functions of space) be the space of `first quantized' 
states. The {\it antisymmetric Fock Hilbert space} over 
$\clh$, denoted $\clf_{\wedge}(\clh) \equiv \oplus_{n=0}^{\infty} \wedge^n 
\clh$, is the space of superpositions of all possible Slater determinants of first 
quantized states. The time-dependent `first quantized' Hamiltonian $\hat{H}_1(t):\clh \rightarrow \clh$ is defined by:
\begin{equation}  \hat{H}_1(t) \psi(\bx,t) = \left(- i \sum_{k=1}^{3} \bar{\sig}_k \grad_k 
+ m \gamma^0 \right) \psi(\bx,t) \label{eq:diss2.3} \end{equation} 
where $\bar{\sig}_k \equiv \gamma^0 \gamma^k = \gamma_k \gamma_0$ 
and $\grad_{\mu} \psi(x) \equiv \partial_{\mu} \psi(x) + i e A_{\mu}(x)  \psi(x)$. The  Dirac equation can be written in 
terms of this Hamiltonian as $i \grad_0 \psi(\bx,t) =  \hat{H}_1(t) \psi(\bx,t)$.
Its expectation value is the 
spatial integral of the `00-component' of the energy-momentum tensor: 
$$ 
\la \psi(t_0) | \hat{H}_1(t_0) | \psi(t_0) \ra \equiv  \int_{t=t_0} {\rm d}^3 \bx \ T^{0}_{\hs{.2} 0}(\psi(x))
$$ 
 We can use $\hat{H}_1(t_0)$ to split $\clh$ into $\clh = \clh^+(t_0) \oplus \clh^-(t_0)$ where:
\begin{align} \clh^+(t_0) & \mbox{ is the span of the positive spectrum of } 
	\hat{H}_1(t_0) \notag \\
\clh^-(t_0) & \mbox{ is the span of the negative spectrum of }
	\hat{H}_1(t_0) \notag \end{align}
$\clh^+(t_0)$ is the set of all
	positive energy states, and $\clh^-(t_0)$ is the set of all
	negative energy states, as defined in the 
background $A_{\mu}(t_0)$ at time $t_0$. In the Canonical formalism this 
definition would correspond to `Hamiltonian diagonalisation'~\cite{GM1,GM2,MMS,GMM} of the 
second-quantized Hamiltonian $$\hat{H}(t_0) \equiv  \int_{t=t_0} {\rm d}^3 \bx 
\ T^{0}_{\hs{.2} 0}(\hat{\psi}(x))\ ,$$ where $\hat{\psi}(x)$ is the field 
operator. In the past, Hamiltonian diagonalisation has been
criticised~\cite{Full2} for its  reliance on
an apparently arbitrary choice of hypersurface $\Sigma_{t_0}$, (given by $t = t_0$
here) and time-translation vector field $k^{\mu}$ ($= \delta^{\mu}_{0}$
here). This arbitrariness has been a particularly difficult problem when
considering
QFT in a gravitational background, but the difficulty can be satisfactorily
resolved~\cite{Me1,mythesis} by specifying
$\Sigma_{\tau_0}$ and $k^{\mu}$ in terms of the worldline of
the observer (or particle detector).
The choice $t=t_0$, $k^{\mu} = \delta^{\mu}_{0}$ then emerges uniquely
as appropriate to an inertial observer stationary in the frame of the
electric field. When restricted to this frame,  this definition is also similar to
that of Fradkin et al.~\cite{FGS} or  Greiner et al.~\cite{GMR}. However, in \cite{FGS,GMR} 
the Hamiltonian is taken to be $\hat{H}'(t) = \hat{H}_1(t) + 
e A_{\mu}(\bx,t)$ so that the Dirac equation becomes $i \partial_{t} 
\psi(\bx,t) =  \hat{H}'(t) \psi(\bx,t)$. Sriramkumar and 
Padmanabhan~\cite{SPad} pointed out that this leads to a particle definition which is 
gauge dependent. This problem, which has plagued previous `Bogoliubov 
coefficient' approaches to particle creation, is usually patched up by
appealing to the tunnelling interpretation: if a gauge $A =
(A(\bx),0,0,0)$ is chosen, particle creation is described in terms
of particles `tunnelling through the barrier that separates
particle/antiparticle states' (see \cite{Dam,GMR} or \cite{BMPS} for a
description of these methods). This leads to 
two theories, each succeeding where
the other fails~\cite{SPad}. This is not the case here, however, since we use eigenstates of the
gauge-covariant Hamiltonian $\hat{H}_1(t_0)$ to define
particle states. It will emerge that particle creation in the background
$A^{\mu}(x)$ requires either: 1) that $A^0(x) \neq 0$, so that $A^0(x)$ can act
as a potential, w.r.t. which tunnelling may occur, or 2) 
$\bA(x) = (A^1(x),A^2(x),A^3(x))$ is
time-dependent, so that eigenstates of $\hat{H}_1(t)$ mix during
evolution. Both effects contribute to the calculation
of Bogoliubov coefficients~\cite{Me1,mythesis}, and ensure their gauge invariance. An immediate 
consequence is that if a
gauge can be chosen such that $A^0 = 0$ and $\bA$ is time-independent,
then particle creation will not occur. That is, {\it time-independent
magnetic fields cannot create particles}. This result agrees
with Schwinger's calculation \cite{Schw} in terms of an effective
Lagrangian, and successfully resolves the inconsistency problems
raised by Sriramkumar and Padmanabhan \cite{SPad}. 

	Having defined $\clh^{\pm}(t_0)$, we can now define 
the {\it vacuum at time $t_0$}, $| {\rm vac}_{t_0} \ra$, to be the 
Slater determinant of any basis of 
$\clh^-(t_0)$ (normalised so that $\la {\rm vac}_{t_0} | {\rm vac}_{t_0} 
\ra = 1$). This 
specifies $| {\rm vac}_{t_0} \ra$ up to an arbitrary phase factor. It is 
the state in which all negative energy degrees of freedom are full, and hence 
	is a concrete manifestation of the Dirac Sea.

	To illustrate this, let $\{ u_{i,t_0} ; i \in I\}$, $\{ v_{i,t_0} ; i \in I \}$ 
be orthonormal bases for $\clh^+(t_0)$ and $\clh^-(t_0)$ respectively, where $I$ is some index set, assumed countable for convenience (the uncountable case introduces no complications). The vacuum at time $t_0$ can be written as:
\begin{equation} |{\rm vac}_{t_0} \ra = v_{1,t_0} \wedge v_{2,t_0} \wedge \dots 
\end{equation}
and is independent of the choice of basis for $\clh^-(t_0)$ 
(up to a phase factor) because of the complete antisymmetry of 
the Slater determinant. The vacuum at some time $t_1 > t_0$ is:
\begin{equation} |{\rm vac}_{t_1} \ra = v_{1,t_1} \wedge v_{2,t_1} \wedge\dots 
\end{equation} 

	By considering fermionic QFT `in a background' we are ignoring direct electron-electron interactions, 
so that the evolved state 
$| {\rm vac}_{t_0}(t_1) \ra$ obtained by evolving $|{\rm vac}_{t_0} \ra$ from time $t_0$ to time $t_1$ is simply:

\begin{equation} |{\rm vac}_{t_0}(t_1) \ra = v_{1,t_0}(t_1) \wedge v_{2,t_0}(t_1) \wedge \dots \end{equation}
 where $v_{i,t_0}(t_1)$ denotes the state obtained from
$v_{i,t_0}$ by evolution to time $t_1$, and will not, in general, be 
contained in $\clh^-(t_1)$. We will often refer to this state as 
the `evolved vacuum', although it is not in fact a vacuum state. 

	Since the inner product of Slater determinants is simply the 
determinant of the `first quantized' inner products,  the 
vacuum-vacuum S-matrix element is simply:

\begin{equation} \la {\rm vac}_{t_1} | {\rm vac}_{t_0}(t_1) \ra = \det [ 
\la v_{i,t_1} | v_{j,t_0}(t_1) \ra ] \label{Svacvac1} \end{equation}
	The probability that $|{\rm vac}_{t_0}(t_1) \ra$ will be vacuum 
at time $t_1$ is then $\clp_{{\rm vac} \rightarrow {\rm vac}}  = 
|\la {\rm vac}_{t_1} | 
{\rm vac}_{t_0}(t_1) \ra |^2$. Although once a particle interpretation is specified this result can be derived by a number of methods~\cite{FGS,Keif2,Sch2}, we believe 
that the derivation presented here (and in \cite{Me1,mythesis}) is 
the shortest and clearest.

	Similarly, the general S-matrix element can be written as:
\begin{align} & \la \mbox{$\binom{i'_1 i'_2 \dots 
i'_{m'}}{j'_1 j'_2 \cdots j'_{n'}}$}_{t_1} | \mbox{$\binom{i_1 i_2 \dots 
i_m}{j_1 j_2 \dots j_n}$}_{t_0}(t_1) \ra \notag \\
 & = (-)^{J-J'} \det \hs{-.1} \left[ \hs{-.1} \begin{array}{cc}  
\left[ \begin{array}{ccc}
  \alpha_{i'_1 i_1} & \cdots & \alpha_{i'_1 i_m} \\ \vdots & & \vdots
  \\ \alpha_{i'_{m'} i_1}  & \cdots & \alpha_{i'_{m'} i_m} \end{array}
  \right] &  \left[ \begin{array}{ccc}  \beta_{i'_1 1} & \cdots & \binom{j_1
  \dots j_n}{missing} \\ \vdots & & \vdots \\
  \beta_{i'_{m'} 1}  & \dots & \binom{j_1 \dots j_n}{missing} \end{array} \right] \\ 
\hs{-.1} \left[ \hs{-.1} \begin{array}{ccc}  \gamma_{1 i_1}
  & \hs{-.1} \cdots \hs{-.1} & \gamma_{1 i_m} \\ 
\vdots &  &  \vdots \\
\binom{j'_1 \dots j'_{n'}}{missing} & \cdots &
  \binom{j'_1 \dots j'_{n'}}{missing}  \end{array} \hs{-.1} \right] \hs{-.2} &  \hs{-.2} \left[ \hs{-.1} \begin{array}{ccc}
  \epsilon_{1 1} & \cdots & \hs{-.1} \binom{j_1 \dots j_n}{missing} \hs{-.1} \\
\vdots &  &  \\
  \binom{j'_1 \dots j'_{n'}}{missing} & &  \end{array} \hs{-.1} \right] \hs{-.1} \end{array} \hs{-.1} \right] \hs{-.1}
  \label{eq:SSBSmat}\end{align} (if $m-n = m'-n'$, and zero otherwise) where 
\bea \alpha_{i j}(t_1,t_0) = \la u_{i,t_1} | u_{j,t_0}(t_1) \ra & \hs{1}
	\gamma_{i j}(t_1,t_0) =  \la v_{i,t_1} | u_{j,t_0}(t_1) \ra
	\label{eq:SSB39}\\ \beta_{i j}(t_1,t_0) = \la u_{i,t_1} |
	v_{j,t_0}(t_1) \ra & \hs{1} \epsilon_{i j}(t_1,t_0) = \la v_{i,t_1} |
	v_{j,t_0}(t_1) \ra \label{eq:SSB40}\eea
are the {\it time-dependent Bogoliubov coefficients}, $(-)^{J-J'}$ is an unimportant sign convention, and $| \mbox{$\binom{i_1 i_2 \dots i_m}{j_1 j_2 
\dots j_n}$}_{t_0} \ra$ denotes a state comprised of $m$ particles 
(in states $u_{i_1,t_0} \dots u_{i_m,t_0}$ with $i_1 < i_2 < \dots i_m$ by 
convention), and $n$ antiparticles (corresponding to 
the absence of states $v_{j_1,t_0} \dots v_{j_n,t_0}$), prepared at time $t_0$.

\subsubsection*{Expectation Values}

	Given an operator $\hat{A}_1(t):\clh \rightarrow \clh$, we can define its 
{\it physical extension} $\hat{A}_{{\rm phys}}(t): \clf_{\wedge}(\clh) \rightarrow 
\clf_{\wedge}(\clh)$ by:
\begin{align} \hat{A}_{{\rm phys}}(t) & = \hat{A}_H(t) - \la {\rm vac}_{t} | \hat{A}_H(t) | {\rm vac}_{t} \ra \hat{1} \label{eq:vacsub1} \\
\mbox{ where } \hat{A}_H : \psi_1 \wedge \psi_2 \wedge \dots 
\wedge \psi_N  & \rightarrow  \sum_{i=1}^{N} \psi_1 \wedge \dots (\hat{A}_1 \psi_i) \wedge 
\psi_{i+1} \dots \wedge \psi_N \label{eq:SSB8} \end{align}
	The expectation value of $\hat{A}_{{\rm phys}}(t)$ in the physical 
vacuum at time $t$, $| {\rm vac}_{t} \ra$ is zero by construction.  
Its expectation value in the `evolved vacuum' $| {\rm vac}_{t_0}(t) \ra$ 
is in general  non-zero, and takes the form 
\begin{align} \la {\rm vac}_{t_0}(t_1) & | \hat{A}_{{\rm phys}}(t_1) 
| {\rm vac}_{t_0}(t_1) \ra \label{eq:vac1} \\
 & \hs{-.5} = \sum_{i=1}^{N} \la v_{i,t_0}(t_1) | \hat{A}_1(t_1) 
| v_{i,t_0}(t_1) \ra - \sum_{i=1}^{N} 
\la v_{i,t_1} | \hat{A}_1(t_1) | v_{i,t_1} \ra  \label{eq:vacsub} \\
	& \hs{-.5} = {\rm Trace}(\bbeta \bbeta^{\dagger} \bA^{++} - \bgamma
	\bgamma^{\dagger} \bA^{--} + \bepsilon \bbeta^{\dagger}
	\bA^{+-} + \bbeta \bepsilon^{\dagger} \bA^{-+})
	\label{eq:SSB40.1} \end{align} 
where we have defined:
\begin{align} \bA^{++}_{j k} \equiv \la u_{j,t_1} | \hat{A}_1(t_1) | u_{k,t_1} \ra 
 \hs{1} & \bA^{--}_{j k} \equiv \la v_{j,t_1} | \hat{A}_1(t_1) | v_{k,t_1} \ra \notag \\ 
\bA^{+-}_{j k} \equiv \la u_{j,t_1} | \hat{A}_1(t_1) |
  v_{k,t_1} \ra \hs{.2} \mbox{ and } & \bA^{-+}_{j k} \equiv \la
  v_{j,t_1} | \hat{A}_1(t_1) | u_{k,t_1} \ra \label{eq:SSB40.2} \\
 & \hs{.8} = \overline{\bA^{+-}_{k j}} \mbox{ if } \hat{A}_1 \mbox{ is Hermitian
  } \notag \end{align} 
	The operator $\hat{N}_{{\rm phys}}(t)$ that represents the number of 
particles (including antiparticles) present at time $t$ is the physical 
extension of $\hat{N}_1(t) = \hat{P}^+(t) - \hat{P}^-(t)$, where 
$\hat{P}^{\pm}(t):\clh \rightarrow \clh^{\pm}(t)$ are the projection operators onto $\clh^{\pm}(t)$. It is easy to verify that $\hat{N}_{{\rm phys}}(t_0)| 
\mbox{$\binom{i_1 i_2 \dots i_m}{j_1 j_2 \dots j_n}$}_{t_0} \ra = (m+n) 
| \mbox{$\binom{i_1 i_2 \dots i_m}{j_1 j_2 \dots j_n}$}_{t_0} \ra$, as required. 
The expectation value of $\hat{N}_{{\rm phys}}(t_1)$ in the `evolved vacuum' $| {\rm vac}_{t_0}(t_1) \ra$ is given (from (\ref{eq:SSB40.1})) by:
\begin{align} N_{{\rm vac},t_0}(t_1) & \equiv \la {\rm vac}_{t_0}(t_1)| \hat{N}_{{\rm phys}}(t_1) | {\rm vac}_{t_0}(t_1) \ra \notag \\
 & = {\rm Trace}( \bbeta \bbeta^{\dagger} + \bgamma \bgamma^{\dagger}) \label{eq:ntot1} \\
 & = \sum_i \{ N^+_{i,t_0}(t_1) + N^-_{i,t_0}(t_1) \} \notag \end{align}
where $N^+_{i,t_0}(t_1) = (\bbeta \bbeta^{\dagger})_{i i}$ is the expectation 
value of the physical extension of $| u_{i,t_1} \ra \la u_{i,t_1}|$ and 
represents 
the probability that the degree of freedom $u_{i,t_1}$ is occupied in 
$| {\rm vac}_{t_0}(t_1) \ra$, i.e. that particle $i$ is present. 
$N^-_{i,t_0}(t_1) = (\bgamma \bgamma^{\dagger})_{i i}$ is the expectation 
value of the physical extension of $- | v_{i,t_1} \ra \la v_{i,t_1}|$;
it represents the probability that the degree of freedom $v_{i,t_1}$ is 
unoccupied in $| {\rm vac}_{t_0}(t_1) \ra$,  i.e. that antiparticle $i$ is present. From the 
unitarity of the `first quantized' evolution matrix $ \bS_1(t_1,t_0) = \left[ \begin{array}{cc}  \balpha(t_1,t_0) & 
\bbeta(t_1,t_0) \\ \bgamma(t_1,t_0) & \bepsilon(t_1,t_0) \end{array} \right]$,
it is easy to show that ${\rm Trace}( \bbeta \bbeta^{\dagger}) = {\rm Trace}( \bgamma \bgamma^{\dagger})$, expressing charge conservation.

	The derivation of $\la F_{t_0}(t_1) |
  \hat{A}_{{\rm phys}}(t_1) | F_{t_0}(t_1) \ra$ for an arbitrary state $| F_{t_0}(t_1) \ra$
  is identical to the derivation of (\ref{eq:SSB40.1}), and gives:

\begin{align} & \la \mbox{$\binom{i_1 i_2 \dots i_m}{j_1 j_2 \dots j_n}$}_{t_0}(t_1)|
 \hat{A}_{{\rm phys}}(t_1) | \mbox{$\binom{i_1 i_2 \dots i_m}{j_1 j_2 \dots j_n}$}_{t_0}(t_1)
 \ra \notag \\
 & \hs{1} = \sum_{k=1}^{m} \la u_{i_k,t_0}(t_1) | \hat{A}_1(t_1) |
 u_{i_k,t_0}(t_1) \ra - \sum_{k=1}^{n} \la v_{j_k,t_0}(t_1) | \hat{A}_1(t_1)
 | v_{j_k,t_0}(t_1) \ra \notag \\
 & \hs{2} + \la {\rm vac}_{t_0}(t_1) | \hat{A}_{{\rm phys}}(t_1)
 | {\rm vac}_{t_0}(t_1) \ra \label{eq:SSB40.12} \end{align}
	More details of this approach to fermionic QFT are given in \cite{Me1,mythesis}.

\section{SPATIALLY UNIFORM ELECTRIC FIELDS}

	A spatially uniform electric field in 
the $z$-direction can be represented by $A_{\mu}(t) = (0,0,0,A(t))$, so that 
$\bE(t) = (0,0,\dot{A}(t))$ and $\bB(t) = \bbo$. To find solutions 
to the Dirac equation in this background, we try separating $\psi(x)$ as
\begin{equation} \psi_{\bp}(\bx,t) = \psi_{\bp}(t) e^{i \bp \cdot \bx} 
\label{eq:CE1.5} \end{equation}
	where $\bp \cdot \bx = \sum_{k=1}^3 p^k x^k$. Then 
$\psi_{\bp}(t)$ must satisfy:

\begin{equation} i \frac{d \psi_{\bp}(t)}{d t} = m \gamma_{0} \psi_{\bp}(t) + 
\bar{\bp}_{\perp} \psi_{\bp}(t) + (p^3 + e A(t)) \bar{\sig}_3 \psi_{\bp}(t) 
\label{eq:CE2} \end{equation}

	where $\bar{\bp}_{\perp} \equiv p^1 \bar{\sig}_1 + p^2 
\bar{\sig}_2$. By introducing a set of basis spinors 
$\phi_{\pm \pm}(\bp)$ satisfying:
\begin{equation}
\begin{array}{rll} \phi^{\dagger}_{s \lambda}(\bp) \phi_{s' \lambda'}(\bp) & = 
\delta_{s s'} \delta_{\lambda \lambda'} & 
\hs{.5} \mbox{ for all } \bp \notag \\
\bar{\sig}_3 \phi_{\pm \lambda}(\bp) & = \pm \phi_{\pm \lambda}(\bp) & 
\hs{.5} \mbox{ for } \lambda = \pm \mbox{ and all } \bp \label{eq:diss6.2}\\
\gamma_{0} \phi_{\pm \lambda}(\bp) & = \phi_{\mp \lambda}(\bp) & \hs{.5} \mbox{ for } 
\lambda = \pm \mbox{ and all } \bp \notag \\
\bar{\bp}_{\perp} \phi_{s \lambda}(\bp) & = s \lambda | \bp_{\perp} | 
\phi_{-s -\lambda}(\bp) & \hs{.5} \mbox{ for } s = \pm, \lambda = \pm, 
\mbox{ and all } \bp \notag\\
\end{array}
 \end{equation}
where $|\bp_{\perp}|^2 \equiv (p^1)^2 + (p^2)^3$, we can  
expand $\psi_{\bp}(t)$ as:
\begin{equation} \psi_{\bp}(t) = \sum_{\lambda} \{ \phi_{+ \lambda}(\bp) 
f_{\lambda}(t) + \phi_{- \lambda}(\bp) g_{\lambda}(t) \} \label{eq:CE3}\end{equation}
	The inner product of two solutions $\psi^1_{\bp}(\bx,t),
 \psi^2_{\bq}(\bx,t)$ expanded as in (\ref{eq:CE1.5}) and (\ref{eq:CE3}) 
is given by:
\begin{equation} \la \psi^1_{\bp}(\bx,t)| \psi^2_{\bq}(\bx,t) \ra = (2 \pi)^3 
\delta(\bp - \bq) \sum_{\lambda}(\bar{f}^1_{\lambda}(t)
f^2_{\lambda}(t) + \bar{g}^1_{\lambda}(t) g^2_{\lambda}(t))
\label{eq:CE5} \end{equation} 
	Substitution of (\ref{eq:CE3}) into (\ref{eq:CE2}) gives:

\begin{align} i \frac{d f_{\pm}}{d t} - (p^3 + e A(t)) f_{\pm} & = m g_{\pm} \pm  
|\bp_{\perp}| g_{\mp} \notag \\
 i \frac{d g_{\pm}}{d t} + (p^3 + e A(t)) g_{\pm} & = m f_{\pm} \mp
|\bp_{\perp}| f_{\mp} \label{eq:CE6} \end{align} 
Define now the dimensionless
quantities
\begin{equation} \tau \equiv m t \hs{.5} p_z \equiv \frac{p^3}{m} \hs{.5} p 
\equiv \frac{|\bp_{\perp}|}{m} \hs{.2} \mbox{ and } a(\tau) = \frac{e A}{m}
\label{eq:CE7}\end{equation}
This means we are measuring distances in multiples of the Compton wavelength $\frac{\lambda_c}{2 \pi} = \frac{\hbar}{m c} \approx 3.9 \times 10^{-13}$ metres, and times in multiples of $\frac{\lambda_c}{2 \pi c} \approx 1.3 \times 10^{-21}$ seconds
 (where we have used the mass and charge of the
	electron). Equations (\ref{eq:CE6}) become:
\begin{align} 
\left(i \frac{d}{d \tau} - (p_z + a(\tau))\right)f_{\pm}(\tau) & = g_{\pm}(\tau) \pm p g_{\mp}(\tau) \notag \\
 \left(i \frac{d}{d \tau} + (p_z + a(\tau))\right)g_{\pm}(\tau) & = f_{\pm}(\tau) \mp p f_{\mp}(\tau) \label{eq:CE8} \end{align} 
from which we find that
\begin{align} \frac{d^2 f_{\lambda}}{d \tau^2} + \left(1 + p^2 + (p_z + a(\tau))^2 + 
i \frac{d a}{d \tau}\right) f_{\lambda} & = 0 \label{eq:CE9.1} \\
 \frac{d^2 g_{\lambda}}{d \tau^2} + \left(1 + p^2 + (p_z + a(\tau))^2 - i \frac{d a}{d \tau}\right) g_{\lambda} & = 0 \label{eq:CE9.2}
	\end{align} for both $\lambda$. 
To construct a
	complete set of solutions to (\ref{eq:CE2}), let $f(\tau)$ be
	any solution of (\ref{eq:CE9.1}). One solution of
	(\ref{eq:CE2}) can then be found by putting $f_{-}(\tau) = 0$, $f_{+}(\tau) =
	f(\tau)$ and using (\ref{eq:CE8}) to get:
\begin{gather} g_{+}(\tau)  = \frac{1}{1 + p^2} h(\tau) \hs{.5} \mbox{ and } 
g_{-}(\tau) = \frac{p}{1 + p^2} h(\tau)  \label{eq:CE26} \end{gather}
	where $h(\tau) \equiv (i
	\frac{d}{d \tau} - (p_z + a(\tau)))f(\tau)$. From this we obtain
\begin{equation} \psi'_{\bp,1}(\bx,t) = C \left[\phi_{+ +}(\bp) \sqrt{1 + p^2} f(\tau) 
+ \frac{\phi_{- +}(\bp) + p \phi_{- -}(\bp)}{\sqrt{1 + p^2}} h(\tau)\right]e^{i \bp 
\cdot \bx}  \label{eq:CE27}\end{equation}
	where $C \equiv ((1 + p^2)|f|^2 + |h|^2)^{-\half}$ is constant by
 virtue of conservation of the norm. Similarly, $\psi'_{\bp,2}(\bx,t)$
 can be found by putting $f_{+}(\tau) = 0$ and $f_{-}(\tau) =
 f(\tau)$, and then using (\ref{eq:CE8}) to give $g_+(\tau)$ and $g_-(\tau)$ (in terms 
of $h(\tau))$. $\psi_{\bp,3}(\bx,t)$ is found by putting $g_{-}(\tau) =0$,
 $g_{+}(\tau) = \bar{f}(\tau)$ (which solves equation (\ref{eq:CE9.2}))
 and then using equations (\ref{eq:CE8}) to specify $f_{\pm}(\tau)$ in terms of 
$\bar{h}(\tau)$,
 while $\psi_{\bp,4}(\bx,t)$ is similarly found by putting $g_{+}(\tau) =0$,
 $g_{-}(\tau) = \bar{f}(\tau)$. This finally allows us to generate the complete
 set of solutions: 

\begin{align} \psi_{\bp,1}(\bx,t) & =
\frac{1}{\sqrt{1 + p^2}}(\psi'_{\bp,1}(\bx,t) - p \psi'_{\bp,2}(\bx,t)) \notag \\
 & = C \left[(\phi_{+ +}(\bp) - p
\phi_{+ -}(\bp))f(\tau) + \phi_{- +}(\bp) h(\tau) \right]e^{i \bp \cdot \bx} \notag \\ 
\psi_{\bp,2}(\bx,t) & = \frac{1}{\sqrt{1 + p^2}}
(p \psi'_{\bp,1}(\bx,t) + \psi'_{\bp,2}(\bx,t)) \notag \\ 
& = C  \left[(p \phi_{+ +}(\bp) +
 \phi_{+ -}(\bp))f(\tau) + \phi_{- -}(\bp) h(\tau) \right]e^{i \bp
 \cdot \bx} \label{eq:CE28} \\ 
\psi_{\bp,3}(\bx,t) & = C \left[ \frac{(- \phi_{+ +}(\bp) + p \phi_{+ -}(\bp))}{
\sqrt{1 + p^2}} \bar{h}(\tau) + \phi_{- +}(\bp) \sqrt{1 + p^2}
 \bar{f}(\tau)\right]e^{i \bp \cdot \bx}  \notag \\
 \psi_{\bp,4}(\bx,t) & = C \left[\frac{(- p \phi_{+ +}(\bp) -
 \phi_{+ -}(\bp))}{\sqrt{1 + p^2}} \bar{h}(\tau) +
 \phi_{- -}(\bp) \sqrt{1 + p^2} \bar{f}(\tau)\right]e^{i \bp \cdot \bx} \notag \end{align} 

It is easy to verify that these solutions are mutually
 orthogonal and are each normalised to $(2 \pi)^3
 \delta(\bbo)$. The choice of $\psi_{\bp,1}$ and $\psi_{\bp,2}$
 here (in terms of $\psi'_{\bp,1}$ and $\psi'_{\bp,2}$) is for later
 convenience.

\subsection{Energy Eigenstates}  

	We now calculate $\clh^{\pm}(t_0)$ at 
	each time $t_0$. We must first decompose the space of 
	spinor-valued functions $\psi(\bx,t_0)$ (of space, at time
	$t_0$) in terms of the spectrum of the operator $\hat{H}(t_0)$
	given by:

\begin{equation} \hat{H}(t_0): \psi(\bx,t_0) \rightarrow - i \sum_k \bar{\sig}_k 
\partial_k \psi(\bx,t_0) + e A(t_0) \bar{\sig}_3 \psi(\bx,t_0) + m \gamma_{0} \psi(\bx,t_0) 
\end{equation} 

	Consider the action of $\hat{H}(t_0)$ on plane wave
	states of the form (\ref{eq:CE1.5}). It is easy to see
	that: 
\begin{gather} \hat{H}(t_0)\psi_{\bp}(\bx,t_0) =
	E_{\bp,t_0} \psi_{\bp}(\bx,t_0) \mbox{ if and only if } \notag
	\\ m \gamma_{0} \psi_{\bp}(t_0) + \bar{\bp}_{\perp} \psi_{\bp}(t_0) + 
(p^3 + e A(t_0)) \bar{\sig}_3 \psi_{\bp}(t_0) = E_{\bp,t_0} \psi_{\bp}(t_0) 
\label{eq:CE15}\end{gather} 

	By substituting $\psi_{\bp,t_0} = \sum_{\lambda} \{ \phi_{+ \lambda}(\bp) 
f_{\lambda,t_0} + \phi_{- \lambda}(\bp) g_{\lambda,t_0} \}$ and using 
(\ref{eq:CE7}) we transform this into the eigenvalue problem: 
\begin{equation}\begin{array}{r} m\left[ \begin{array}{cccc}
p_z + a(\tau_0) & 0 & 1 & p \\
0 & p_z + a(\tau_0) & -p & 1 \\
1 & -p & - p_z - a(\tau_0) & 0 \\
p & 1 & 0 & - p_z - a(\tau_0) \end{array} \right] 
\left[ \begin{array}{c} f_{+,t_0} \\ f_{-,t_0} \\ g_{+,t_0} \\ g_{-,t_0} \end{array} 
\right]\\
 = E_{\bp,t_0} 
\left[ \begin{array}{c} f_{+,t_0} \\ f_{-,t_0} \\ g_{+,t_0} \\ g_{-,t_0} \end{array} 
\right]\qquad \label{eq:diss6.1}\\
\end{array}
\end{equation}
	This matrix squares to $(m^2 + |\bp_{\perp}|^2 + (p^3 + e E A(t_0))^2)I_4$. Its eigenvalues are therefore 
$E_{\bp,t_0} = \pm \sqrt{m^2 + |\bp_{\perp}|^2 + (p^3 + e E A(t_0))^2} = 
\pm m E_{\tau_0}$  in terms of the dimensionless quantity 
$E_{\tau} \equiv \sqrt{1 + p^2 + (p_z + a(\tau))^2}$. We consider the following 
cases:

\begin{itemize}
\item[{\bf Case 1:}] $E_{\bp,t_0} = m E_{\tau_0}$ \newline 

	We seek two independent solutions of (\ref{eq:diss6.1}) with 
$E_{\bp,t_0} = m E_{\tau_0}$. 
So that these are similar to 
the $\psi_{\bp,i}(\bx,t)$, we choose $u_{\bp,1,t_0}(\bx)$ 
such that $g_{+,t_0} = 1$ and $g_{-,t_0} = 0$. 
In 
(\ref{eq:diss6.1}) this gives:
\begin{equation} u_{\bp,1,t_0}(\bx) = D\frac{\phi_{+ +}(\bp) - p \phi_{+ -}(\bp) + (E_{\tau_0} - p_z - a(\tau_0)) 
\phi_{- +}(\bp)}{E_{\tau_0} - p_z - a(\tau_0)} 
 e^{i \bp \cdot \bx} \end{equation}
	where $D = $ constant, and we have used the identity 
$(E_{\tau_0} - p_z - a(\tau_0))(E_{\tau_0} + p_z + a(\tau_0)) = 1 + p^2$. 
By requiring $u_{\bp,1,t_0}(\bx)$ to have norm $(2 \pi)^3 \delta(\bbo)$ 
we obtain $D = 
\sqrt{\frac{E_{\tau_0} - p_z - a(\tau_0)}{2 E_{\tau_0}}}$. 

	Similarly, $u_{\bp,2,t_0}(\bx)$ can be found by putting $g_{+,t_0} = 0$, 
$g_{-,t_0} = 1$, and then normalising to $(2 \pi)^3 \delta(\bbo)$. This gives:
\begin{align} u_{\bp,1,t_0}(\bx) & = \frac{ \phi_{+ +}(\bp) - p \phi_{+ -}(\bp) + 
(E_{\tau_0} - p_z - a(\tau_0)) \phi_{- +}(\bp)}{\sqrt{2 E_{\tau_0}(E_{\tau_0} - 
p_z - a(\tau_0))}} e^{i \bp \cdot \bx} \label{eq:CE32}\\
 u_{\bp,2,t_0}(\bx) & = \frac{ p \phi_{+ +}(\bp) + \phi_{+ -}(\bp) + (E_{\tau_0} - 
p_z - a(\tau_0)) \phi_{- -}(\bp)}{\sqrt{2 E_{\tau_0}(E_{\tau_0} - p_z - a(\tau_0))}} 
e^{i \bp \cdot \bx} \label{eq:CE33}\end{align} 

\item[{\bf Case 2:}] $E_{\bp,t_0} = - m E_{\tau_0}$ \newline
	A similar procedure gives:
\begin{align} v_{\bp,1,t_0}(\bx) & = \frac{ - \phi_{+ +}(\bp) + p \phi_{+ -}(\bp) +
 (E_{\tau_0} + p_z + a(\tau_0)) \phi_{- +}(\bp)}{\sqrt{2 E_{\tau_0}(E_{\tau_0} + p_z
 + a(\tau_0))}} e^{i \bp \cdot \bx} \label{eq:CE34}\\
 v_{\bp,2,t_0}(\bx) & = \frac{ - p \phi_{+ +}(\bp) - \phi_{+ -}(\bp) + (E_{\tau_0}
 + p_z + a(\tau_0)) \phi_{- -}(\bp)}{\sqrt{2 E_{\tau_0}(E_{\tau_0} + p_z + 
a(\tau_0))}} e^{i \bp \cdot \bx} \label{eq:CE35}
\end{align} 
\end{itemize}
These states are eigenstates of the
gauge invariant 3-momentum operator $- i \partial_k + e A_{k}$ of
momentum $\bp\prime = (p^1,p^2,p^3 + e A(t_0))$. Hence the 1-electron state
$| \mbox{$\binom{\bp,\lambda}{}$}_{t_0}\ra \equiv u_{\bp,\lambda,t_0} \wedge 
| {\rm vac}_{t_0} \ra$ has physical (gauge
invariant) momentum $\bp\prime$. The  1-positron state
$| \mbox{$\binom{}{\bp,\lambda}$}_{t_0}\ra \equiv i_{v_{\bp,\lambda,t_0}} \wedge 
| {\rm vac}_{t_0} \ra$ (where $i_{v_{\bp,\lambda,t_0}}$ removes the $v_{\bp,\lambda,t_0}$ degree of freedom from $| {\rm vac}_{t_0} \ra$ \cite{Me1,mythesis}) has physical
momentum $- \bp\prime$.

Suppose now that $A(t_0) \rightarrow \mp \infty$ as $t_0
\rightarrow \pm \infty$. This holds for the constant
electric field $A(t) = E t$ ($E < 0$) and  many other
electric fields. In this case we have:
\begin{align} \bar{\sig}_3 u_{\bp,\lambda,-\infty}(\bx) = 
u_{\bp,\lambda,-\infty}(\bx) \hs{2} & \bar{\sig}_3 v_{\bp,\lambda,-\infty}(\bx) = 
- v_{\bp,\lambda,-\infty}(\bx) \notag \\
 \bar{\sig}_3 u_{\bp,\lambda,\infty}(\bx) = - 
u_{\bp,\lambda,\infty}(\bx) \hs{2} & \bar{\sig}_3 v_{\bp,\lambda,\infty}(\bx) 
= v_{\bp,\lambda,\infty}(\bx) \notag \end{align}
It follows that any electron (of charge $e<0$) in this
	electric field, that has a finite momentum at finite time,
	will at late times be moving in the negative $x^3$ direction
	with infinite momentum (i.e. at speeds approaching that of
	light),  whereas a positron with charge $-e$ will be moving in the positive $x^3$ direction with infinite momentum. 
At very early times this situation
	is reversed. Clearly it is important to keep track of
	what constitutes particles/antiparticles at different times,
	since not only does this change with time in an electric
	field, but  here the
	spaces $\clh^{\pm}$ have completely reversed  between early and late times;
	$\clh^{\pm}(-\infty) = \clh^{\mp}(\infty)$! This fact 
 is also mentioned
	in \cite{Keif2}.

\subsection{Finite Time Transition Probabilities} 

We now have the positive/negative energy eigenstates $\{ |u_{\bp,\lambda;t_0}\ra, |v_{\bp,\lambda;t_0}\ra \}$ at each time $t_0$, and the evolution operator: 

\begin{equation} \hat{U}_1(t_1,t_0) = \int \frac{d^3 \bp\prime}{(2 \pi)^3} \sum_{i=1}^{4}
 |\psi_{\bp\prime,i}(t_1) \ra \la \psi_{\bp\prime,i}(t_0)|  \notag \end{equation}
	Hence we can write $|u_{\bp,\lambda;t_0}(t_1)\ra = \hat{U}_1(t_1,t_0) |u_{\bp,\lambda;t_0}\ra$, 
\newline $|v_{\bp,\lambda;t_0}(t_1)\ra = \hat{U}_1(t_1,t_0) |v_{\bp,\lambda;t_0}\ra$ 
and the time-dependent Bogoliubov coefficients become
\begin{align} S_{\bp,\lambda;\bq,\sigma}(t_1,t_0) & = \left[ \begin{array}{cc} 
 \alpha_{\bp,\lambda;\bq,\sigma}(t_1,t_0) &
 \beta_{\bp,\lambda;\bq,\sigma}(t_1,t_0) \\
 \gamma_{\bp,\lambda;\bq,\sigma}(t_1,t_0) &
 \epsilon_{\bp,\lambda;\bq,\sigma}(t_1,t_0) \end{array} \right]
 \notag \\ & \hs{-2} =  \left[ \begin{array}{cc}  \la
 u_{\bp,\lambda;t_1}(\bx)|\hat{U}_1(t_1,t_0)|u_{\bq,\sigma;t_0}(\bx)
 \ra  & \la
 u_{\bp,\lambda;t_1}(\bx)|\hat{U}_1(t_1,t_0)|v_{\bq,\sigma;t_0}(\bx)
 \ra \\ \la
 v_{\bp,\lambda;t_1}(\bx)|\hat{U}_1(t_1,t_0)|u_{\bq,\sigma;t_0}(\bx)
 \ra & \la
 v_{\bp,\lambda;t_1}(\bx)|\hat{U}_1(t_1,t_0)|v_{\bq,\sigma;t_0}(\bx)
 \ra \end{array} \right] \notag \\
& = 
\int \frac{d^3 \bp\prime}{(2 \pi)^3} \sum_{i=1}^{4} 
M_{\bp,\lambda;\bp\prime,i}(\tau_1) \bar{M}_{\bq,\sigma;\bp\prime,i}(\tau_0) 
\label{eq:CE23} \end{align}
 where:
\begin{equation} M_{\bp,\lambda;\bp\prime,i}(\tau)  =  \left[ \begin{array}{c} 
 \la u_{\bp,\lambda;t}(\bx)|\psi_{\bp\prime,i}(\bx,t)\ra \\ \la
 v_{\bp,\lambda;t}(\bx)|\psi_{\bp\prime,i}(\bx,t) \ra \end{array} \right]
 \label{eq:CE24} \end{equation} which is unitary for all time, since
 it describes the overlap between two orthonormal sets of states.

	We can now use (\ref{eq:CE32}) - (\ref{eq:CE35}) and 
	(\ref{eq:CE28}) to calculate $\bM(\tau)$ as
\begin{equation} M_{\bp,\lambda;\bp\prime,i}(\tau) 
 = (2 \pi)^3 \delta(\bp - \bp\prime) \left[ \begin{array}{cc}  A_{p,p_z}(\tau)
  I_2 & -\bar{B}_{p,p_z}(\tau) I_2 \\ B_{p,p_z}(\tau) I_2 & \bar{A}_{p,p_z}(\tau) I_2
  \end{array} \right] \label{eq:CE36}\end{equation} where
\begin{align} A_{p,p_z}(\tau) & = \frac{C}{\sqrt{2 E_{\tau} (E_{\tau} - p_z - a(\tau))}} 
((E_{\tau} - p_z - a(\tau)) h(\tau) + (1 + p^2) f(\tau)) \\
 & = C \sqrt{\frac{E_{\tau} - p_z - a(\tau)}{2 E_{\tau}}} \left(i \frac{d f}{d \tau} + E_{\tau} f\right) \label{eq:CE37}\\ 
B_{p,p_z}(\tau) & = \frac{C}{\sqrt{2
E_{\tau} (E_{\tau} + p_z + a(\tau))}} ((E_{\tau} + p_z + a(\tau)) h(\tau) -
(1 + p^2) f(\tau)) \\ & = C \sqrt{\frac{E_{\tau} + p_z + a(\tau)}{2
E_{\tau}}} \left(i \frac{d f}{d \tau} - E_{\tau} f\right) \label{eq:CE38}\end{align}
 
and $I_2$ is the $2 \times 2$ identity matrix. From (\ref{eq:CE36}) we have
\begin{equation} S_{\bp,\lambda;\bq,\sigma}(\tau_1,\tau_0) 
 = (2 \pi)^3 \delta(\bp - \bq) \left[ \begin{array}{cc}
  \alpha_{p,p_z}(\tau_1,\tau_0) I_2 & \beta_{p,p_z}(\tau_1,\tau_0) I_2 \\
  -\bar{\beta}_{p,p_z}(\tau_1,\tau_0) I_2 &
  \bar{\alpha}_{p,p_z}(\tau_1,\tau_0) I_2 \end{array} \right]
  \label{eq:CE39}\end{equation} where:
\begin{align} \alpha_{p,p_z}(\tau_1,\tau_0) & \equiv A_{p,p_z}(\tau_1) \bar{A}_{p,p_z}(\tau_0) + 
\bar{B}_{p,p_z}(\tau_1) B_{p,p_z}(\tau_0) \label{eq:CE40}\\
 \beta_{p,p_z}(\tau_1,\tau_0) & \equiv A_{p,p_z}(\tau_1) \bar{B}_{p,p_z}(\tau_0)  - \bar{B}_{p,p_z}(\tau_1)
 A_{p,p_z}(\tau_0) \label{eq:CE41}\end{align}

	We can now use the formulae in Section 2.3 to extract
	from (\ref{eq:CE39}) whatever information we wish about this
	system of electrons evolving in a background $\bE = (0,0,\dot{A}(t))$, 
expressed in terms of an arbitrary solution to equation
	(\ref{eq:CE9.1}). 
	For example, suppose we start the system at some initial time $t_0$ in the vacuum state $| {\rm vac}_{t_0} \ra$,
 and let this state evolve until some later time $t_1$. 
Then, from (\ref{eq:ntot1}), the expected number of particles per unit volume in this evolved vacuum 
(not including antiparticles) 
with spin $\lambda$ and physical (gauge-invariant) 3-momentum $\bp\prime = (p^1,p^2,p^3 + e A(t_1))$ in a range $d^3 \bp$ 
 about $\bp\prime$ is given by:
\begin{equation} \frac{N_{\bp,\lambda,t_0}(t_1) d^3 \bp}{V} = |\beta_{p,p_z}(\tau_1,\tau_0)|^2 d^3 \bp \end{equation}
(where $V = (2 \pi)^3 \delta(\bbo)$). 
The the same number of antiparticles is also created with physical momenta $-\bp\prime$. 
The total number of particles per unit volume in this evolved vacuum (including antiparticles) is therefore
\begin{equation} 
\begin{array}{rcl}
\displaystyle\frac{N_{{\rm vac},t_0}(t_1)}{V} &=& \displaystyle 2 \sum_{\lambda} \int \frac{d^3 \bp}{(2 \pi)^3}
 |\beta_{p,p_z}(\tau_1,\tau_0)|^2\\
\\
& =&
 \displaystyle\frac{m^3}{\pi^2} \int_{-\infty}^{\infty} \ {\rm d} p_z  \int_{0}^{\infty} {\rm d}p\ p\ |\beta_{p,p_z}(\tau_1,\tau_0)|^2 
\end{array}
\end{equation}
	The probability that this state is vacuum at time $t_1$ is determined from (\ref{Svacvac1}),
using $\det(e^A) = e^{{\rm Trace}(A)}$, to give
\begin{equation} \clp_{|{\rm vac}_{t_0} \ra \rightarrow |{\rm vac}_{t_1} \ra} = \exp\left(\frac{m^3 V}{2 \pi^2}
 \int_{-\infty}^{\infty} \ {\rm d} p_z\  \int_{0}^{\infty} {\rm d}p\ p\  \log(1
 - |\beta_{p,p_z}(\tau_1,\tau_0)|^2)\right) \label{eq:Svacvac2} \end{equation}
From (\ref{eq:CE32}) - (\ref{eq:CE35}), the total current density in this evolved vacuum is  
(\ref{eq:SSB40.1}) and $\hat{J}_{1,k} = e \bar{\sig}_{k}$) 
\begin{align} \frac{J_{{\rm vac},3,t_0}(t_1)}{V} & = \frac{e m^3}{\pi^2} \int_{-\infty}^{\infty}{\rm d} p_z 
\int_{0}^{\infty} \frac{ {\rm d} p\ p \ }{E_{\tau_1}} \{ (p_z + a(\tau_1)) |\beta_{p,p_z}(\tau_1,\tau_0)|^2 \notag \\
 & \hs{1} - \sqrt{1 + p^2}\ \Re(\beta_{p,p_z}(\tau_1,\tau_0) 
 \alpha_{p,p_z}(\tau_1,\tau_0)) \} 
 \label{eq:diss6.8} \end{align}
	where $J_{{\rm vac},1,t_0}(t_1) = 0 = J_{{\rm vac},2,t_0}(t_1)$ by symmetry. This expression will 
generally contain a logarithmic divergence proportional to $\ddot{a}(\tau)$, which 
must be isolated and absorbed into a renormalisation of the coupling constant $e$. In 
Section 4, $\ddot{a}(\tau)=0$ so that no renormalisation is necessary. For most of 
Section 5 we use a cut-off in the integrals  such that the 
cut-off-dependent renormalisation term is small, and can be ignored.

\subsubsection*{Non-Vacuum Initial Conditions}

	Consider an initial state
	at time $t_0$ consisting of N particles having momenta $\bp_1,
	\dots \bp_N$ and spins $\lambda_1, \dots \lambda_N$. 
From equations (\ref{eq:SSB40.12}) and (\ref{eq:CE39}), the expected
	number of out particles at some later time $t_1$ is:
\begin{equation} N_{{\rm out},t_0}(t_1) = \sum_{i=1}^{N} (|\alpha_{p_i,p_{z,i}}(\tau_1,\tau_0)|^2
 - |\beta_{p_i,p_{z,i}}(\tau_1,\tau_0)|^2) + N_{{\rm vac},t_0}(t_1) \end{equation}
	which is less than $N_{{\rm vac},t_0}(t_1) + N$ due to Pauli
	exclusion (recall that  
$$|\alpha_{p,p_{z}}(\tau_1,\tau_0)|^2 + |\beta_{p,p_{z}}(\tau_1,\tau_0)|^2 = 1).$$ 
The probability that the out state contains
	exactly N particles is:
\begin{equation} \clp_{N \rightarrow N,t_0}(t_1) = 
\frac{\clp_{|{\rm {\rm vac}}_{t_0} \ra \rightarrow |{\rm vac}_{t_1} \ra}}{\prod_{i=1}^{N}
|\alpha_{p_i,p_{z,i}}(\tau_1,\tau_0)|^2} \end{equation} 
This expression highlights the well-known fact that particle creation is less 
likely in a state that already contains particles. The 
current density of this evolved state is:
\begin{equation} J_{{\rm out},t_0}(t_1) = \sum_{i=1}^{N} J(u_{\bp_i,\lambda_i,t_0}(\bx,t_1)) + 
J_{{\rm vac},t_0}(t_1) \end{equation}
	where from (\ref{eq:CE39})
\begin{equation} u_{\bp,\lambda,t_0}(\bx,t_1) = 
\alpha_{p,p_z}(\tau_1,\tau_0) u_{\bp,\lambda,t_1}(\bx) - 
\bar{\beta}_{p,p_z}(\tau_1,\tau_0) v_{\bp,\lambda,t_1}(\bx) \notag \end{equation}
	so that
\begin{align} J(u_{\bp,\lambda,t_0}(\bx,t_1)) & = \frac{e}{E_{\tau_1}} 
\{(1 - 2 |\beta_{p,p_z}(\tau_1,\tau_0)|^2) (p_z + a(\tau_1)) \notag \\
& \hs{2} + 2 \sqrt{1 + p^2} \Re(\alpha_{p,p_z}(\tau_1,\tau_0)\beta_{p,p_z}(\tau_1,\tau_0)) \} \label{eq:jpart} \end{align}

\newpage

\subsubsection*{Simplification}

	Although $\bM(t)$ depends on the initial
	conditions chosen for $f(\tau)$ in (\ref{eq:CE9.1}), the
	elements of $\bS(t_1,t_0)$ do not depend on them. 
We can, however, by an 
appropriate choice of initial conditions,  simplify the 
expression for $\bS(t_1,t_0)$.
Impose temporarily the conditions  $f(\tau_0) = 1$ and $h(\tau_0) = 
E_{\tau_0} - p_z - a(\tau_0)$. (This second condition is 
equivalent  $i \frac{d f}{d \tau}|_{\tau=\tau_0} = 
E_{\tau_0} f(\tau_0)$,  revealing a connection between our
particle definition  and that motivated by 
 `instantaneous frequency'.) These 
conditions yield $A_{p,p_z}(\tau_0) = 1$ and 
$B_{p,p_z}(\tau_0) = 0$. Hence we can identify 
$\psi_{\bp,i}(t) = u_{\bp,i,t_0}(t)$ and $\psi_{\bp,i+2}(t) = 
v_{\bp,i,t_0}(t)$ for $i = 1,2$, so that 
$\alpha_{p,p_z}(\tau_1,\tau_0) = A_{p,p_z}(\tau_1)$ and $\beta_{p,p_z}(\tau_1,\tau_0) = - \bar{B}_{p,p_z}(\tau_1)$ in this case.
The current density in the evolved vacuum then simplifies to:
\begin{equation}
\begin{array}{rcl}
 J_{{\rm vac},3,t_0}(t_1) &=& \displaystyle\frac{e m^3}{2
 \pi^2} \int_{-\infty}^{\infty} {\rm d}p_z\ \int_{0}^{\infty} 
 {\rm d}p\ p \ \left\{ \frac{E_{\tau_1} + p_z + a(\tau_1))}{E_{\tau_1}}\right.\\
 &-& \displaystyle\left.\frac{E_{\tau_0} + p_z + a(\tau_0))}{E_{\tau_0}}|f(\tau_1)|^2 \right\}\\
\end{array}
 \end{equation}
	and $J(u_{\bp_i,\lambda_i,t_0}(\bx,t_1))$ becomes: 
\begin{equation} J(u_{\bp_i,\lambda_i,t_0}(\bx,t_1)) = e \left\{ \frac{E_{\tau_0} + p_z + a(\tau_0))}{E_{\tau_0}}
|f(\tau_1)|^2 - 1 \right\} \end{equation}
	These simplifications will be used in Section 5.

\section{CONSTANT ELECTRIC FIELD}

	For constant electric field we have $a(\tau) = \frac{e E}{m^2} \tau \equiv a_0 \tau$. 
Since $e$ is negative,  positive $a_0$ corresponds to an 
electric field in the $-z$ direction. 
By writing $a_0 = -\frac{E}{E_c}$ we identify the `natural' electric field strength $E_c = \frac{m^2 c^3}{|e| \hbar} \approx 1.3 \times 10^{18}
	$~V~m$^{-1}$ (where we have substituted the mass and absolute charge of the
	electron). This  corresponds to an electromagnetic
	energy density of approximately 137 electron rest mass energies
	per Compton volume $\left(\frac{\hbar}{mc}\right)^3$,
and is too large to be produced on macroscopic
	scales in the laboratory,  although astrophysical examples have been considered~\cite{Ast1,Ast3}. 
Also, fields of this strength would be present on a
	microscopic scale in
	superheavy nuclei with charges greater than $137 |e|$.

	Equation (\ref{eq:CE9.1}) can now be written as:
\begin{equation} \frac{d^2 f}{d \tau^2} + (1 + p^2 + (p_z + a_0 \tau)^2 + i a_0) f = 0 \end{equation}
	Comparison with equation 8.2(1) of \cite{Bat} reveals that solutions of
	this equation are linear combinations of $D_{\nu_p -1}(\pm \frac{1-i}{\sqrt{a_0}} (p_z + a_0 \tau))$
	and $D_{-\nu_p}(\pm \frac{1+i}{\sqrt{a_0}} (p_z + a_0 \tau))$, where $\nu_p \equiv \frac{i (1 + p^2)}{2 a_0}$
 and $D_{\nu}(z)$ are parabolic cylinder
	functions. We choose $ f(\tau) \equiv D_{\nu_p
	-1}(\frac{i - 1}{\sqrt{a_0}}(p_z + a_0 \tau)) $, which gives (after exploiting properties of the
	parabolic cylinder functions) $$h(\tau) = (1+i) D_{\nu_p}\left(\frac{i - 1}{\sqrt{a_0}}(p_z + a_0 \tau)\right) \mbox{\ and\ } 
C = \frac{e^{\frac{- \pi (1 + p^2)}{8 a_0}}}{\sqrt{2}}$$. 

	The first  point  about these expressions is that $f(\tau)$ (and hence $A_{p,p_z}(\tau)$ and 
$B_{p,p_z}(\tau)$)  depends only on $p_z$ and $\tau$ through the combination $p_z + a_0 \tau$, 
which is the physical $z$-momentum (in units of mass) at time $\tau$ of the state labelled by the subscript $\bp$. 
This allows us to write $A_{p,p_z}(\tau_1) = A_{p,p^{\rm{out}}_z}(0)$ where $p^{\rm{out}}_z \equiv p_z + a_0 \tau_1$, 
so that the evolution matrix can be written as:
\begin{equation} S_{\bp,\lambda;\bq,\sigma}(\tau_1,\tau_0) 
 = (2 \pi)^3 \delta(\bp - \bq) \left[ \begin{array}{cc}
  \alpha_{p,p^{\rm{out}}_z}(0,-\tau) I_2 & \beta_{p,p^{\rm{out}}_z}(0,- \tau) I_2 \\
  -\bar{\beta}_{p,p^{\rm{out}}_z}(0,-\tau) I_2 &
  \bar{\alpha}_{p,p^{\rm{out}}_z}(0,-\tau) I_2 \end{array} \right]
  \notag\end{equation}
	and depends only on the evolution time $\tau \equiv \tau_1 - \tau_0$. 
This follows from the time-translation invariance of the constant electric field. 

We now recognise that the physically relevant infinite-time
	scattering matrix is:
\begin{equation}  \lim_{\tau \rightarrow \infty} 
S_{\bp,\lambda;\bq,\sigma}(0,-\tau) = (2 \pi)^3 \delta(\bp - \bq) \left[ \begin{array}{cc}  A_{p,p^{\rm{out}}_z}(0)
  I_2 & -\bar{B}_{p,p^{\rm{out}}_z}(0) I_2 \\ B_{p,p^{\rm{out}}_z}(0) I_2 & \bar{A}_{p,p^{\rm{out}}_z}(0) I_2
  \end{array} \right] \end{equation}
where we have used asymptotic properties of $A_{p,p^{\rm{out}}_z}(\tau)$ and $B_{p,p^{\rm{out}}_z}(\tau)$ 
 and discarded unimportant phase factors. 
The 
expected number of particles per unit volume in the infinitely evolved vacuum is now
\begin{align} \frac{N_{\bp,\lambda}(\infty)}{V} & = |B_{p,p^{\rm{out}}_z}(0)|^2 \\
& = \frac{e^{\frac{- \pi (1 + p^2)}{4 a_0}}}{4 E_{0} (E_{0} + p^{\rm{out}}_z)} 
\left| (E_{0} + p^{\rm{out}}_z) (1+i) D_{\frac{i (1 + p^2)}{2 a_0}}\left(\frac{i - 1}{\sqrt{a_0}} p^{\rm{out}}_z\right) \notag\right. \\
 & \left.\hs{2} - 
(1 + p^2) D_{\frac{i (1 + p^2)}{2 a_0} -1}\left(\frac{i - 1}{\sqrt{a_0}} p^{\rm{out}}_z\right) \right\vert^2
\label{eq:CE46}\end{align}

\begin{figure}[htb]
\begin{minipage}[b]{5.5cm}
\epsfig{figure=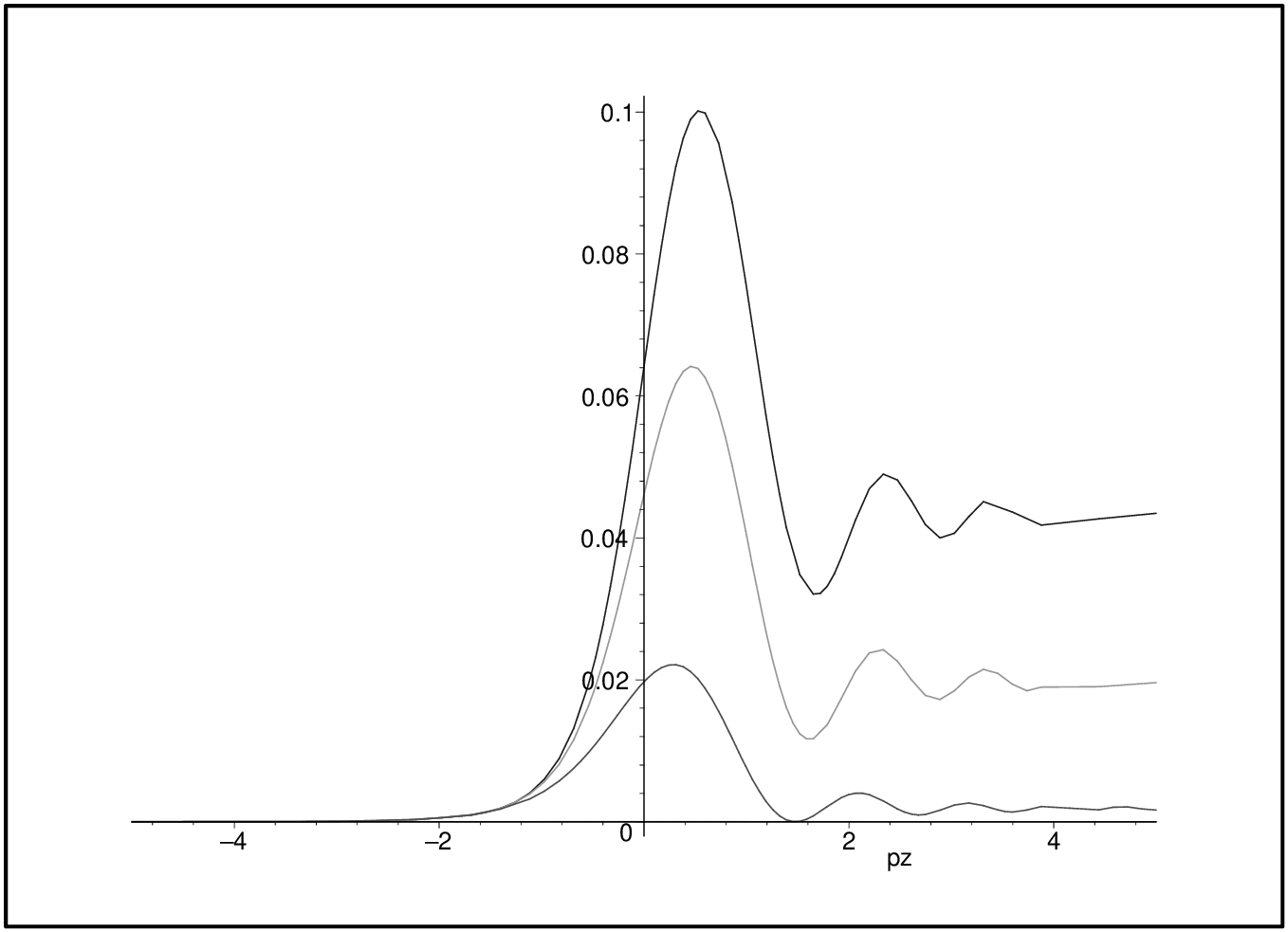,width=5.5cm,angle=-90}
\vs{-1}

{\footnotesize {\bf FIG. 1(A).} Number density in an infinitely evolved vacuum as a
function of $p^{\rm{out}}_z$, for $a_0=1$ and $p =
0$ (top curve), $0.5$ (middle curve), and 1 (bottom curve).}
\vs{1.35}
\end{minipage}\hs{.5}\begin{minipage}[b]{5.5cm}
\epsfig{figure=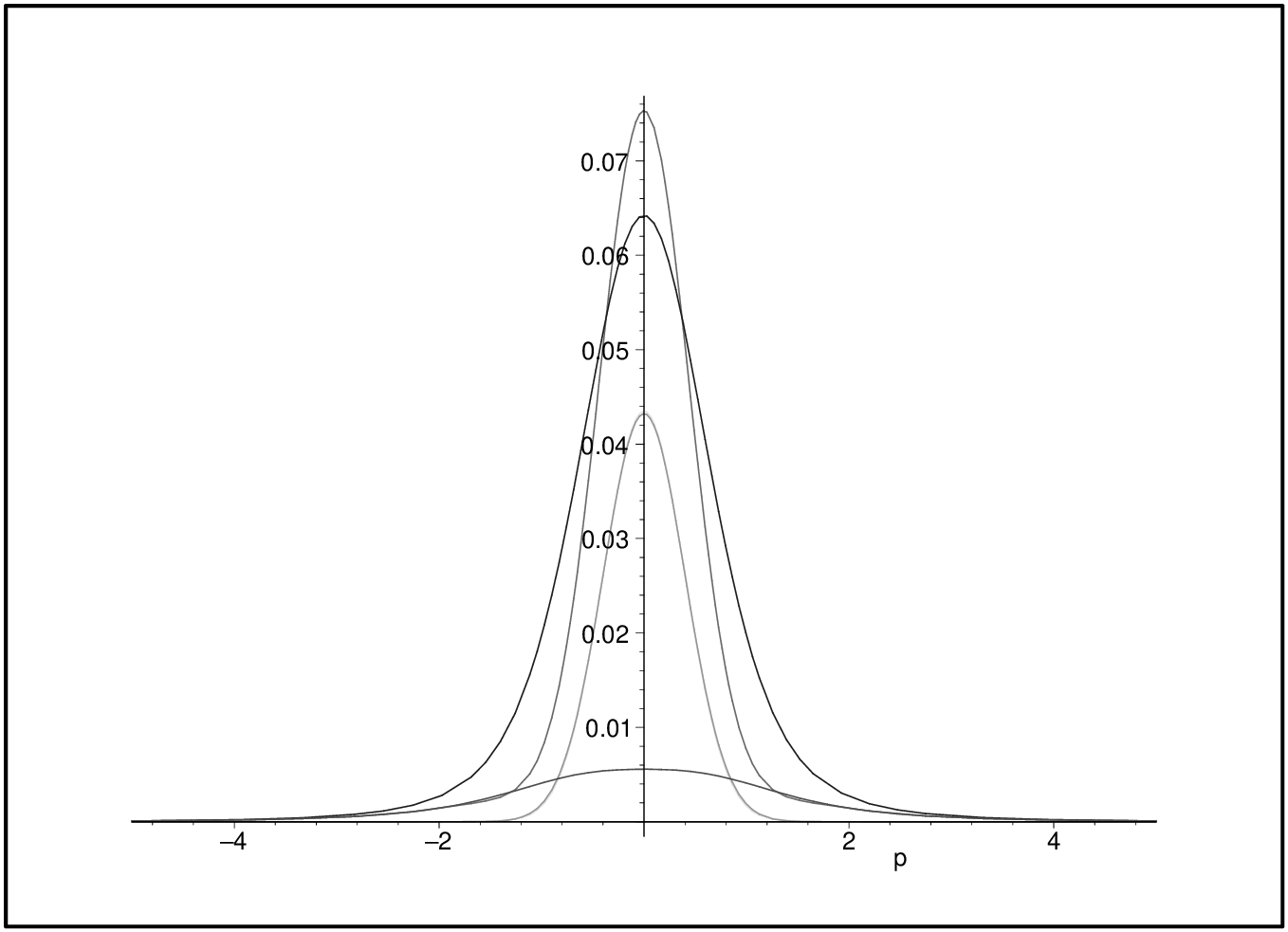,width=5.5cm,angle=-90}
\vs{-1}

{\footnotesize {\bf FIG. 1(B).} Number density in an infinitely evolved vacuum as a
function of $p$ for $a_0=1$ and $p^{\rm{out}}_z = -1$ (bottom curve), $0$ (outermost curve), $1$ (top curve) and $5$ (innermost curve). A plot of $e^{\frac{-\pi (1 + p^2)}{a_0}}$ is also 
included, which coincides almost exactly with the curve for
$p^{\rm{out}}_z = 5$.}
\end{minipage}
\end{figure}

Curves of this expression for various values of $a_0$ are
	shown in Figure 1. We see that $N_{\bp,\lambda}(\infty)$ is
	effectively zero for negative $p^{\rm{out}}_z$, 
and approaches
	$e^{\frac{- \pi (1 + p^2)}{a_0}}$ for large $p^{\rm{out}}_z$. 
in agreement with~\cite{Mott2}.
The result is plausible since any particle, once created, will be accelerated
	to ever increasing $p^{\rm{out}}_z$ by the electric field. The number
	density of created antiparticles is identical, but with the
	sign of $p^{\rm{out}}_z$ reversed.

	A common misconception is that $\frac{N_{\bp,\lambda}(\infty)}{V} = e^{\frac{- \pi (1 + p^2)}{a_0}}$ 
independent of $p^{\rm{out}}_z$~\cite{GaGi}. 
This misconception stems from taking $\tau_1 \rightarrow \infty$ and $\tau_0 \rightarrow - \infty$ when 
defining `in' and `out' modes,  inadvertently taking the limit as $p^{\rm{out}}_z \rightarrow \infty$. 
A more relevant result,  which allows Schwinger's formula to be derived
	in the large time limit, is described below.

\begin{figure}[hbt]
\hs{.7}\epsfig{figure=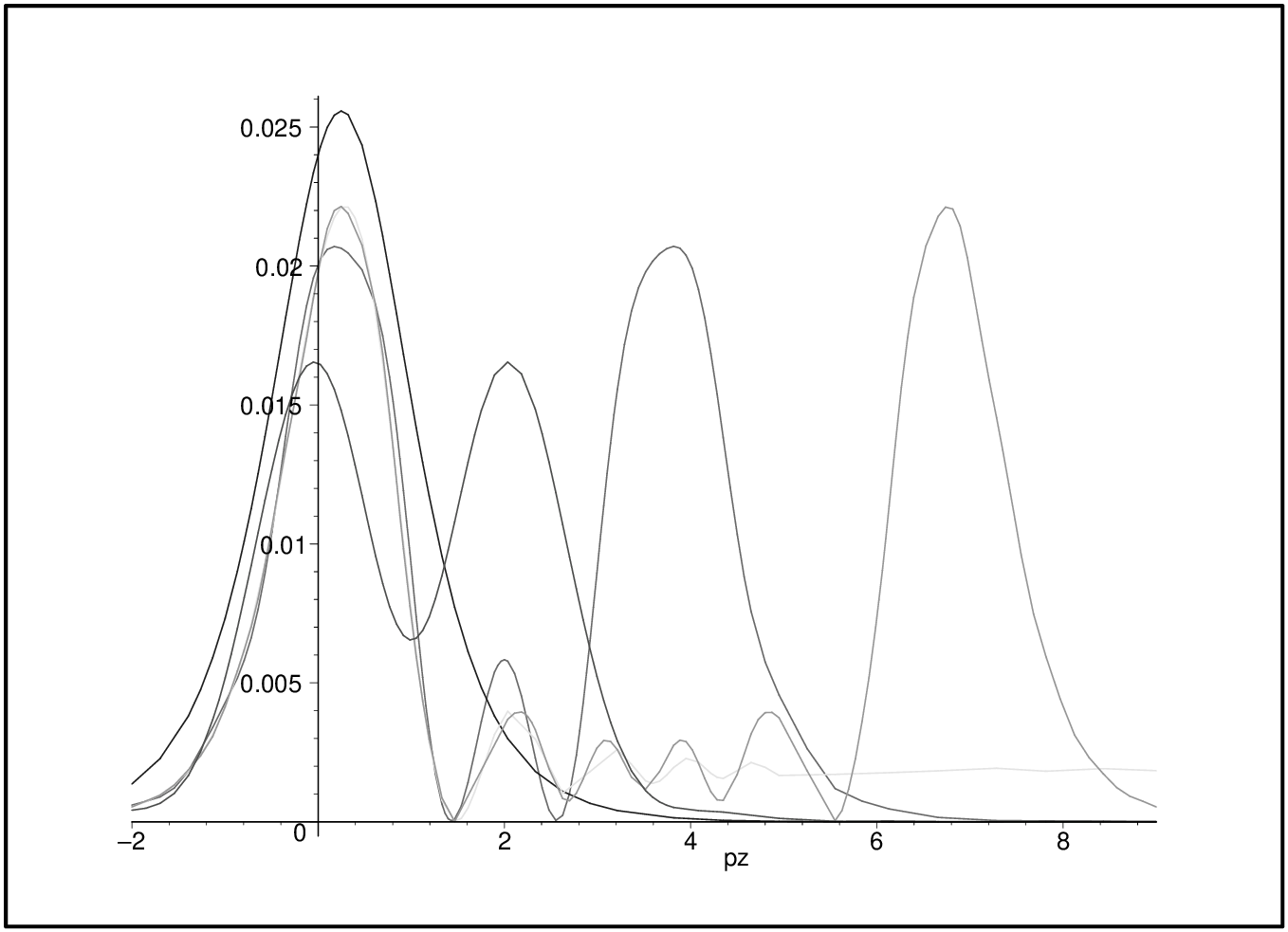,width=10cm,angle=-90}
\vs{-2}

{\footnotesize {\bf FIG. 2.} Number density in the evolved vacuum, as a function
of $p^{\rm{out}}_z$, for $a_0=1$, $p = 1$, and $\tau =  7$
(right), $4$, $2$ and $0.5$ (single peak). A graph for $\tau = \infty$
is included, which is seen to fit the left peak of the  $\tau = 7$
line and becomes constant at large $p^{\rm{out}}_z$.}
\end{figure} 

 Figure 2 shows the expected number density $N_{\bp,\lambda}(\tau)$ in the evolved vacuum for $a_0 = 1$ 
and $p=1$, as a function of
	$p^{\rm{out}}_z$, after an evolution time $\tau = 7$ (right), 
$4$, $2$ and $0.5$ 
(single peak). A graph
	for $\tau = \infty$ is also included, which is seen to fit
	the left peak of the  $\tau = 7$ line, and
	becomes constant at large $p^{\rm{out}}_z$. The finite time behaviour
	demonstrated here agrees with that presented in \cite{Hal}. 
Figure~2 has an interesting
	physical interpretation. 
At early times, particles are created mainly with
	$p^{\rm{out}}_z$ close to zero (Figure 1(B) shows that small  $p$ is
	also favoured), as  expected, since these pairs
	require the least creation energy. At later times, the right
	peak represents the same early-time particles (which have 
	been accelerated by the electric field), and the left
	peak represents recently created pairs, again at low
	momenta. Figure 3(A) shows
	$\frac{N_{{\rm vac}}(\tau)}{V m^3}$ for $a_0 =
	\half$, $1$ and $2$.	

\begin{figure}[hbt]
\begin{minipage}[b]{5.5cm}
\vs{.5}
\epsfig{figure=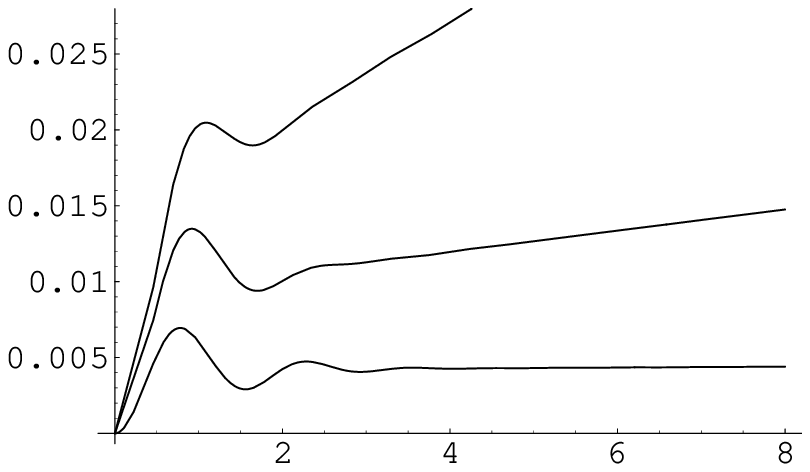,width=6cm,angle=0}

{\footnotesize {\bf FIG. 3(A).} Total number of particles in a Compton volume $m^{-3}$ of
the evolved vacuum as a function of $\tau$ for $a_0 = \half$ 
(bottom curve), $1$ and $2$ (top curve).}
\vs{.5}
\end{minipage}\hs{.5}\begin{minipage}[b]{5.5cm}
\epsfig{figure=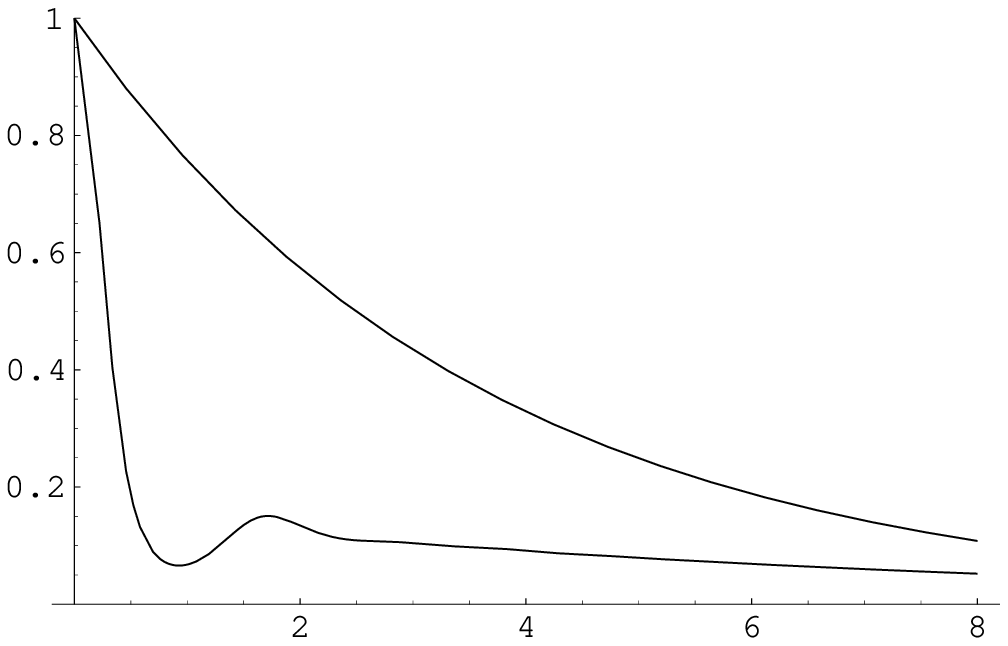,width=5.5cm,angle=0}

{\footnotesize {\bf FIG. 3(B).} Vacuum-vacuum transition
	probability for $a_0 = 1$ and $V = 40 \pi^2 m^{-3}$. The 
initial burst of pair creation reduces the exact transition probability 
(bottom curve) below Schwinger's prediction 
(top curve) at early times.}
\end{minipage}
\end{figure}

	From Figure 2 we see that for. large $\tau$,
	$N_{\bp,\lambda}(\tau)$ can be well approximated by
\begin{equation} N_{\bp,\lambda}(\tau) \approx e^{\frac{- \pi (1 + p^2)}{a_0}} (\theta(p^{\rm{out}}_z) - 
\theta(a_0 \tau - p^{\rm{out}}_z)) \label{eq:CE100}\end{equation}
where $\theta(p^{\rm{out}}_z)$ is the Heavyside step function. This allows us to
approximate:
\begin{equation} \int N_{\bp,\lambda}(\tau) d p^{\rm{out}}_z \approx a_0 \tau 
e^{\frac{- \pi (1 + p^2)}{a_0}} \end{equation}
	from which we obtain the total number of pairs created:
\begin{equation} N_{TOT}(\tau) \approx \frac{(e E)^2 V t}{4 \pi^3} 
e^{\frac{- \pi m^2}{|e E|}} \mbox{ for large } t \equiv \frac{\tau}{m} \end{equation}
	This analysis correctly predicts the final gradient of the graphs in
	Figure 3(A), and is the result most commonly referred to in the
	literature~\cite{Hal,GaGi,GMR}. However, we see in
	Figure 3(A) that for $|E| \ll E_c$ the dominant contribution to
	the total number of produced pairs comes not from this final gradient, but from the initial
	burst of pair creation. Even for $|E| = \half E_c$,  
approximately $ 0.44$ particles are created in a volume $\lambda_c^3$ in the first 
$10^{-20}$ seconds, with only $\sim 0.005$ particles 
every $10^{-20}$ seconds thereafter. For $|E| = \frac{E_c}{10}$ approximately 
$0.014$ particles are created in a volume $\lambda_c^3$ in the first 
$2.5 \times 10^{-20}$ seconds, with only $8.6 \times 10^{-15}$ 
particles created every $2.5 \times 10^{-20}$ seconds thereafter. This 
difference becomes exponentially more pronounced as the ratio $|\frac{E_c}{E}|$ 
increases.

	Similarly, substitution of (\ref{eq:CE100}) into (\ref{eq:Svacvac2}) together with the relation
$$\log(1 - e^{\frac{- \pi \lambda}{a_0}}) = - \sum_{n=1}^{\infty} \frac{e^{\frac{- \pi n
\lambda}{a_0}}}{n}$$, gives
\begin{equation} \clp_{{\rm vac} \rightarrow {\rm vac}}(\tau) \approx 
\exp\left[\frac{- (e E)^2 V t}{4 \pi^3} \sum_{n=1}^{\infty} 
\frac{1}{n^2} e^{\frac{- \pi n m^2}{e E}}\right] \mbox{ for } \tau \mbox{ large. } 
\label{eq:Schw} \end{equation}
	in agreement with Schwinger's original calculation~\cite{Schw}
	(also in \cite{IZ}, \cite{GaGi}, \cite{Hal}). In Figure 3(B) 
we show the exact vacuum-vacuum transition
	probability, and the large time approximation of (\ref{eq:Schw}) 
for $a_0 = 1$ (we have taken $V = 40 \pi^2$~m$^{-3}$ for convenience).

\subsection*{Vacuum Current, and Induced Electric Field}

Figure 4 shows the integrand in $J_{{\rm vac},3,t_0}(t_1)$ from 
(\ref{eq:diss6.8}) as a
	function of $p^{\rm{out}}_z$ for various evolution times, and 
	includes  a graph of the $|\beta|^2$ contribution.

\begin{figure}[hbt]
\begin{minipage}[b]{5.5cm}
\epsfig{figure=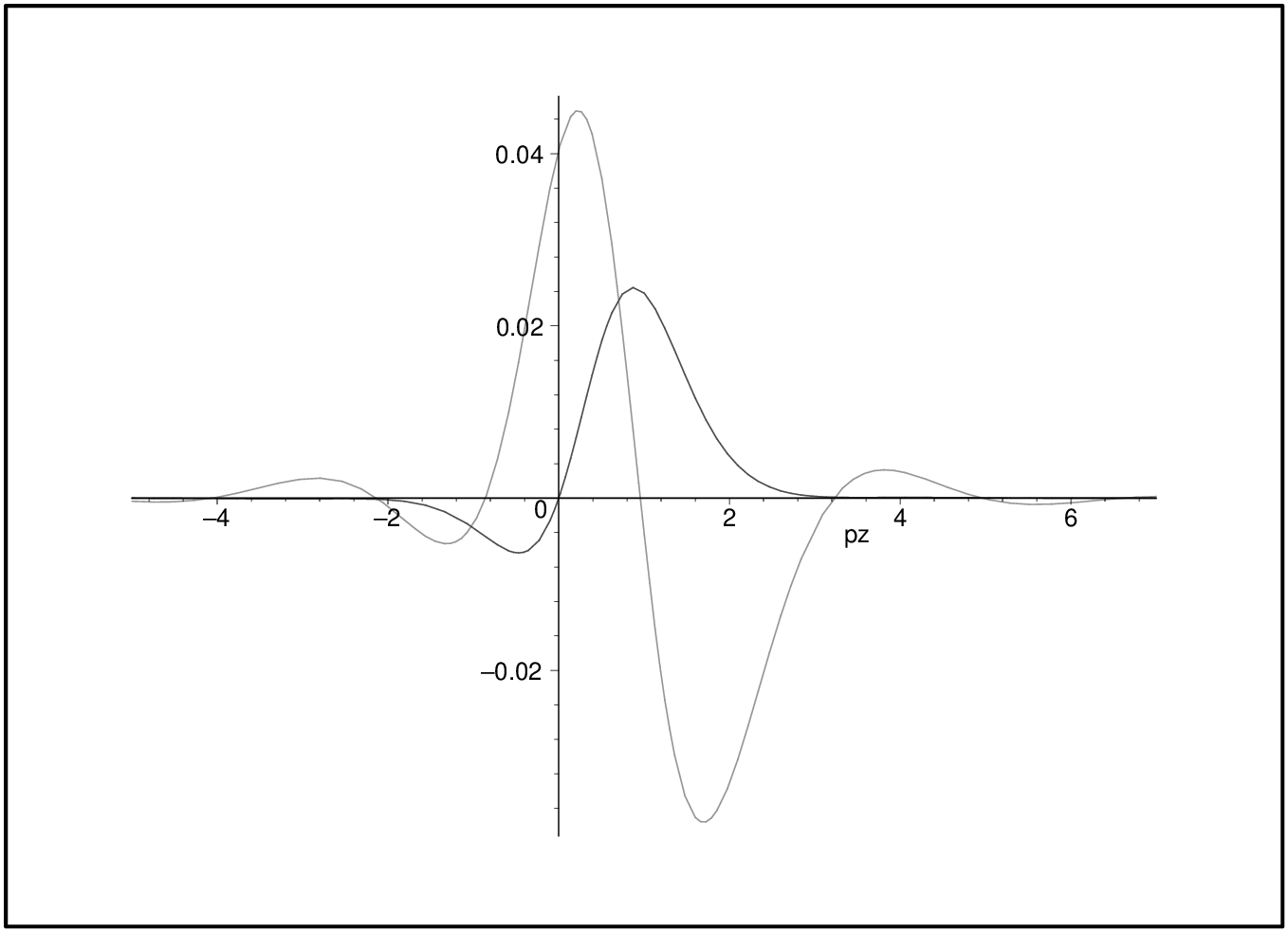,width=5.5cm,angle=-90}

\vs{-1}
{\footnotesize {\bf (A).} $\triangle t = \sqrt{e E}$}
\end{minipage}\hs{.5}\begin{minipage}[b]{5.5cm}
\epsfig{figure=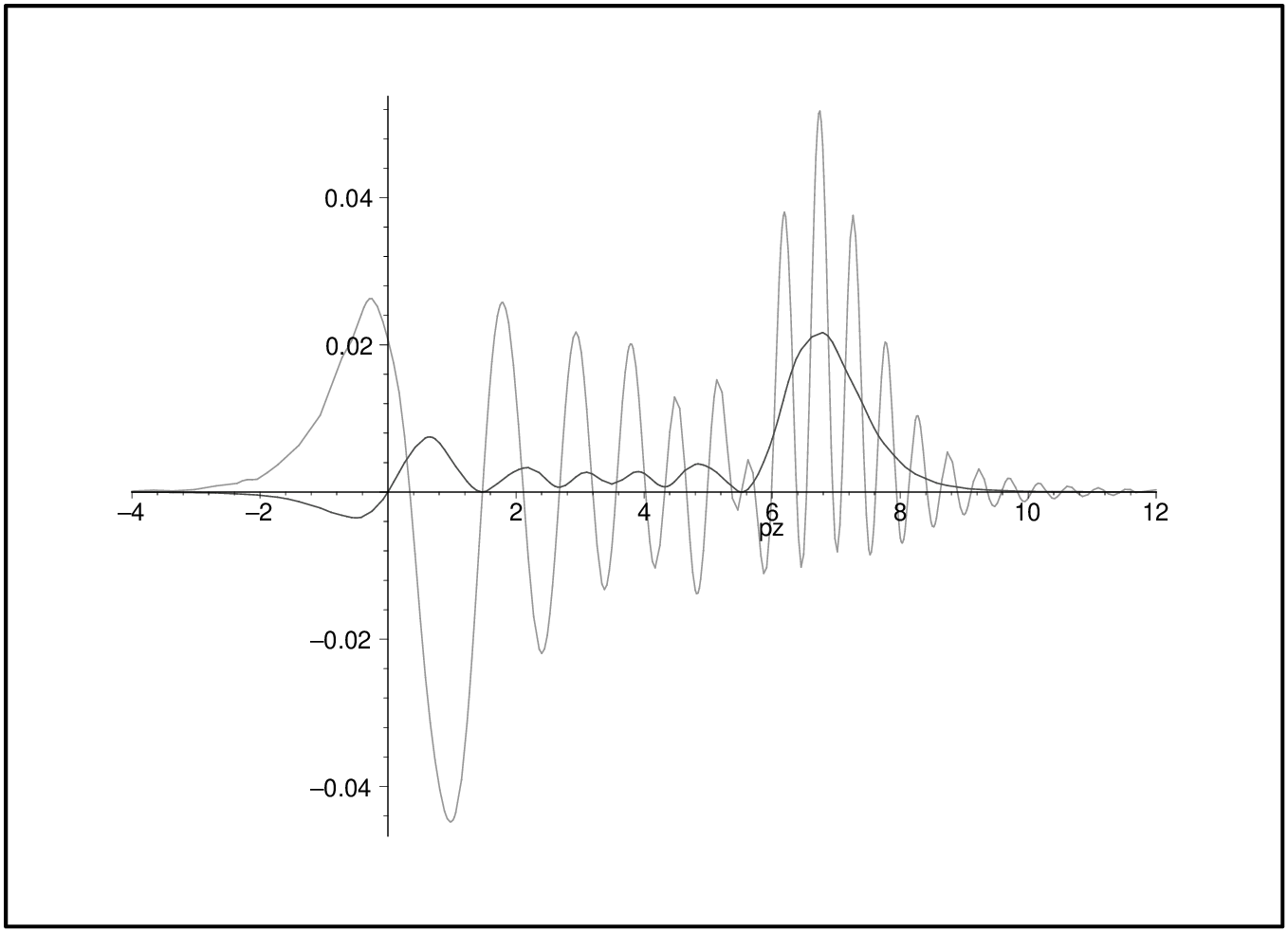,width=5.5cm,angle=-90}

\vs{-1}
{\footnotesize {\bf (B).} $\triangle t = 7 \sqrt{e E}$}
\end{minipage}

\hspace{3cm}\begin{minipage}[b]{5.5cm}
\epsfig{figure=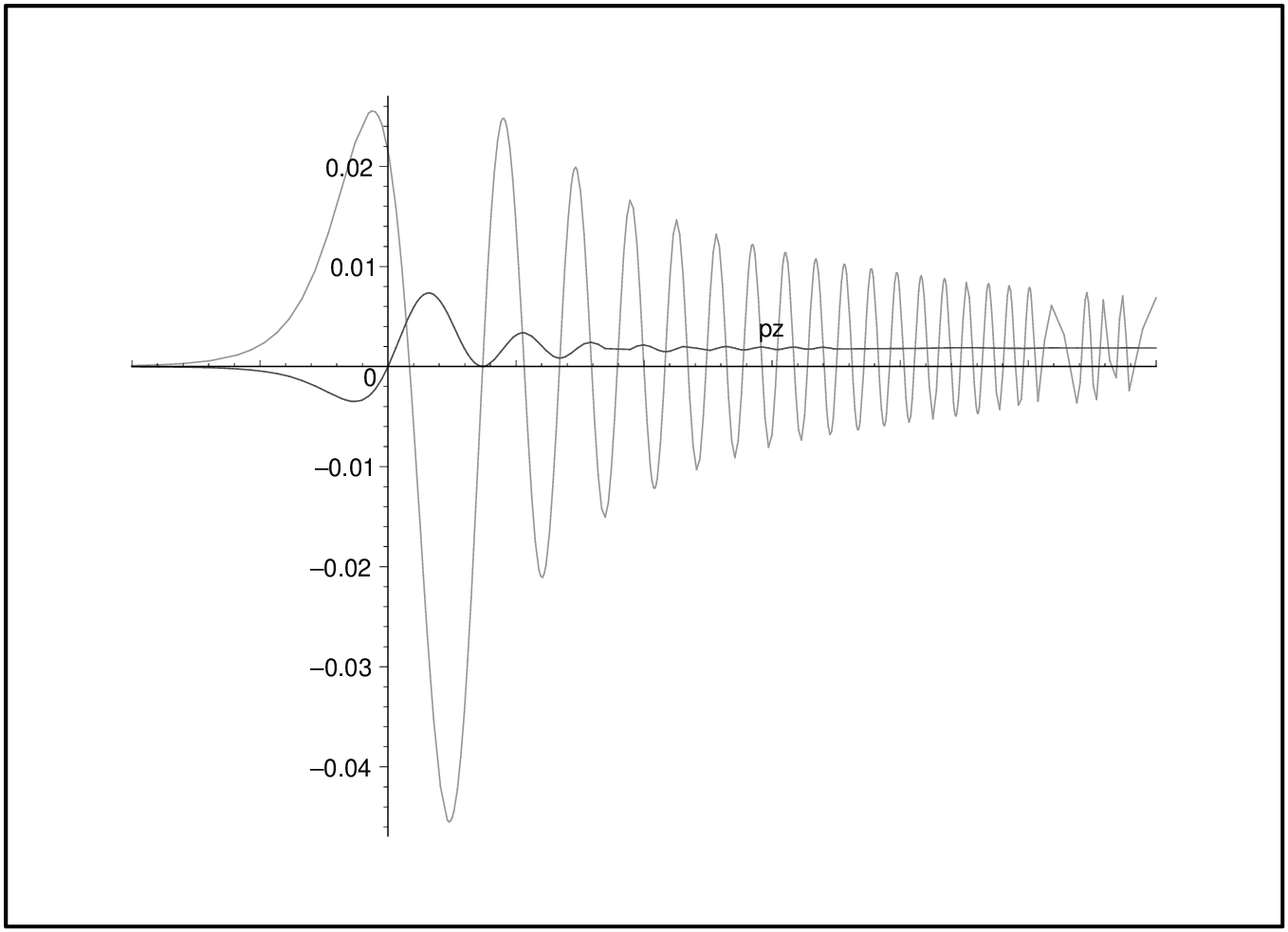,width=5.5cm,angle=-90}

\vs{-1}
{\footnotesize {\bf (C).} $\triangle t = \infty$}
\end{minipage}

{\footnotesize {\bf FIG. 4.} Current density in an evolved vacuum
after time $\triangle t = \sqrt{e E}$ (A), $7 \sqrt{e E}$ (B) and $\infty$ (C), as a function of $p_z$, for $a_0 = 1$ and
$p = 1$. The $|\beta|^2$ 
contribution is also shown (darker curve). In (B) and (C) this term  provides the dominant contribution, with the 
$\Re(\beta \alpha)$ term producing
rapid oscillations about it.}
\end{figure}

 We recognise that at late times the dominant
	contribution to the integral comes from the $|\beta|^2$ term;
	 the other term generates rapid oscillations about
	this contribution. The $|\beta|^2$ term is the 
 `classical' contribution of the form `sum
	over created particles of the current of each particle', while
	the oscillatory $\Re(\beta\alpha)$ term represents quantum interference
	effects. These oscillations are on a time scale $t_{qu} \sim
	\frac{\lambda_c}{2 \pi c} = \frac{\hbar}{m}$ and a length
	scale of $\triangle z \sim \lambda_c$, as are the smaller, more strongly 
damped oscillations already present in the $|\beta|^2$ term. 
Oscillations on
	the time scale of $t_{qu}$ can be identified with those
	mentioned in \cite{Mott2} and \cite{Mott1}
as significant in quantum decoherence and
	effective dissipation processes. 
 Figure 5 shows the
	total vacuum current density as a function of $\tau$, together
	with the $|\beta|^2$ contribution, and the approximation that
	results from  ignoring the $\Re(\beta\alpha)$ terms
	and using (\ref{eq:CE100}).

\begin{figure}[hbt]
\hspace{2cm} \epsfig{figure=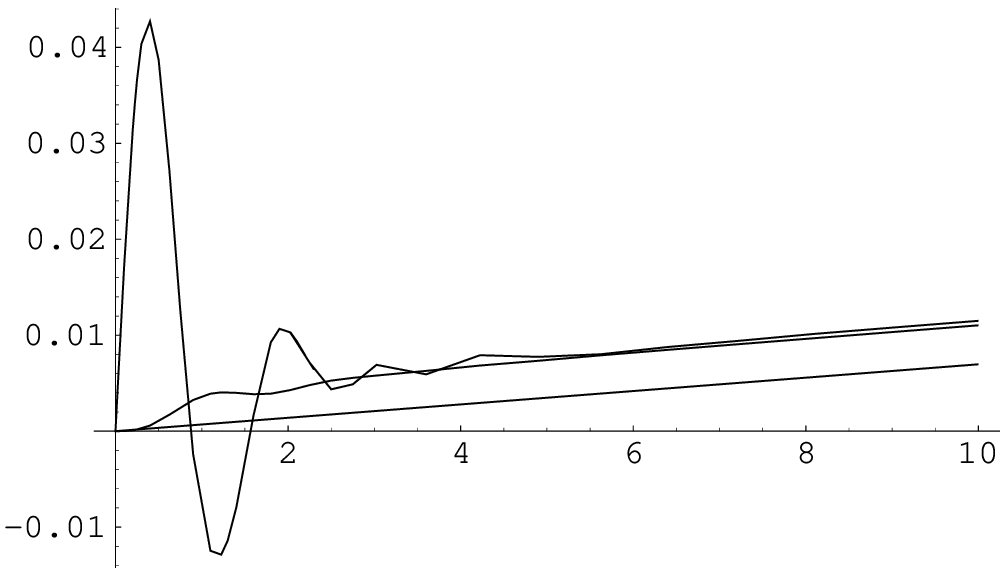,width=8cm}

{\footnotesize {\bf FIG. 5.} Total current in a Compton volume $m^{-3}$  of an 
evolved vacuum as a 
function of $\tau$. The $|\beta|^2$
contribution is also shown,  providing an accurate approximation 
except at very early times. The late time approximation of 
(\ref{eq:Japp}) is also included, which has correct final gradient 
but misses the initial `burst' of creation, as  
with the total number of created particles.}
\label{Jtottau}
\end{figure}

  The $|\beta|^2$ contribution 
shows good agreement with Figure 5 of Kluger, Mottola et al.~\cite{Mott2} 
(though they use a different scale). We also see that, although 
the `semiclassical' $|\beta|^2$ term dominates after the first $10^{-20}$ 
seconds, the `quantum interference' term is significantly more important in 
this first instant. Finally, although we have 
$J_{1,{\rm vac}}(\tau) = 0 = J_{2,{\rm vac}}(\tau)$ by symmetry,  
the integrands of  $J_{1,{\rm vac}}(\tau)$ and $J_{2,{\rm vac}}(\tau)$ still display 
rapid oscillatory dependence on $p^{\rm{out}}_z$.

 The large-time approximation of $J_{3,{\rm vac}}(\tau)$, obtained by ignoring the $\Re(\alpha \beta)$ term and using (\ref{eq:CE100}), is:
\begin{equation} J_{3,{\rm vac}}(\tau) \approx \frac{e (e E)^2 \triangle t}{2 \pi^3} 
e^{\frac{- \pi m^2}{e E}} \label{eq:Japp}\end{equation}
	This would generate an induced electric field in the positive
	z-direction:
\begin{equation} E_{ind}(\tau) \approx \frac{e (e E)^2 (\triangle t)^2}{4 \pi^3} 
e^{\frac{- \pi m^2}{e E}} \end{equation}
	We have not taken back-reaction into account, so these
	results can only be considered realistic at times when
	$E_{ind} \ll |E|$,
 implying $\triangle t \ll
	\sqrt{\frac{\pi}{\alpha}} \frac{e^{\frac{\pi}{2}|\frac{E_c}{E}|}}{\sqrt{e E}}$. 
For small electric fields (where $e E \ll m^2$) this approximation is valid for an exponentially long time.
	It never holds for an infinite time, however, so 
	calculations of infinite time pair creation effects without
	including back-reaction have little meaning. In fact, for $|E| \sim
	E_c$ this approximation is valid only for $\triangle t \sim
	10^{-19}$ seconds!

\subsection*{Non-Vacuum initial conditions}

	As an example of a non-vacuum initial condition, consider the 
state $u_{\bp_{{\rm in}},\lambda,t_0}(t) \wedge |{\rm vac}_{t_0}(t)\ra$, which at  
time $t_0$ contained one particle with 
momentum $\bp_{{\rm in}}$. Recall from 
equation (\ref{eq:jpart}) that $J(u_{\bp_{{\rm in}},\lambda,t_0}(t))$ does not represent 
the current at time $t$ of this particle. Rather, $J(u_{\bp,\lambda,t_0}(t)) 
+ J_{{\rm vac}}(\tau)$ represents the expected current at time $t_0 + 
\frac{\tau}{m}$ of this state,  which can no longer be viewed as 
containing one particle. This current will contain contributions not 
only from the 1-particle component of the evolved state, but also from the entire 
evolved state. 
Notice that, from equation (\ref{eq:diss6.8}), $J_{{\rm vac}}(\tau)$ contains a factor $V$, while 
$J(u_{\bp_{{\rm in}},1,t_0}(t))$ does not. The vacuum contribution is that of a 
uniform current density pervading all space, and the contribution from 
$J(u_{\bp,\lambda,t_0}(t))$ represents a single particle moving on this `sea of 
current'. 

\begin{figure}[hbt]
\begin{minipage}[b]{5.5cm}
\epsfig{figure=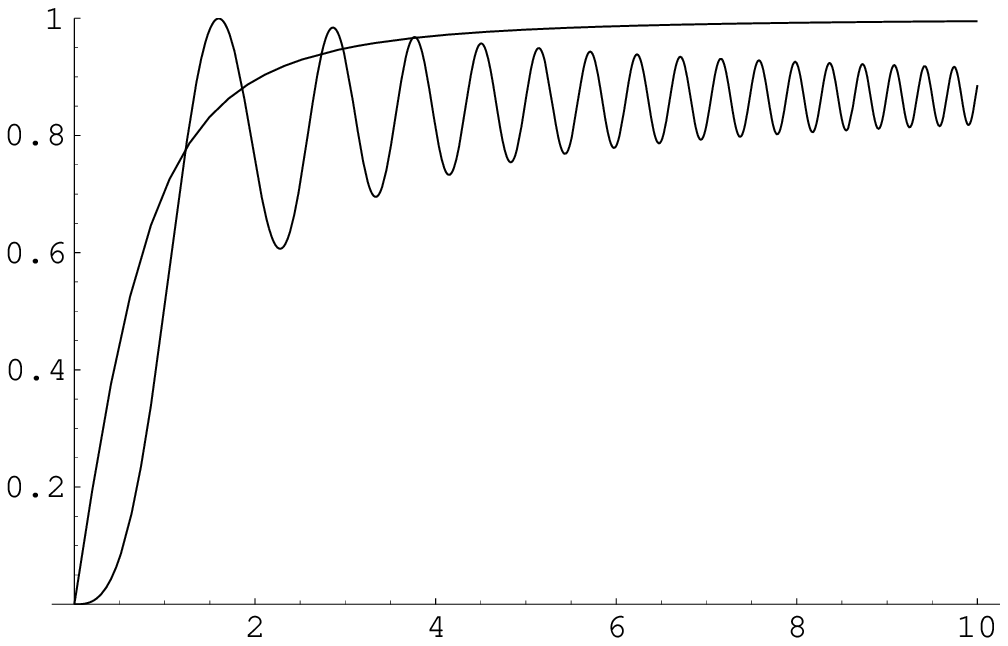,width=5.5cm,angle=0}

{\footnotesize {\bf (A).} $|E| = E_c$}
\end{minipage}\hs{.5}\begin{minipage}[b]{5.5cm}
\epsfig{figure=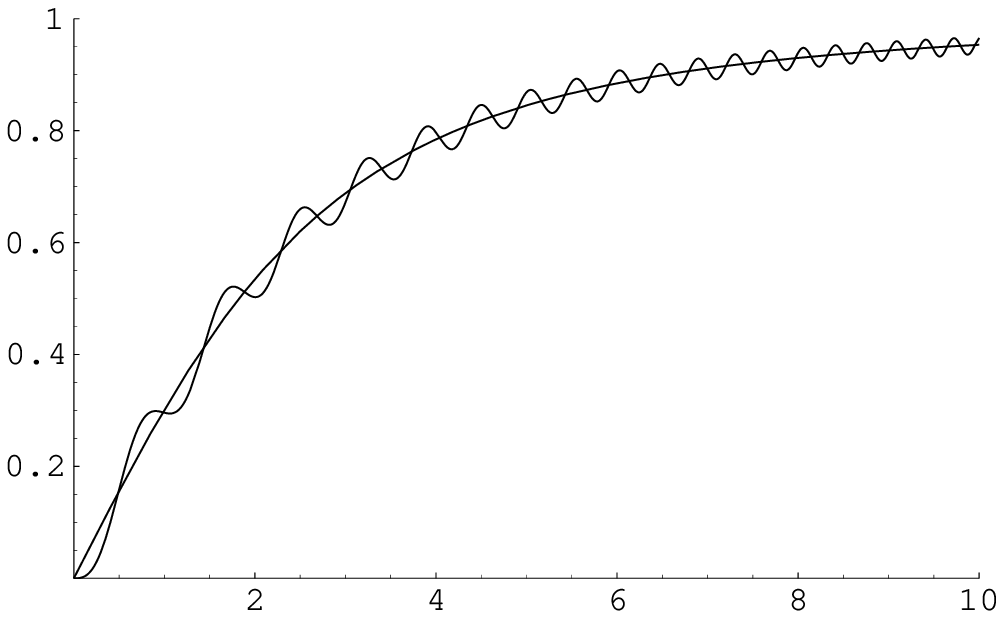,width=5.5cm,angle=0}

{\footnotesize {\bf (B).} $|E| = \frac{E_c}{10}$}
\end{minipage}
{\footnotesize {\bf FIG. 6.} $J(u_{\bp,\lambda;t_0}(\bx,t_0 + \frac{\tau}{\sqrt{e E}}))$ as a 
function of $\tau$, for $\bp_{{\rm in}} = 0$ and $|E| =E_c$ (A) and $\frac{E_c}{10}$ (B). Also included is the 
classically expected current for a particle accelerated from rest in this electric field. As expected, the full result
is closer to the classical result in (B) than in (A).}
\end{figure}

Figure 6 shows the $z$-component of 
$J(u_{\bp,\lambda,t_0}(\bx,t))$ as a function of $\tau$, for $\bp_{{\rm in}} = 0$  
and $|E| =E_c$ (A), and $\frac{E_c}{10}$ (B). Also included 
in these graphs is the classical current for a single particle
accelerating from rest in the relevant electric field. For $|E| =E_c$ the 
classical result differs significantly from the quantum 
result, even when the vacuum contribution $J_{{\rm vac}}(\tau)$ (from 
Figure~9) is included. This can be understood as follows. From 
equation (\ref{eq:vac1})  the vacuum current density is simply 
the sum over all modes of the difference between the two curves on these graphs (with 
the sign reversed, since it is negative energy modes that appear 
in (\ref{eq:vac1})). We know from the curves  in Figure 4 that 
this difference can be expressed as the sum of two contributions. The 
$\Re(\alpha \bar{\beta})$ term, which explains the  `zitterbewegung' 
oscillations in Figure 6, and the $|\beta|^2$ term,  which explains the difference 
in height between the two graphs, and represents the probability that the 
particle falls into a hole in the Dirac Sea vacated by a newly 
freed $v_{\bp}$ degree of freedom. This interpretation is also 
supported by Figures (13), (14) and (16) of the `adiabatic' case. For 
$|E| = \frac{E_c}{10}$, for which
the probability of particle creation is small, the quantum 
result agrees well with the classical prediction. The graph for 
$|E| = \frac{E_c}{100}$ is not included 
since the quantum result becomes indistinguishable from the 
classical prediction. Finally,  if at time $t$ we project out 
the 1-particle component of $u_{\bp_{{\rm in}},1,t_0}(t) \wedge 
|{\rm vac}_{t_0}(t)\ra$ (for example if the wave-function `collapsed' 
during a measurement of particle numbers) and measure the current of 
this projected state, we again obtain the classical 1-particle prediction.

\section{ADIABATIC ELECTRIC FIELD}

Consider now the spatially uniform electric field $E(t) = E \sech^2(\frac{m t}{\rho})$. This is generated by:
$a(\tau) = a_0 \rho \tanh\left(\frac{\tau}{\rho}\right)$.
Hence, (\ref{eq:CE9.1}) becomes
\begin{equation} \frac{d^2 f_{\lambda}}{d \tau^2} + \left[1 + p^2 + \left(p_z + a_0 \rho \tanh\left(\frac{\tau}{\rho}\right)\right)^2 + 
i a_0 \sech^2\left(\frac{\tau}{\rho}\right)\right] f_{\lambda} = 0 \label{eq:70} \end{equation}
	To solve this equation we use the same procedure as in \cite{GaGi}. Define
\begin{align}
y &  \equiv \half\left(1 + \tanh\left(\frac{\tau}{\rho}\right)\right) \notag \\
f(\tau) & =  g(y(\tau)) e^{-i E_{-\infty} \tau} (1 + e^{\frac{2 \tau}{\rho}})^{\frac{- i \rho \triangle E}{2}} \end{align}
	so that we can rewrite (\ref{eq:70}) as:
\begin{equation}
y(1-y) \frac{d^2 g}{d y^2} + \{ 1 - i \rho E_{-\infty} - (2 + i \rho \triangle E) y \}
 \frac{d g}{d y} - \nu_{-} (1 + \nu_{+}) g = 0
\end{equation}
where $\triangle E \equiv E_{+\infty} - E_{-\infty}$ is the change in energy caused by evolution from $\tau = -\infty$
 to $\tau = +\infty$ in this electric field, $\triangle p_z = 2 a_0 \rho$ is the corresponding change in $z$-momentum, and we have defined $\nu_{\pm} \equiv \frac{i \rho}{2} (\triangle E \pm \triangle p_z)$. It follows (see \cite{Bat}, pg 56) that a possible solution is the hypergeometric function 
 $g(y) = F(\nu_{-}, 1 + \nu_+; 1 - i \rho E_{-\infty}; y)$, 
	from which we find that
\begin{equation}
\begin{array}{rl} 
f(\tau) & =  F(\nu_{-}, 1 + \nu_+; 1 - i \rho E_{-\infty}; y(\tau)) e^{-i E_{-\infty} \tau} (1 + e^{\frac{2 \tau}{\rho}})^{\frac{- i \rho \triangle E}{2}} \\
\\
h(\tau) & = \left\{ \frac{2 i \nu_-}{\rho} (1 - y(\tau)) F(1+ \nu_{-}, 1 + \nu_+; 1 - i \rho E_{-\infty}; y(\tau)) 
\right. \\
\\
& \hs{-1.5} \left.+ (E_{+\infty} - p_z - a_0 \rho) F(\nu_{-}, 1 + \nu_+; 1 - i \rho E_{-\infty}; y(\tau)) \right\} e^{-i E_{-\infty} \tau} (1 + e^{\frac{2 \tau}{\rho}})^{\frac{- i \rho \triangle E}{2}} \\
\\
 C^{-2} & = 2 E_{-\infty} (E_{-\infty} - p_z + a_0 \rho) 
\end{array}
\end{equation}

	We have again made this choice of $f(\tau)$ and $h(\tau)$ so that $B_{p,p_z}(\tau) \rightarrow 0$ 
as $\tau \rightarrow -\infty$ and we can write $\bS(\tau_1,-\infty) = \bM(\tau_1)$ (dropping irrelevant phases) 
and  use the simplifications described at the end of Section 3 in this limit.

\subsection*{Number density in an evolved vacuum}

	Consider a state which at asymptotically early times contained no particles. The expected number of particles (not including antiparticles) with physical momentum
 $\bp^{\rm{out}} = \bp\prime = (p^1,p^2,p^3 + e A(t))$
in this state at some later time $t$ is given by
 \begin{equation} \frac{N_{\bp,\lambda,-\infty}(t)}{V} = |B_{p,p_z}(\tau)|^2 \notag \end{equation} 
As $\tau \rightarrow \infty$ the asymptotic properties of hypergeometric functions~\cite{GaGi,NaNi} imply that: \begin{equation} \frac{N_{\bp,\lambda,-\infty}(\infty)}{V} = \frac{\cosh(\pi \rho \triangle E) - \cosh(\pi \rho \triangle p_z)}{\cosh(\pi \rho \triangle E) - \cosh(\pi \rho (E_{+\infty} + E_{-\infty}))} \notag \end{equation}

\begin{figure}[hbt]
\begin{minipage}[b]{5.5cm}
\epsfig{figure=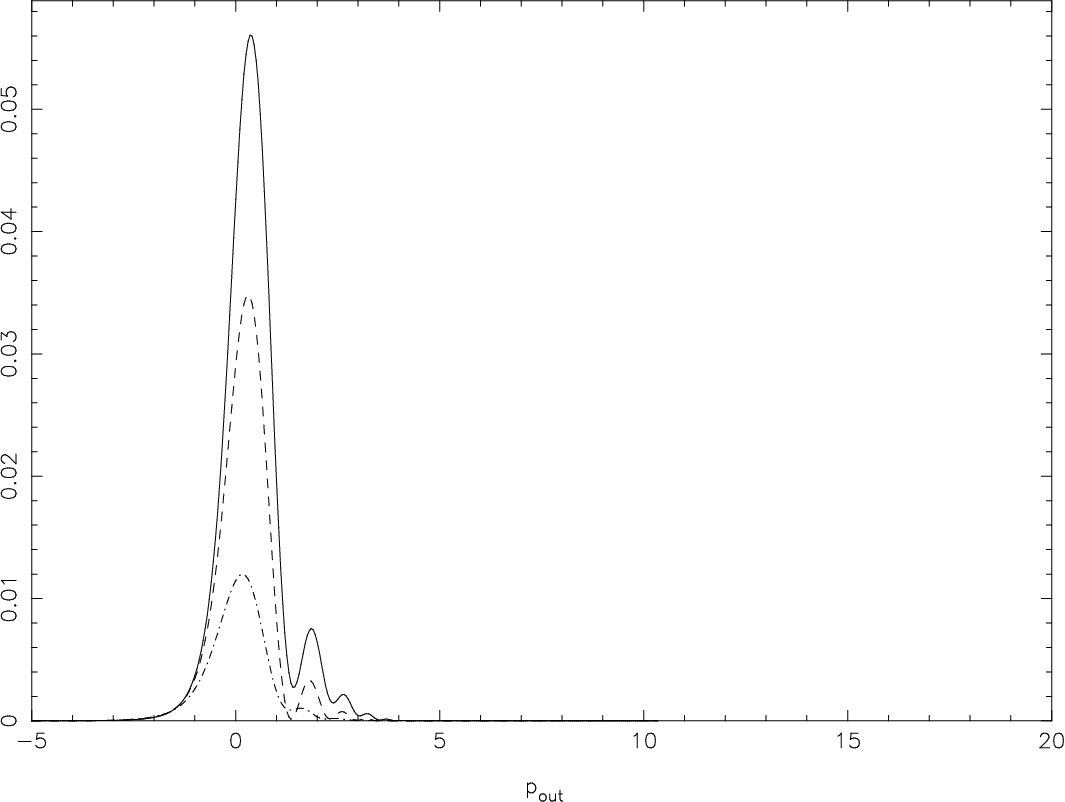,width=4cm,angle=-90}
{\footnotesize \hs{1} {\bf (A).} $\tau = -5$}
\end{minipage}\hs{.5}\begin{minipage}[b]{5.5cm}
\epsfig{figure=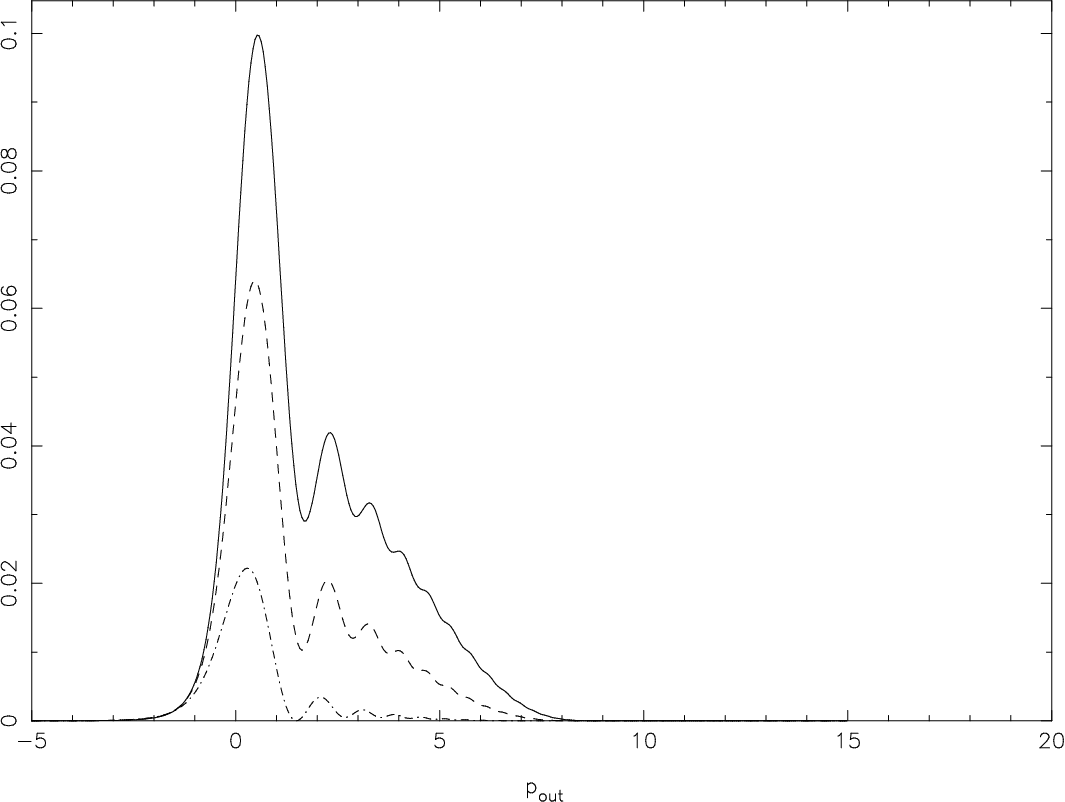,width=4cm,angle=-90}
{\footnotesize \hs{1} {\bf (B).} $\tau = 0$}
\end{minipage}

\begin{minipage}[b]{5.5cm}
\epsfig{figure=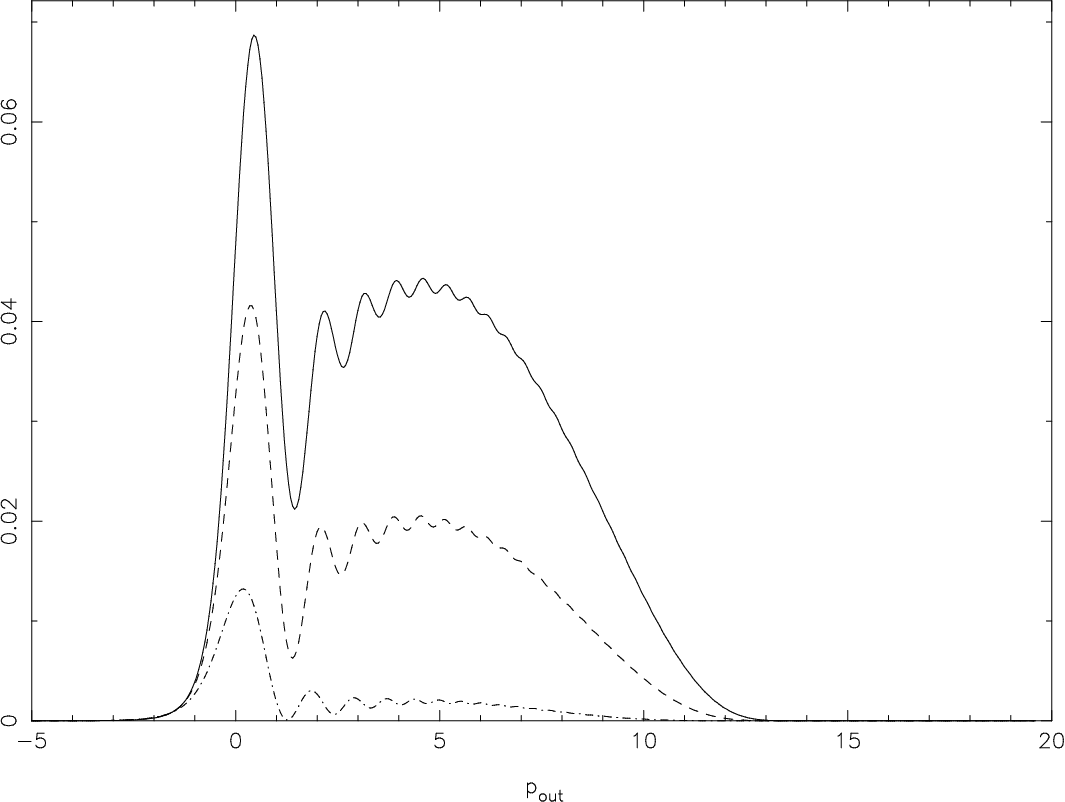,width=4cm,angle=-90}
{\footnotesize \hs{1} {\bf (C).} $\tau = 5$}
\end{minipage}\hs{.5}\begin{minipage}[b]{5.5cm}
\epsfig{figure=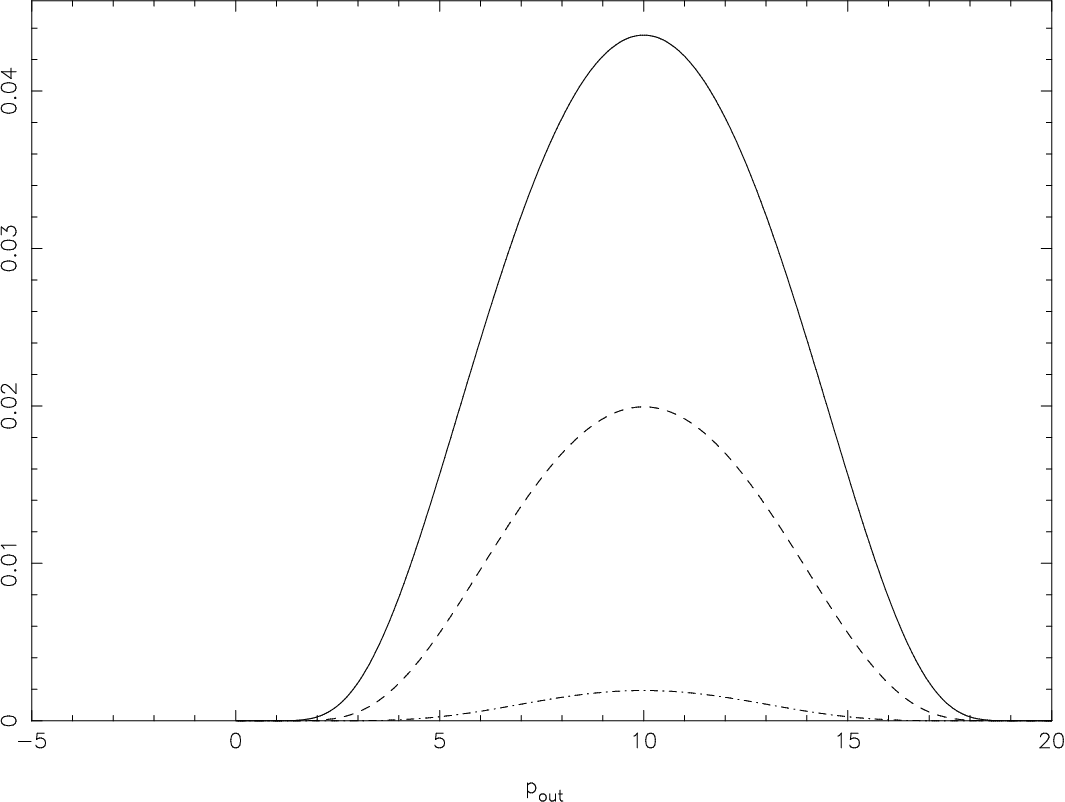,width=4cm,angle=-90}
{\footnotesize \hs{1} {\bf (D).} $\tau = \infty$}
\end{minipage}

{\footnotesize {\bf FIG 7.} The particle density 
$\frac{N_{\bp,\lambda,-\infty}(t)}{V}$ in the evolved `in' 
vacuum as a function of $\frac{p^3_{\rm{out}}}{m}$, for 
$a_0 = \frac{E}{E_c} = 1$, $\rho = 10$, $\frac{|\bp_{\perp}|}{m} = 0, 0.5 $ and $1$, and 
for $\tau = mt = -5$ (A), $0$ (B), $5$ (C) 
and $\infty$ (D).}
\end{figure}

Figure 7 shows $\frac{N_{\bp,\lambda,-\infty}(t)}{V}$ as a function 
of $p_z^{\rm{out}}$ for $a_0 = 1$, $\rho = 10$ and $\tau = -5,0,5$ 
and $\infty$. The three curves in each figure are for 
$\frac{|\bp_{\perp}|}{m} = 0,0.5 $ and $1$. These figures 
show how particle creation proceeds. 
Particles are created with small $p_z$, 
and are then accelerated by the electric field to 
take their place in the final `bulge'. The RHS of this final 
bulge comes from particles created at early times, while 
the LHS comprises particles that were created at late times and 
 have undergone less acceleration since their 
creation. The height of the peak at the origin in Figure 
7 depends on the strength of 
the electric field at that time, since this peak 
represents newly created particles. The corresponding distribution of antiparticles is obtained by reversing the sign of $p_z^{\rm{out}}$.

\setcounter{figure}{7}
\begin{figure}[htb!]
\centerline{
\put(-140,0){\hbox{\psfig{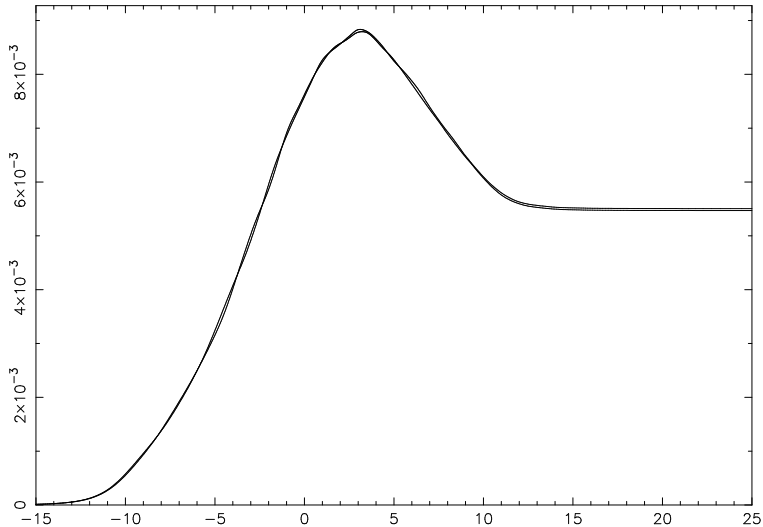}}}
}
\vs{0.2}
\caption[dummy1]{Monte-Carlo calculation of the total number density of particles in a Compton volume $m^{-3}$
for $a_0 = 1$ and $\rho = 10$. }
\label{ntotmc}
\end{figure}

Figure~\ref{ntotmc} shows the total number density of particles in a Compton volume ($m^{-3}$) of the evolved vacuum,
$\frac{N_{{\rm vac},-\infty}(\tau)}{V m^3}$ as a function of $\tau$.
This has been evaluated by Monte Carlo
sampling  over $|p_x| \leq 2$, $|p_y| \leq 2$ and 
over $|p_z| \leq 8$ (which corresponds to $-3 < p_z^{\rm{out}} < 13$). 
This range is large enough to include the dominant contribution to the integral.
The two curves are the result of independent
sets of 100,000 samples, giving an idea of the numerical accuracy.
\begin{figure}[hbt]
\begin{minipage}[b]{5.5cm}
\epsfig{figure=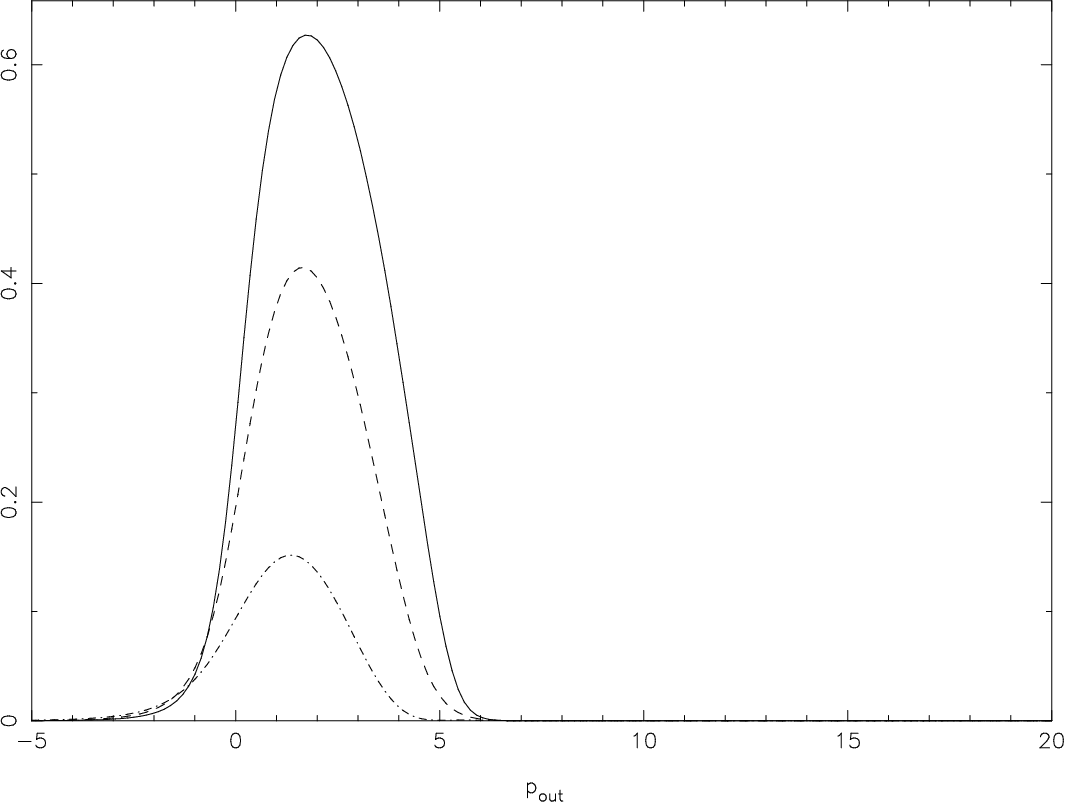,width=4cm,angle=-90}
{\footnotesize \hs{1} {\bf (A).} $\tau = -.5$}
\end{minipage}\hs{.5}\begin{minipage}[b]{5.5cm}
\epsfig{figure=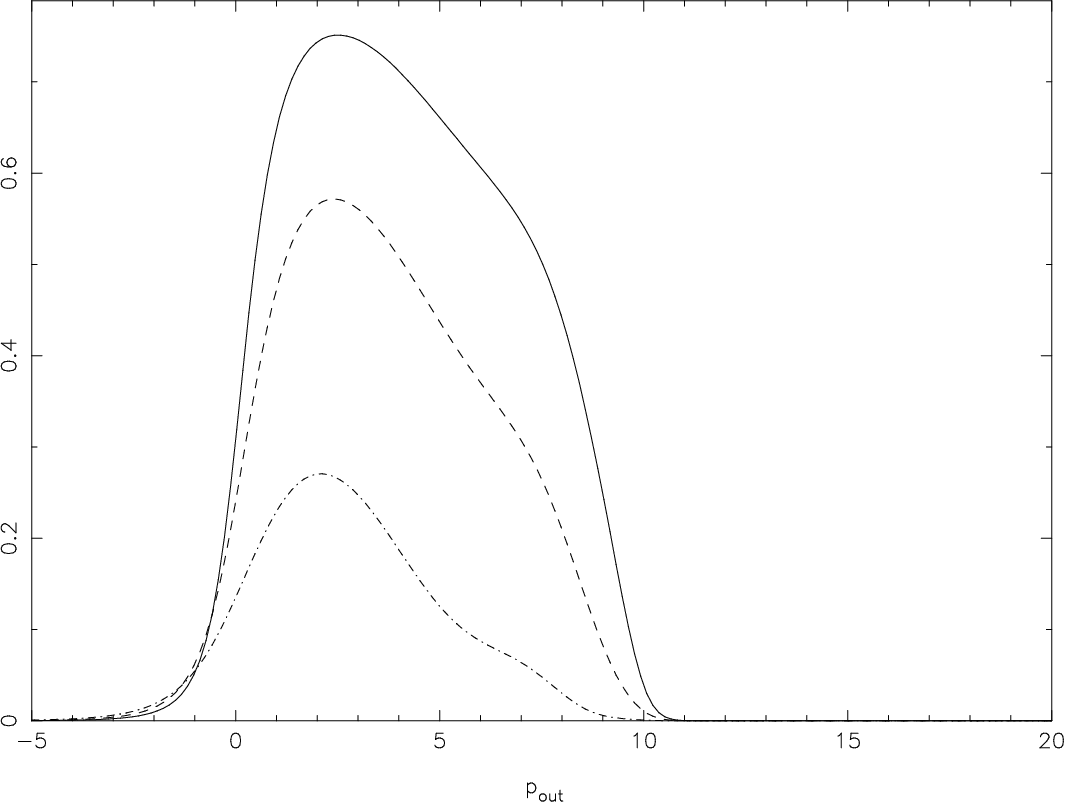,width=4cm,angle=-90}
{\footnotesize \hs{1} {\bf (B).} $\tau = 0$}
\end{minipage}

\begin{minipage}[b]{5.5cm}
\epsfig{figure=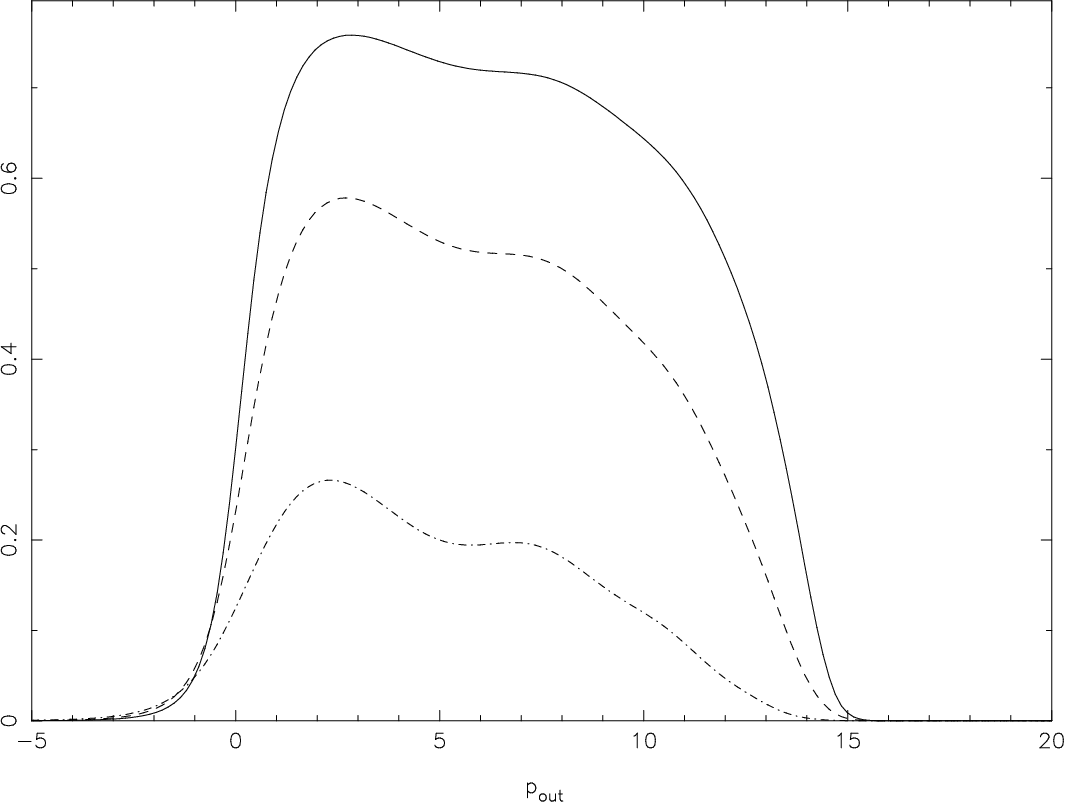,width=4cm,angle=-90}
{\footnotesize \hs{1} {\bf (C).} $\tau = .5$}
\end{minipage}\hs{.5}\begin{minipage}[b]{5.5cm}
\epsfig{figure=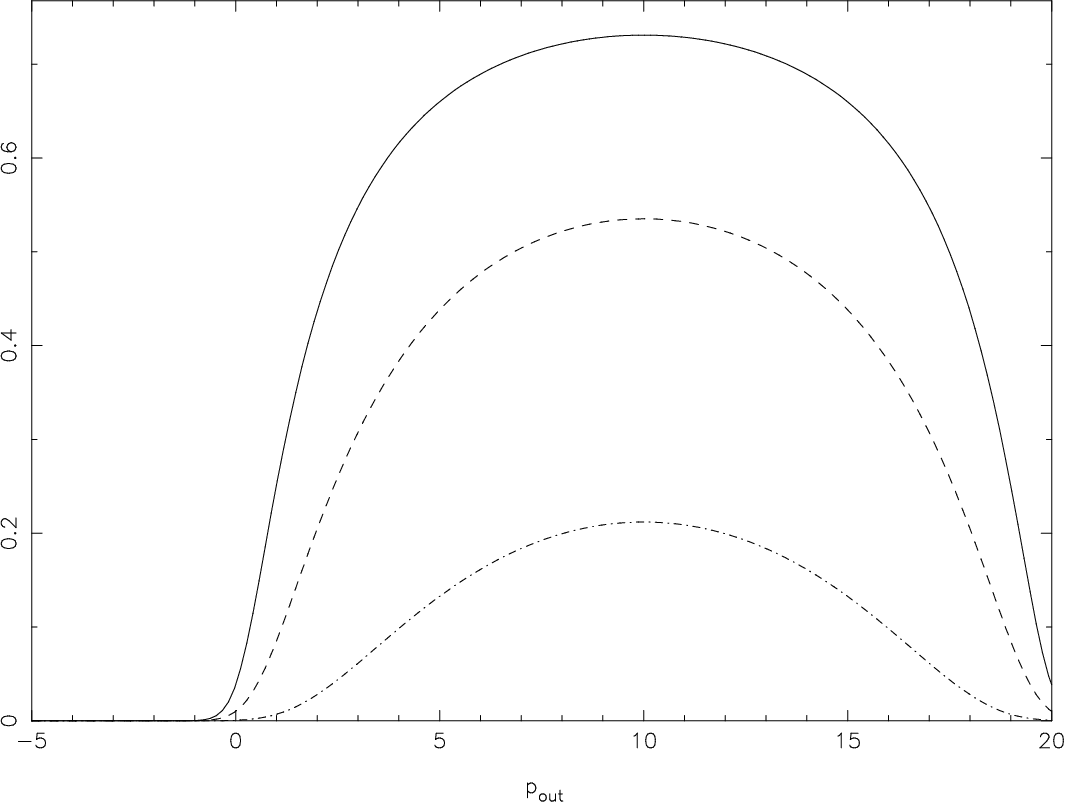,width=4cm,angle=-90}
{\footnotesize \hs{1} {\bf (D).} $\tau = \infty$}
\end{minipage}

{\footnotesize {\bf FIG. 9.} The particle density 
$\frac{N_{\bp,\lambda,-\infty}(t)}{V}$ in the evolved `in' 
vacuum shown as a function of $\frac{p^3_{\rm{out}}}{m}$, for 
$a_0 = \frac{E}{E_c} = 10$, $\rho = 1$, 
$\frac{|\bp_{\perp}|}{m} = 0,1 $ and $2$, and 
for $\tau = mt = -.5$ (A), $0$ (B), $.5$ (C) 
and $\infty$ (D).}
\end{figure}

Figure 9 is a repeat of Figure 7  but with $a_0 = \frac{E}{E_c} = 10$ and $\rho = 1$. 
Particle creation now occurs over a much shorter period of time, so the distinction between the peak at the origin 
and the late-time bulge is much less pronounced.

\begin{figure}[hbt]
\begin{minipage}[b]{5.5cm}
\epsfig{figure=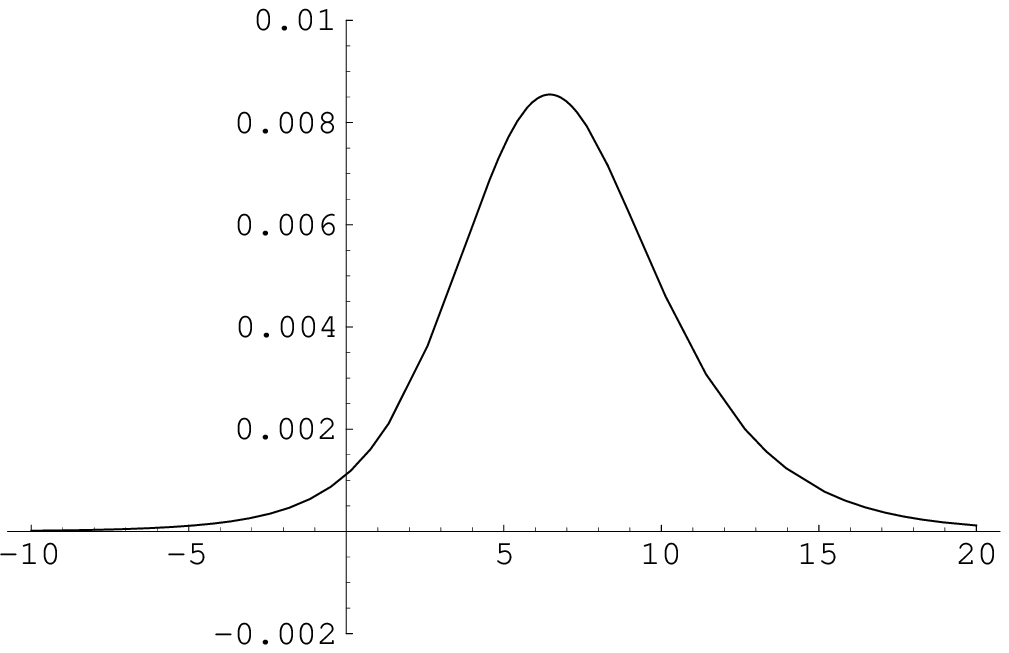,width=5.5cm,angle=0}
{\footnotesize \hs{1} {\bf (A).} $p_z = -6$. The covariant $z$-momentum of this mode is always negative.}
\end{minipage}\hs{.5}\begin{minipage}[b]{5.5cm}
\epsfig{figure=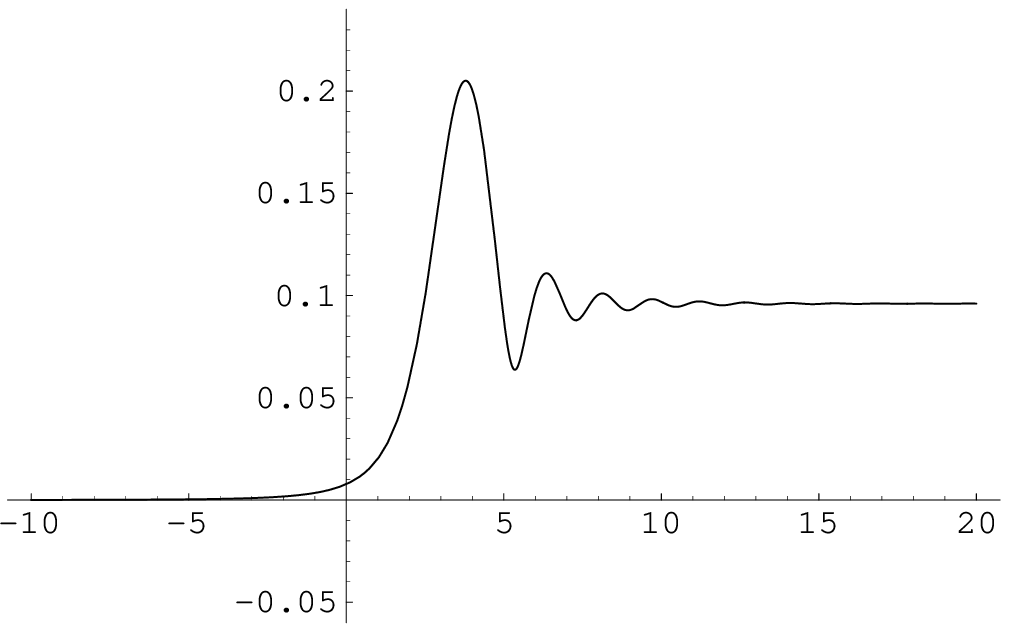,width=5.5cm,angle=0}
{\footnotesize \hs{1} {\bf (B).} $p_z = -3$. The covariant $z$-momentum of this mode passes through zero at $\tau \approx 3.47$.}
\end{minipage}

\vs{.5}

\begin{minipage}[b]{5.5cm}
\epsfig{figure=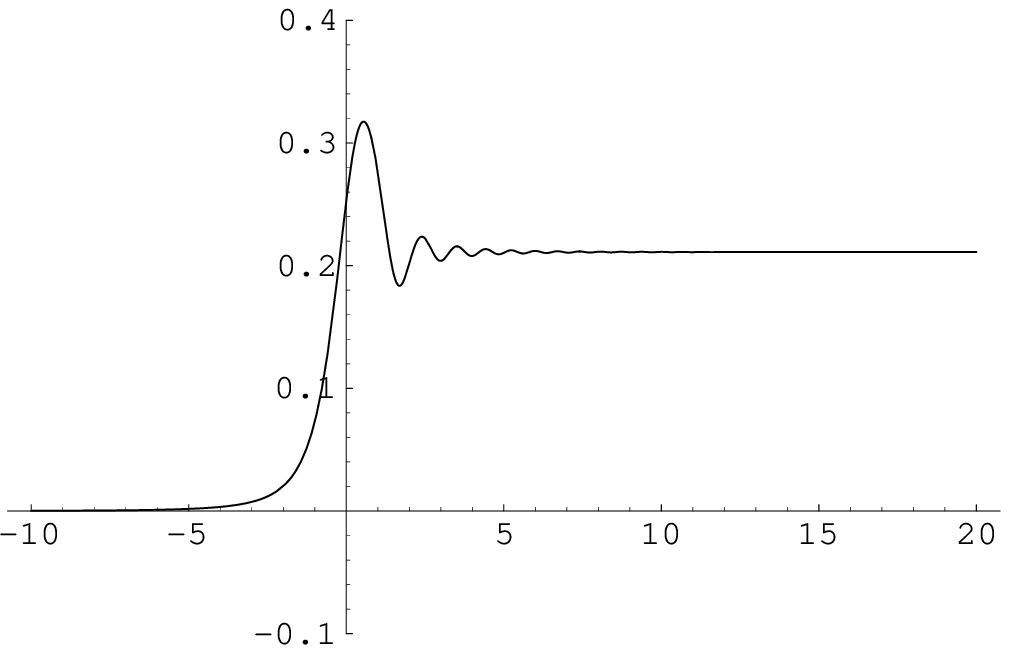,width=5.5cm,angle=0}
{\footnotesize \hs{1} {\bf (C).} $p_z = 0$. The covariant $z$-momentum of this mode passes through zero at $\tau = 0$.}
\end{minipage}\hs{.5}\begin{minipage}[b]{5.5cm}
\epsfig{figure=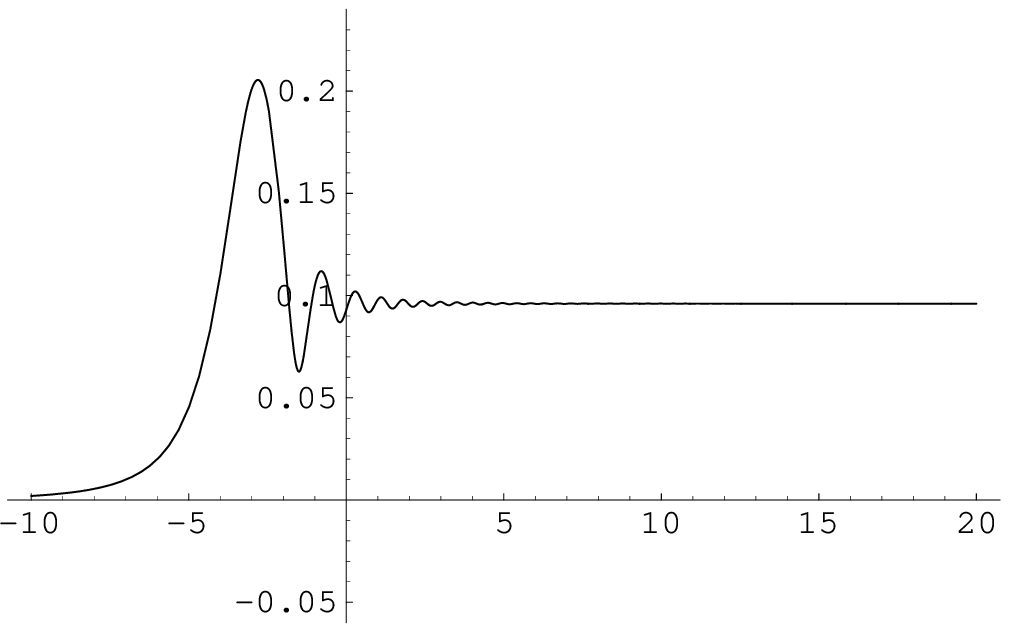,width=5.5cm,angle=0}
{\footnotesize \hs{1} {\bf (D).} $p_z = 3$. The covariant $z$-momentum of this mode passes through zero at $\tau \approx -3.47$.}
\end{minipage}

\vs{.5}

\begin{minipage}[b]{5.5cm}
\epsfig{figure=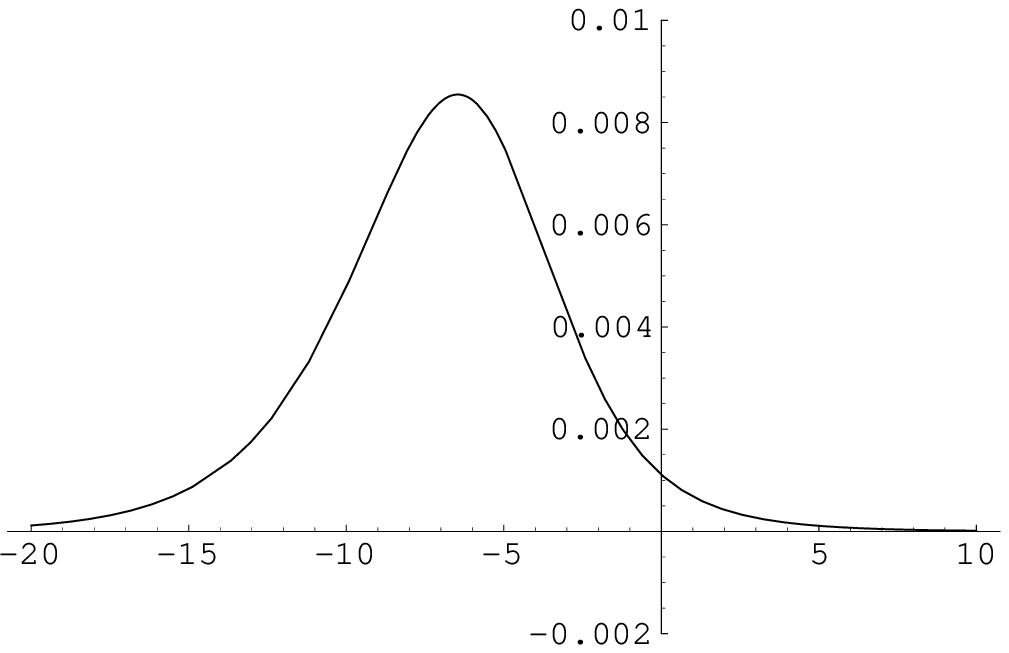,width=5.5cm,angle=0}
{\footnotesize \hs{1} {\bf (E).} $p_z = 6$. The covariant $z$-momentum of this mode is always positive.}
\end{minipage}\hs{.5}\begin{minipage}[b]{5.5cm}
\epsfig{figure=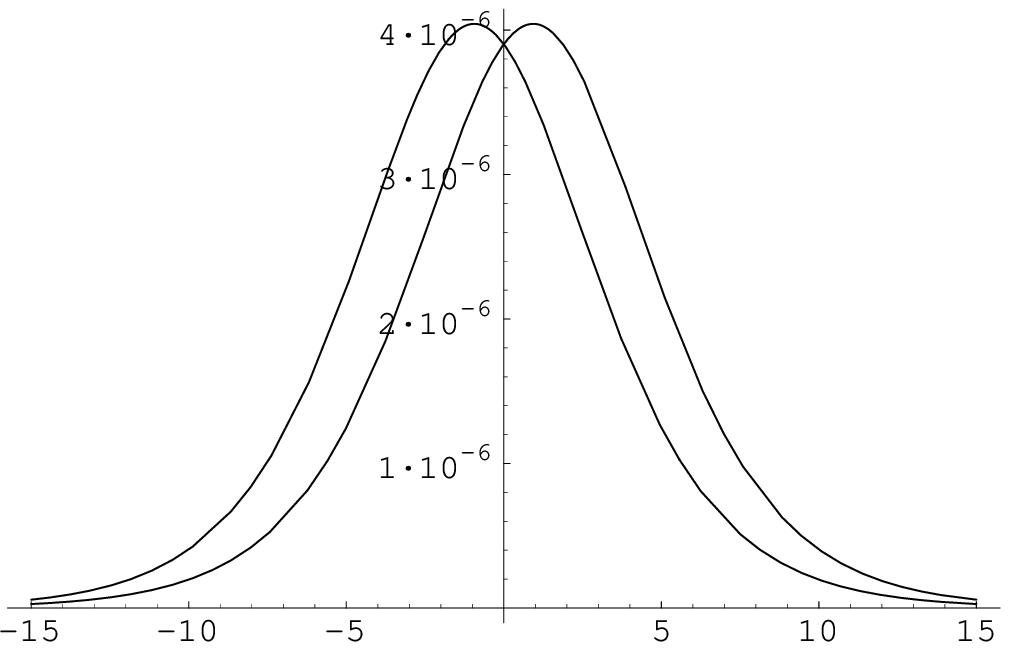,width=5.5cm,angle=0}
{\footnotesize \hs{1} {\bf (F).} $p_z = \pm 40$. 
The position of the peak approaches zero as $p_z \rightarrow \pm \infty$. }
\end{minipage}
\vs{.2}

{\footnotesize {\bf FIG. 10.} The particle density 
$\frac{N_{\bp,\lambda,- \infty}(t)}{V}$ in the evolved `in' vacuum as a function of $\tau$, for 
$a_0 = \frac{E}{E_c} = 1$, $\frac{|\bp_{\perp}|}{m} = 0$, $\rho = 5$, and 
for $p_z = -6$ (A), $-3$ (B), $0$ (C), $3$ (D), $6$ (E) 
and $\pm 40$ (F). }
\end{figure}

To understand this peak at the origin in more detail, examine Figure 10. Here 
we show $\frac{N_{\bp,\lambda,- \infty}(t)}{V}$ as a function of $\tau$, for 
$a_0 = \frac{E}{E_c} = 1$, $\frac{|\bp_{\perp}|}{m} = 0$, $\rho = 5$, and 
for $p_z = -6$ (A), $-3$ (B), $0$ (C), $3$ (D), $6$ (E) 
and $\pm 40$ (F). From (B), (C) and (D), the most significant change 
in the occupation number of any mode occurs at the time when the covariant 
momentum of that mode passes through zero. That is, a mode is most likely to be 
excited from the Dirac Sea when its momentum passes through zero in the frame 
of the electric field. If the mode does not pass through zero (as in (A) (E) 
and (F)) then, although $\frac{N_{\bp,\lambda,- \infty}(t)}{V}$ can become non-zero 
during periods when  $a(\tau)$ changes so that $E(t)$ is significant, this effect 
is small and transient. This transient effect is also influenced slightly by the momentum of the relevant mode, 
with the position of the peak shifted towards times when the momentum was smallest. However, 
when the relative change in the momentum is small,
the peak is displaced much less, and $\frac{N_{\bp,\lambda,- \infty}(t)}{V}$ more closely resembles $E(t)$.

\vs{.5}

\begin{figure}[hbt]
\begin{minipage}[b]{5.5cm}
\epsfig{figure=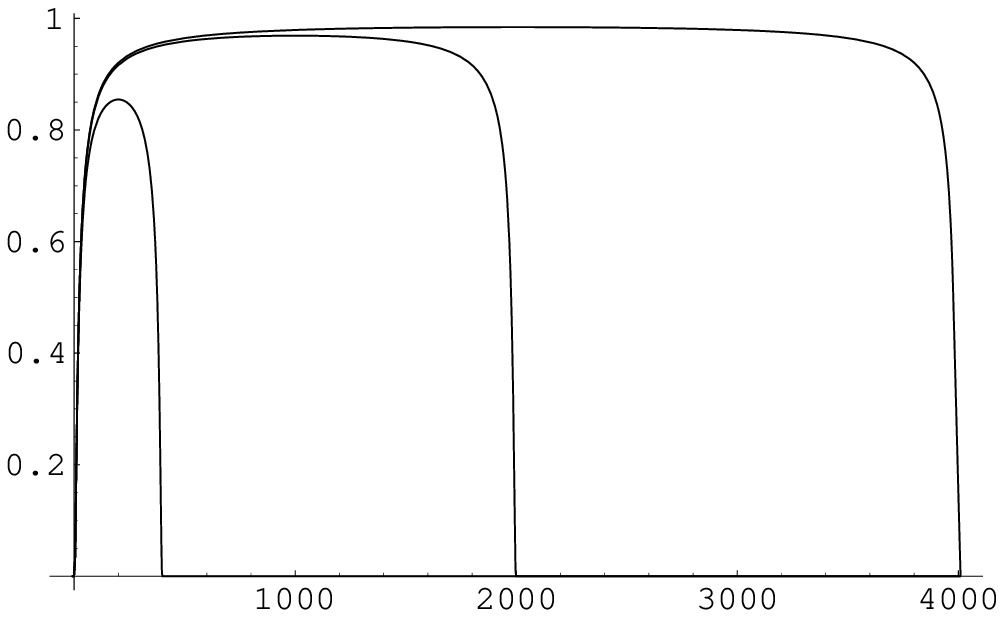,width=5.5cm,angle=0}

{\footnotesize {\bf FIG. 11(A).} The particle density 
$\frac{N_{\bp,\lambda,-\infty}(\infty)}{V}$ in the infinitely 
evolved `in' vacuum shown as a function of $\frac{p^3_{\rm{out}}}{m}$, for 
$\rho = 10$, $\frac{|\bp_{\perp}|}{m} = 0$, and $a_0 = \frac{E}{E_c} = 20, 100$ and $200$.}
\end{minipage}\hs{.5}\begin{minipage}[b]{5.5cm}
\epsfig{figure=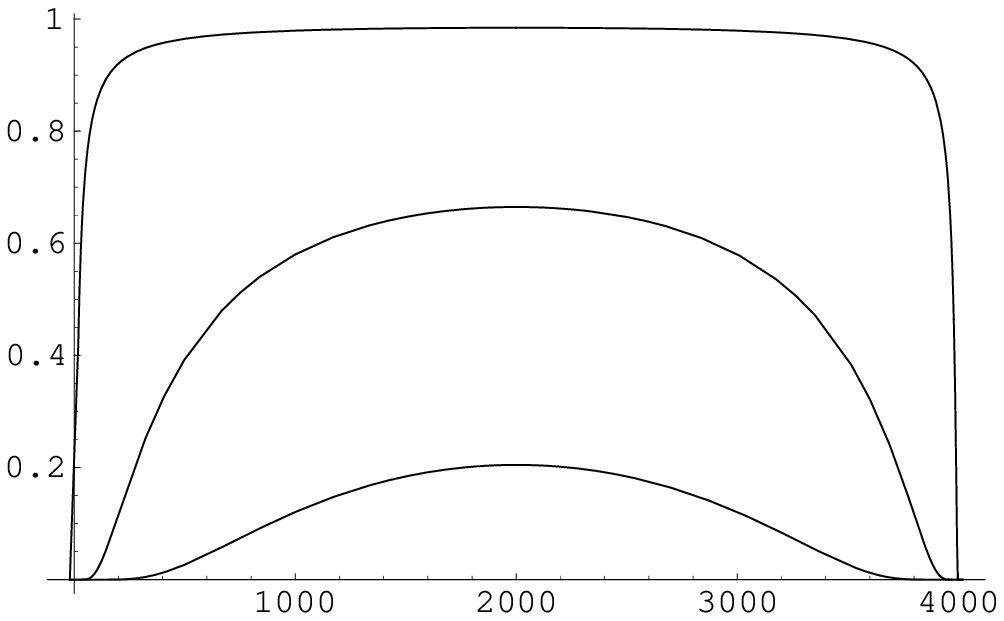,width=5.5cm,angle=0}

{\footnotesize {\bf FIG. 11(B).} The particle density 
$\frac{N_{\bp,\lambda,-\infty}(\infty)}{V}$ in the infinitely 
evolved `in' vacuum shown as a function of $\frac{p^3_{\rm{out}}}{m}$, for 
$\rho = 10$, $a_0 = \frac{E}{E_c} = 200$, and $\frac{|\bp_{\perp}|}{m} = 0, 5$, and $10$.}
\end{minipage}
\end{figure}

\vs{.5}
It is useful to compare Figures such as 7(D) and 9(D) with Figure 11(A).
In Figure 11(A) we see that when $a_0$ is large the bulge becomes much more steep-edged. 
For $a_0 = 200$ we see that (at least for $|\bp_{\perp}| = 0$) any mode for which $p_z^{\rm{out}}$ passes through
 zero will almost certainly be created. 
Figure 11(B) shows how this 
result changes upon increasing $|\bp_{\perp}|$. 
Modes with higher $|\bp_{\perp}|$ require more energy for their creation,
hence an increase in $|\bp_{\perp}|$ decreases the number and smoothes the distribution of the created particles.

\begin{figure}[hbt]
\begin{minipage}[b]{5.5cm}
\epsfig{figure=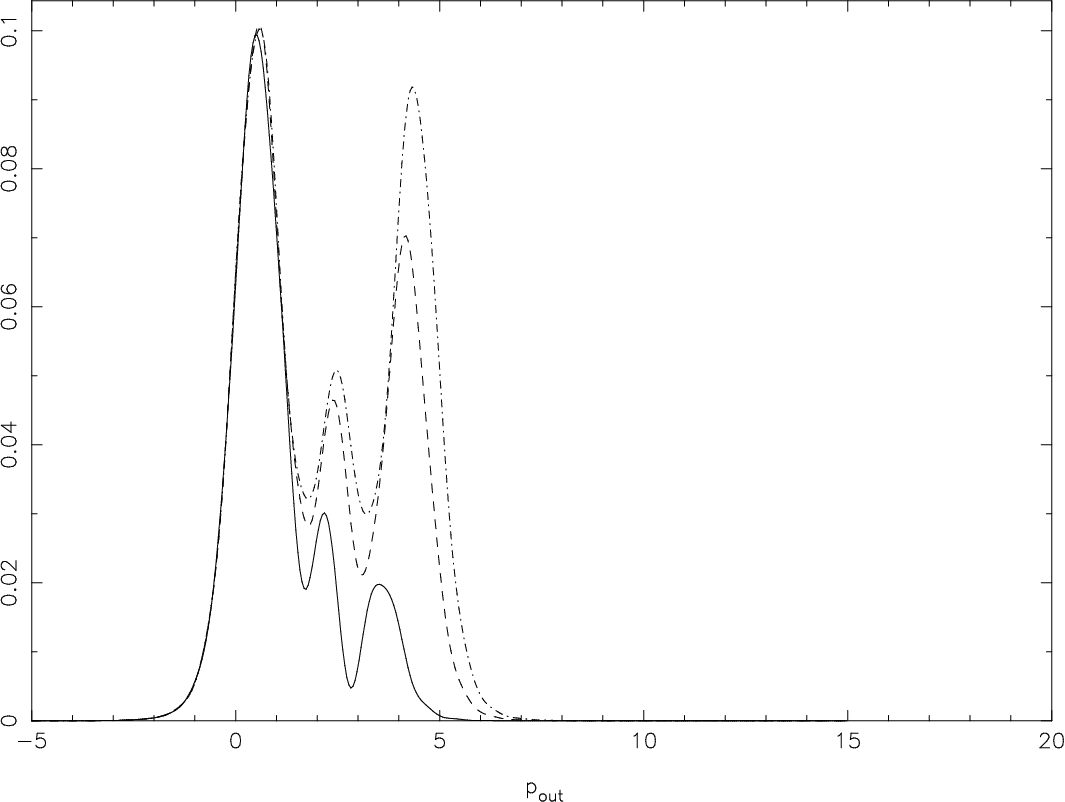,width=4cm,angle=-90}
{\footnotesize \hs{1} {\bf (A).} $\tau = 0$}
\end{minipage}\hs{.5}\begin{minipage}[b]{5.5cm}
\epsfig{figure=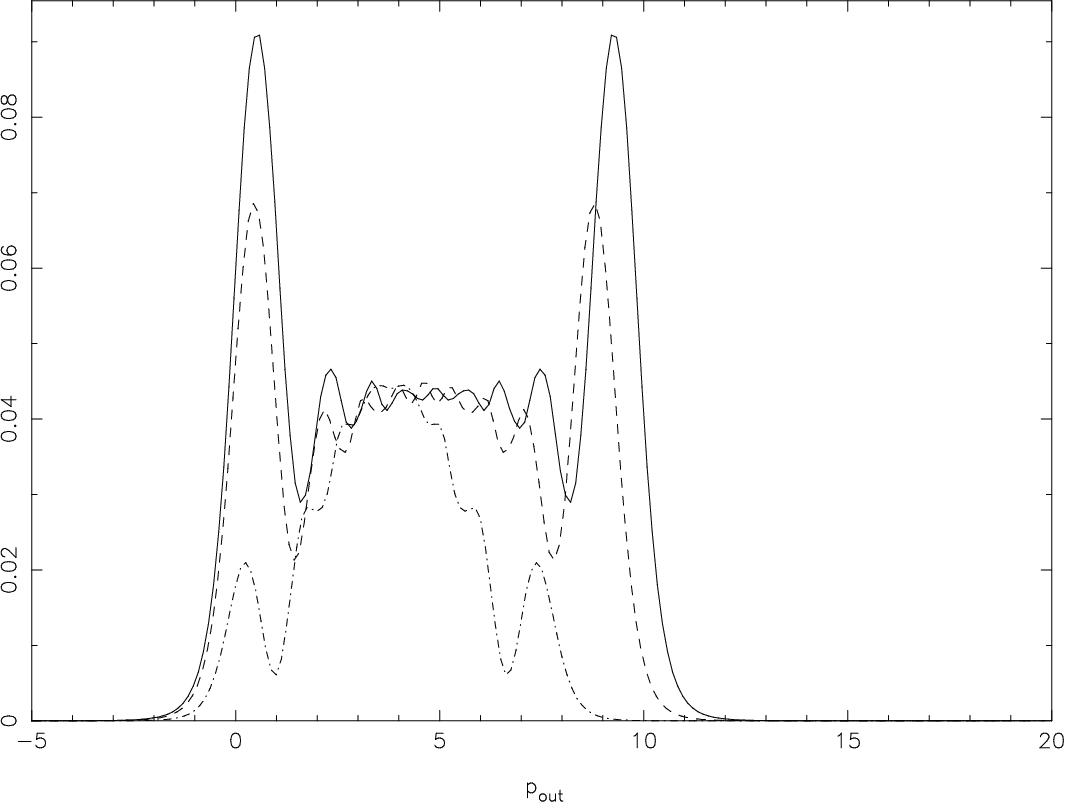,width=4cm,angle=-90}
{\footnotesize \hs{1} {\bf (B).} $\tau = 5$}
\end{minipage}

\begin{minipage}[b]{5.5cm}
\epsfig{figure=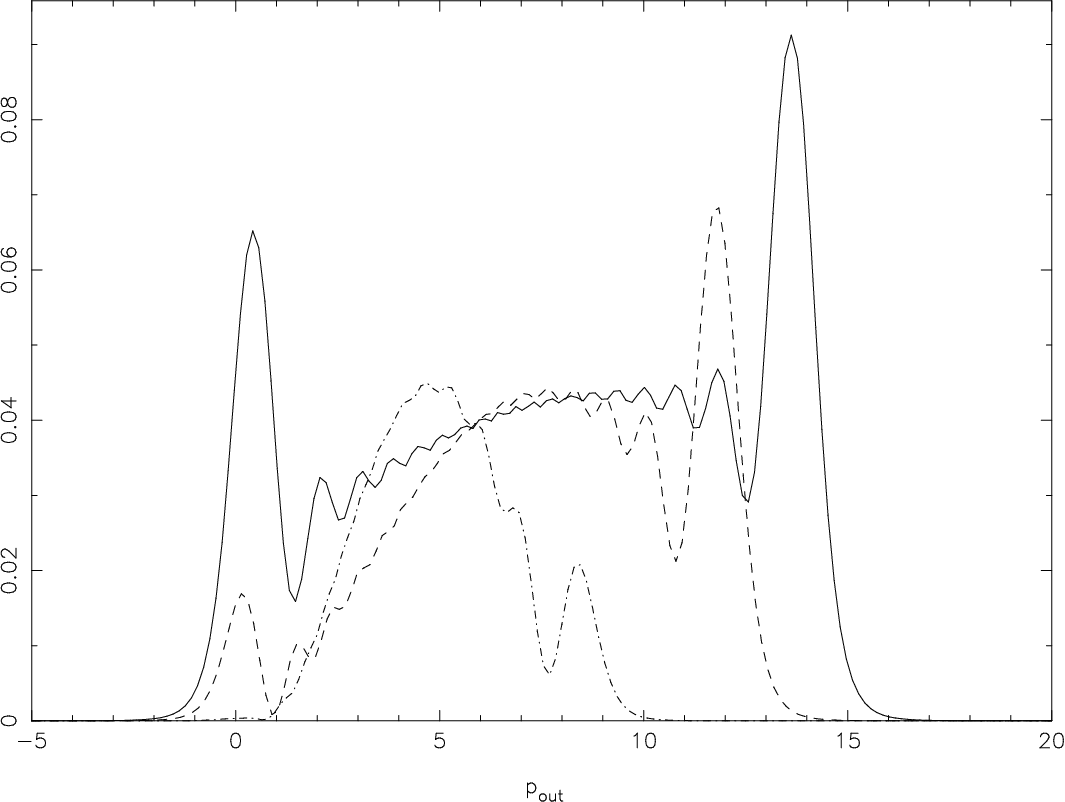,width=4cm,angle=-90}
{\footnotesize \hs{1} {\bf (C).} $\tau = 10$}
\end{minipage}\hs{.5}\begin{minipage}[b]{5.5cm}
\epsfig{figure=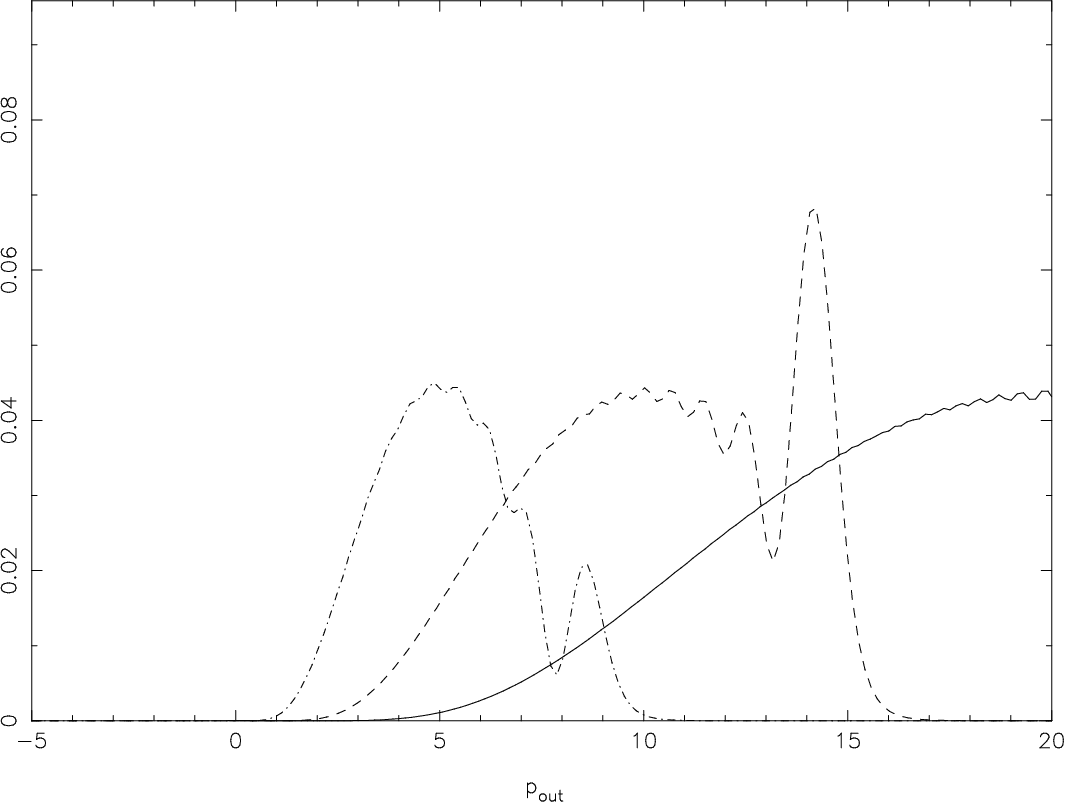,width=4cm,angle=-90}
{\footnotesize \hs{1} {\bf (D).} $\tau = \infty$}
\end{minipage}

{\footnotesize {\bf FIG. 12.} The particle density 
$\frac{N_{\bp,\lambda,t_0}(t)}{V}$ in an evolved state that was  
vacuum at time $\tau_0 = mt = -5$ shown as a function of $\frac{p^3_{\rm{out}}}{m}$, for 
$a_0 = \frac{E}{E_c} = 1$, $\frac{|\bp_{\perp}|}{m} = 0$, $\rho = 5, 10$ and $20$, and 
for $\tau = mt = 0$ (A), $5$ (B), $10$ (C) 
and $\infty$ (D).}
\end{figure}

 Figure 12 shows the particle density 
$\frac{N_{\bp,\lambda,t_0}(t)}{V}$ in an evolved state that was  
vacuum at time $\tau_0 = mt = -5$ as a function of $\frac{p^3_{\rm{out}}}{m}$ for 
$a_0 = \frac{E}{E_c} = 1$, $\frac{|\bp_{\perp}|}{m} = 0$, and for various $\rho$ and $\tau$. 
The curves further confirm the present interpretation of the particle creation process.
 All have a peak at the right hand side (even for $\tau = \infty$), which comes from the 
initial burst of particle creation, much as in the constant electric field case. 
The size of this burst is related to the strength of the electric field
at time $t_0$, such that the graphs in 12(D) are the reflection of $\frac{N_{\bp,\lambda,-\infty}(t)}{V}$ 
about $p_z^{\rm{out}} = a_0 \rho$. 
Figures (A), (B) and (C) are qualitatively similar to the curves in Figure 2, especially for $\rho = 20$. 
This is  expected, since $a(\tau)$ does not differ significantly from $a_0$ for $\rho = 20$ $-5<\tau<10$.

\subsection*{Current in an evolved vacuum}

Having seen how the process of particle creation takes place, the predicted current in the evolved vacuum
 provides no further surprises.
\begin{figure}[hbt]
\begin{minipage}[b]{5.5cm}
\epsfig{figure=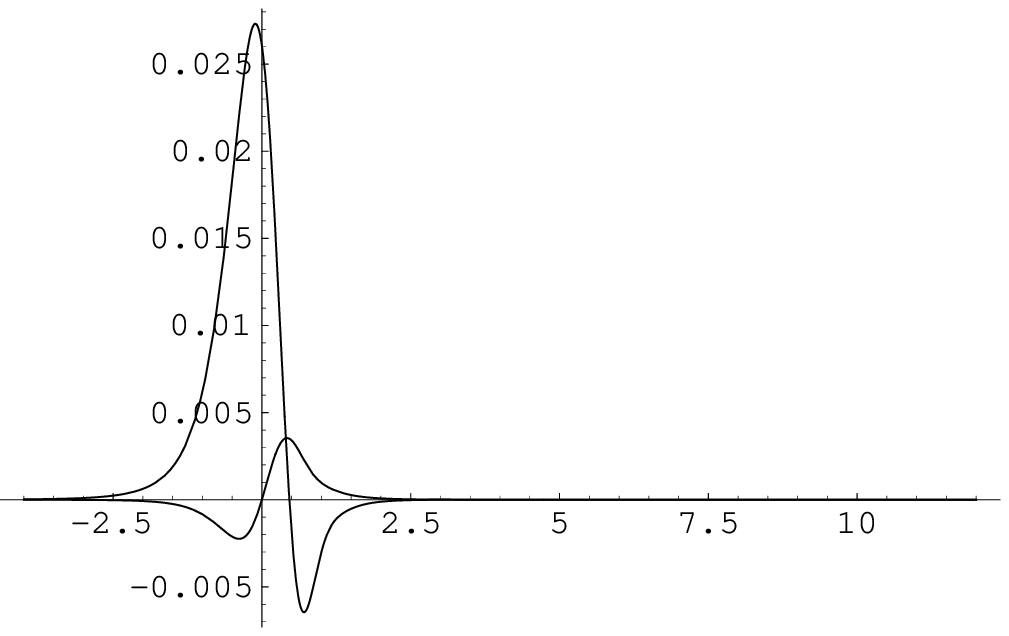,width=5.5cm,angle=0}
{\footnotesize \hs{1} {\bf (A).} $\tau = -5$. }
\end{minipage}\hs{.5}\begin{minipage}[b]{5.5cm}
\epsfig{figure=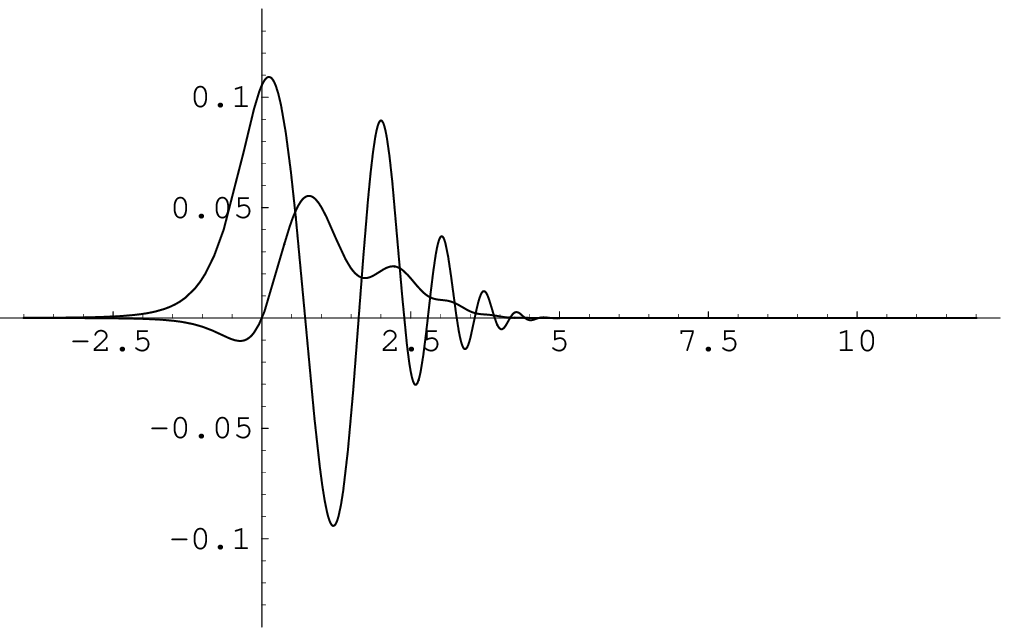,width=5.5cm,angle=0}
{\footnotesize \hs{1} {\bf (B).} $\tau = 0$. }
\end{minipage}

\vs{.5}

\begin{minipage}[b]{5.5cm}
\epsfig{figure=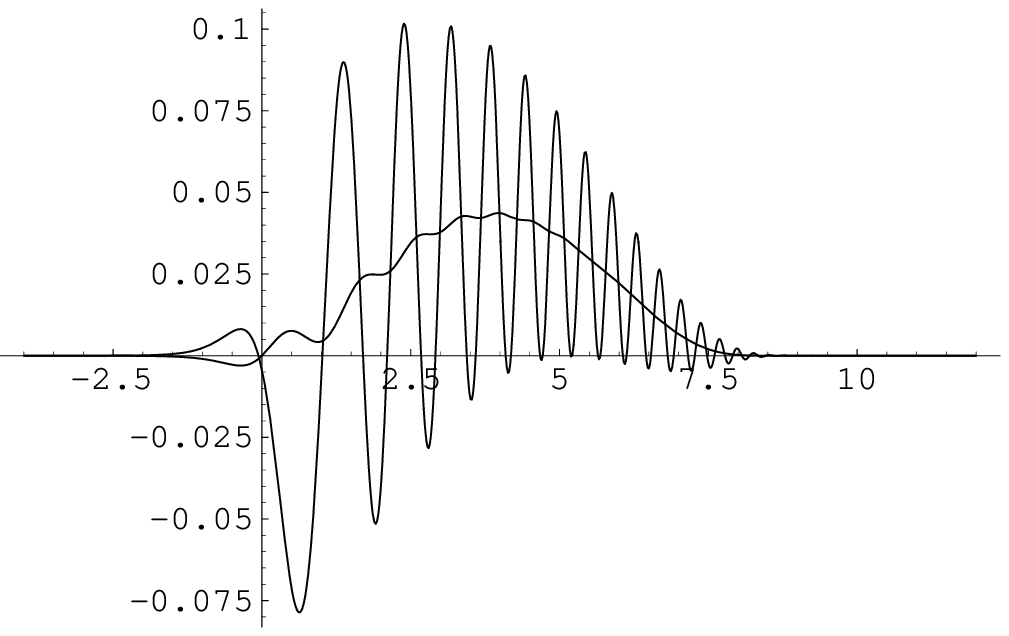,width=5.5cm,angle=0}
{\footnotesize \hs{1} {\bf (C).} $\tau = 5$. }
\end{minipage}\hs{.5}\begin{minipage}[b]{5.5cm}
\epsfig{figure=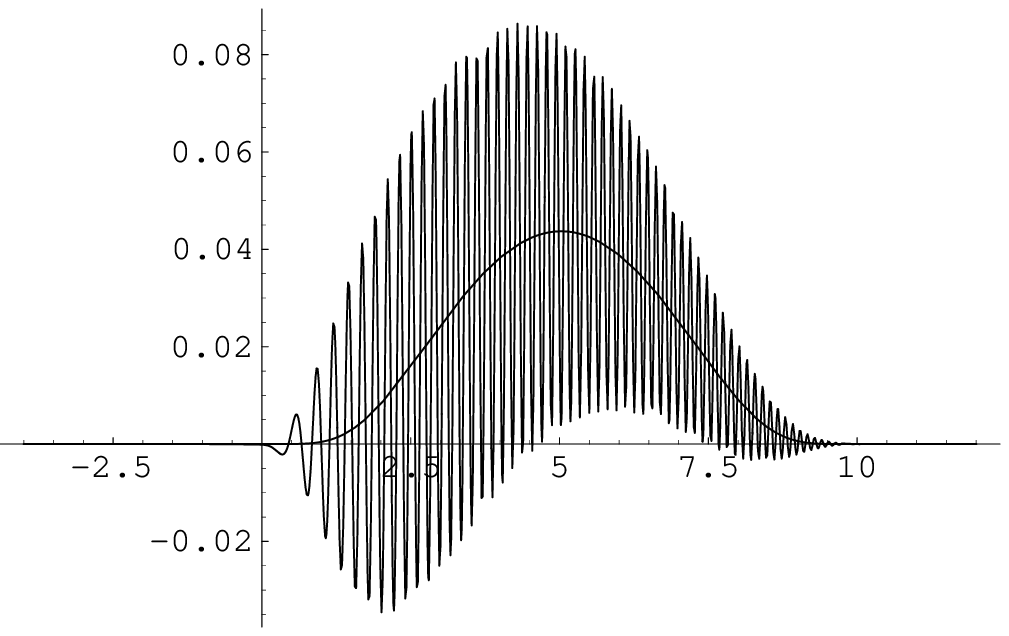,width=5.5cm,angle=0}
{\footnotesize \hs{1} {\bf (D).} $\tau = 20$. }
\end{minipage}

{\footnotesize {\bf FIG. 13.} Current density in the evolved `in' vacuum shown as a function of $p_{z}^{\rm{out}}$ 
for $a_0 = \frac{E}{E_c} = 1$, $\frac{|\bp_{\perp}|}{m} = 0$, $\rho = 5$ and $\tau = -5$ (A), $0$ (B), $5$ (C) and $20$ (D). The $|\beta|^2$ 
contribution is also shown (the darker curve). The  
$\Re(\beta \alpha)$ term at late times  produces
rapid oscillations about the $|\beta|^2$ term, in contrast to early times (as shown in A).}
\end{figure}

Figure 13 shows the integrand of $\frac{J_{{\rm vac},3,-\infty}(t)}{V}$ as a 
function of $p_z^{\rm{out}}$, for $a_0 = \frac{E}{E_c} = 1$, $\frac{|\bp_{\perp}|}{m} = 0$, 
$\rho = 5$, and for various $\tau$;  Figure 14 shows the same quantity as a 
function of $\tau$ for various $p_z$. 
The $|\beta|^2$ contribution has also been 
included. 
As in Figure 4, we see that at late times the $\Re(\beta \alpha)$ term simply provides rapid oscillations
 about the $|\beta|^2$ contribution.
Also,  the contribution to $\frac{J_{{\rm vac},3,-\infty}(t)}{V}$ from the $p_z = 6$ mode is very small in Figure~14. 
This is because the covariant momentum of this mode never passes through zero.

\begin{figure}[hbt]
\begin{minipage}[b]{5.5cm}
\epsfig{figure=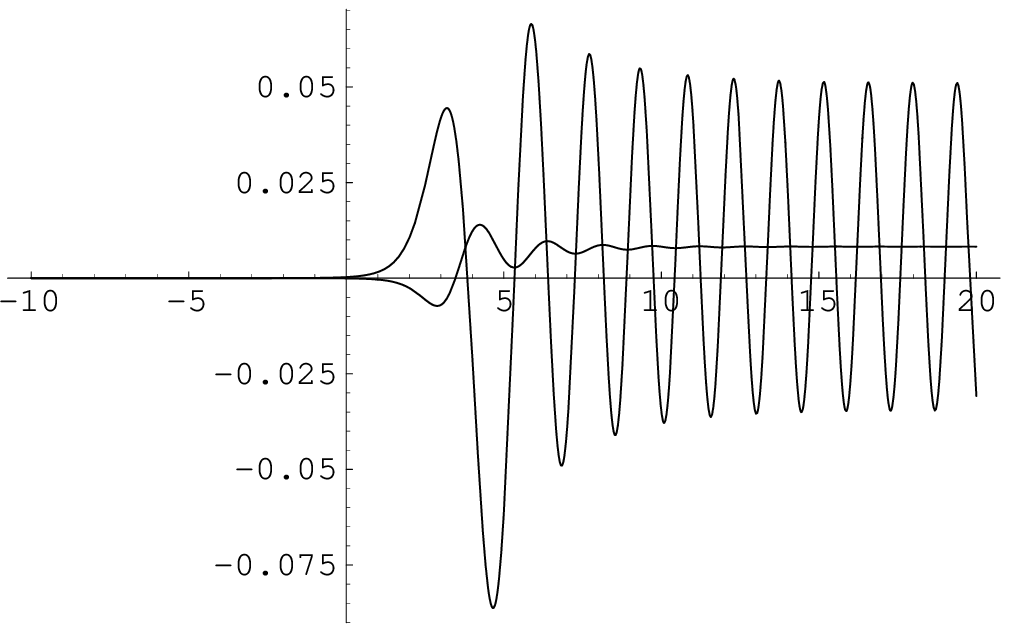,width=5.5cm,angle=0}
{\footnotesize \hs{1} {\bf (A).} $p_z = -3$. }
\end{minipage}\hs{.5}\begin{minipage}[b]{5.5cm}
\epsfig{figure=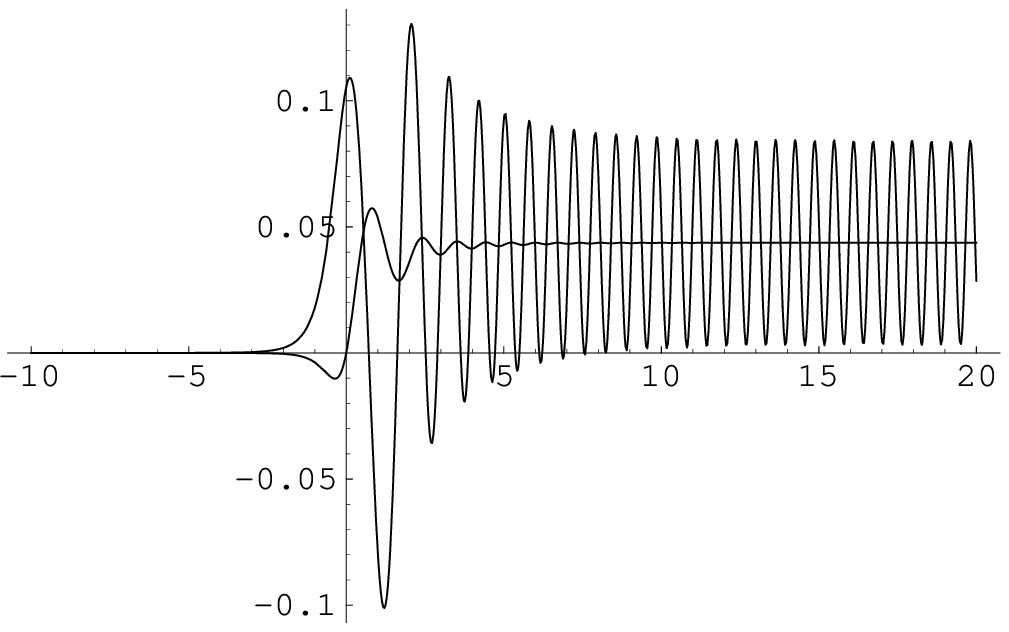,width=5.5cm,angle=0}
{\footnotesize \hs{1} {\bf (B).} $p_z = 0$. }
\end{minipage}

\vs{.5}

\begin{minipage}[b]{5.5cm}
\epsfig{figure=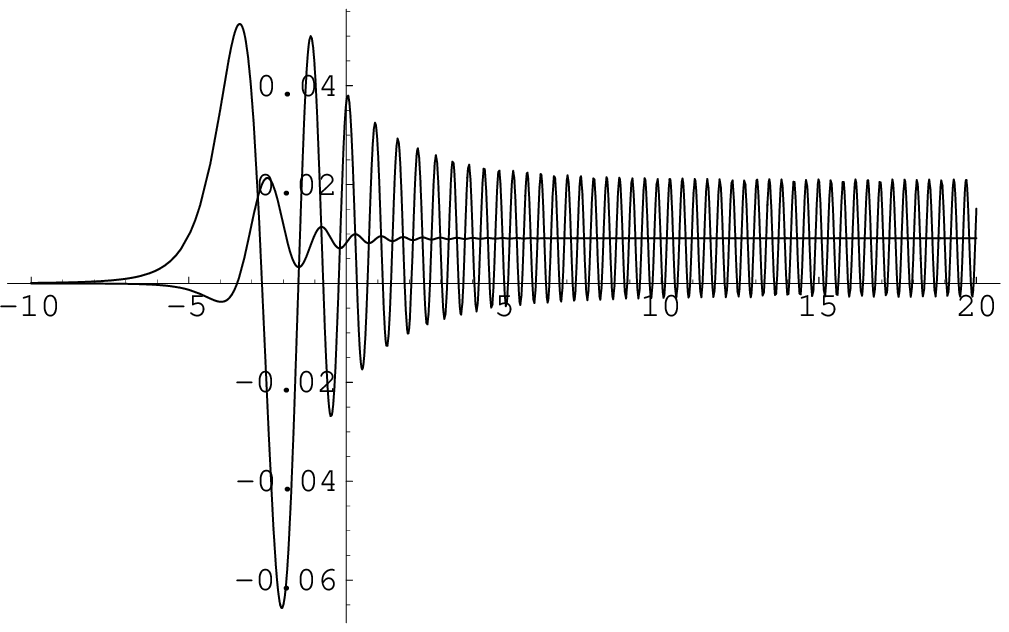,width=5.5cm,angle=0}
{\footnotesize \hs{1} {\bf (C).} $p_z = 3$. }
\end{minipage}\hs{.5}\begin{minipage}[b]{5.5cm}
\epsfig{figure=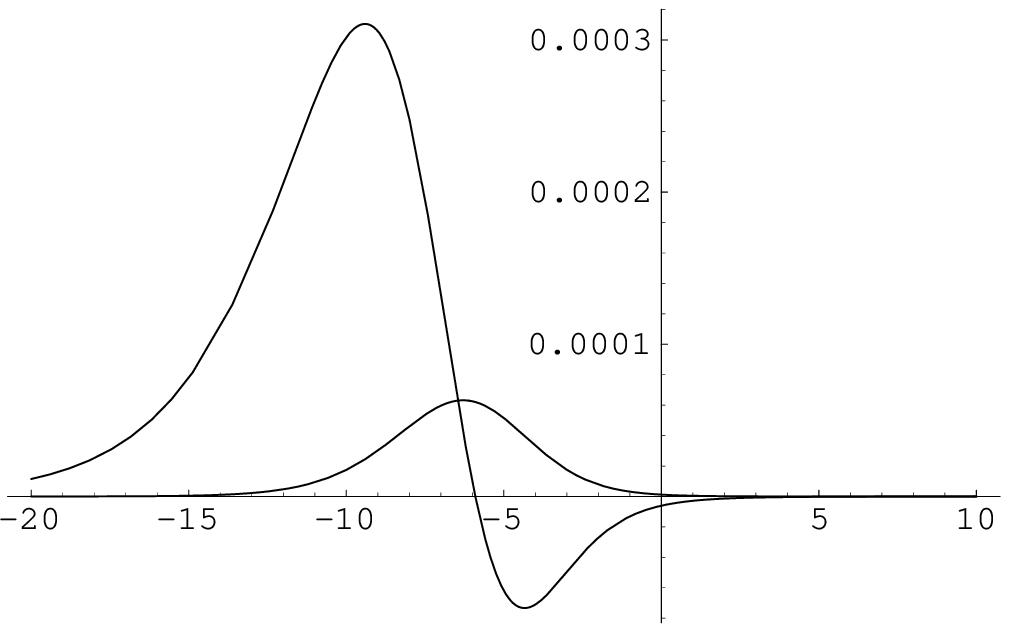,width=5.5cm,angle=0}
{\footnotesize \hs{1} {\bf (D).} $p_z = 6$. }
\end{minipage}

{\footnotesize {\bf FIG. 14.} Current density in the evolved `in' vacuum shown as a function of $\tau$ for $a_0 = \frac{E}{E_c} = 1$, $\frac{|\bp_{\perp}|}{m} = 0$, $\rho = 5$ and $p_z = -3$ (A), $0$ (B), $3$ (C) and $20$ (D). The $|\beta|^2$ 
contribution is also shown (darker curve).}
\end{figure}
\setcounter{figure}{14}
\begin{figure}[htb!]
\centerline{
\put(-140,0){\hbox{\psfig{figure=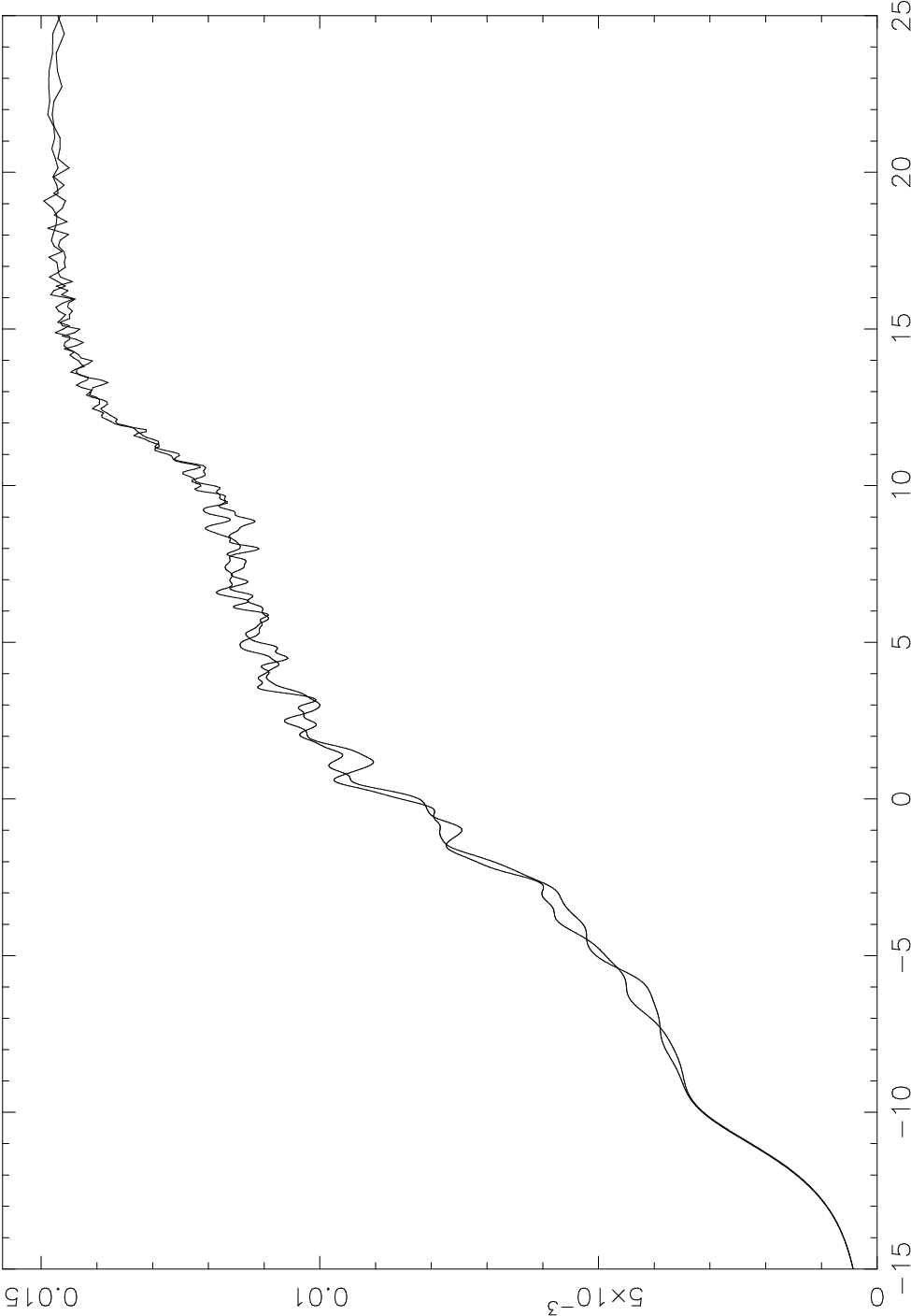,height=10cm,angle=-90}}}
}
\vs{0.2}
\caption[dummy1]{Monte-Carlo calculation of the total current in a Compton volume $m^{-3}$
for $a_0 = 1$ and $\rho = 10$. The two curves are independent sets of 100,000 samples, giving an idea of the 
numerical accuracy.}
\label{jtotmc}
\end{figure}
Figure~\ref{jtotmc}
 shows the total current  $J_{{\rm vac},3,-\infty}(t)$ in a Compton volume
$m^{-3}$ of the evolved `in' vacuum, as a function of $\tau$, for $a_0 = 1$ and 
$\rho = 5$. 
	This current has been evaluated by integrating $J_{{\rm vac},3,-\infty}(t)$ over $|p_x| \leq 2$, $|p_y| \leq 2$ and 
over $|p_z| \leq 8$ (which corresponds to $-3 < p_z^{\rm{out}} < 13$). 
This range is large enough to include the dominant contribution to the integral, but ensures that the cut-off-dependent 
logarithmic divergence term is small enough to be ignored. 
By analysing the high-momentum behaviour of the integrand we can confirm the existence of a logarithmic divergence 
proportional to $\ddot{a}(\tau)$, as expected. The curves are not as smooth as those of Figure~\ref{ntotmc} because
the rapid oscillations have not been completely averaged out.

\subsubsection*{Non-Vacuum initial conditions}

Consider now a state which at asymptotically early times contained exactly 1 particle, with momentum $\bp_{{\rm in}}$. 
As in Section 4, the current of this state at later times contains two contributions, one from $J_{{\rm vac},3,-\infty}(t)$ 
and one from $J(u_{\bp_{{\rm in}},\lambda,-\infty}(\bx,t))$. 
The $z$-component of $J(u_{\bp_{{\rm in}},\lambda,-\infty}(\bx,t))$ is shown in Figure~16 as a function of $\tau$, 
for $a_0 = 1$, $\rho = 5$, $|\bp_{\perp}^{{\rm in}}| = 0$, and $p_{z}^{{\rm in}} = -8, -5, -2$ and $1$.
 Also included on each graph is the classically  expected current
for a particle accelerated from rest in this electric field. 
For (A), (B) and (C) these curves are qualitatively similar to Figure 6, the only difference being that 
$\tau$ now runs  from $-\infty$. 
In Figure 16(D), for which the physical momentum of the mode never passes through zero, 
the quantum result is indistinguishable from the classical result.

\begin{figure}[hbt]
\begin{minipage}[b]{5.5cm}
\epsfig{figure=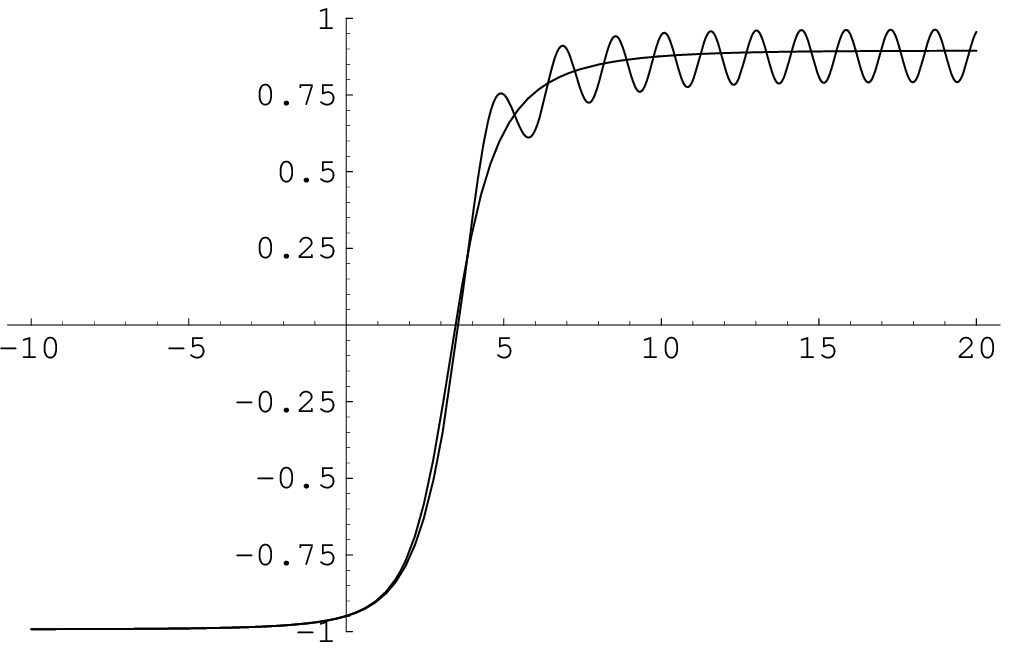,width=5.5cm,angle=0}
{\footnotesize \hs{1} {\bf (A).} $p_z^{{\rm in}} = -8$. }
\end{minipage}\hs{.5}\begin{minipage}[b]{5.5cm}
\epsfig{figure=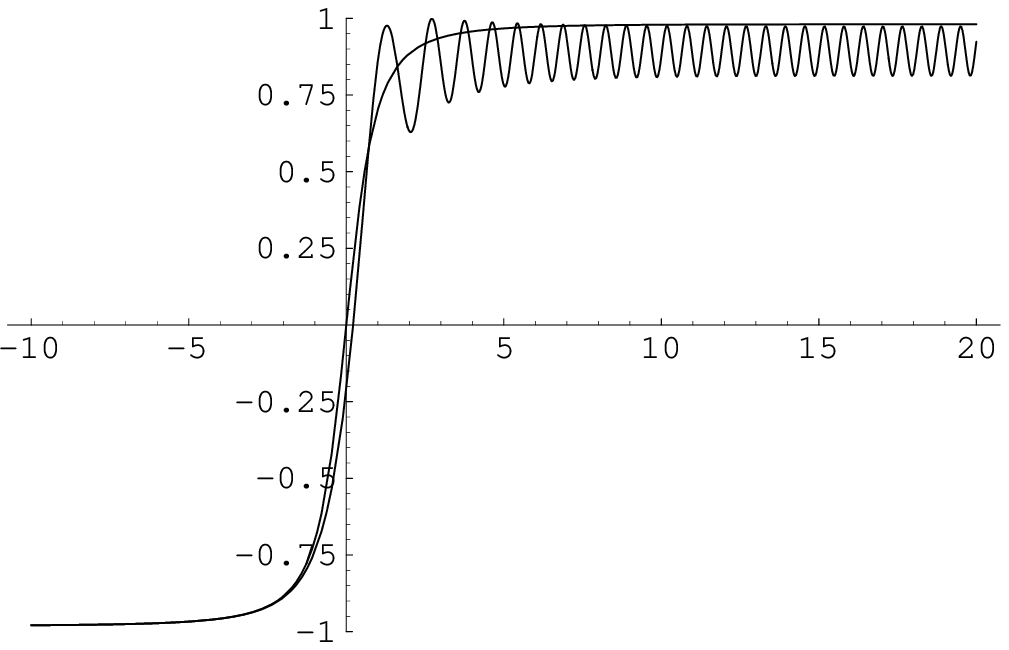,width=5.5cm,angle=0}
{\footnotesize \hs{1} {\bf (B).} $p_z^{{\rm in}} = -5$. }
\end{minipage}

\vs{.5}

\begin{minipage}[b]{5.5cm}
\epsfig{figure=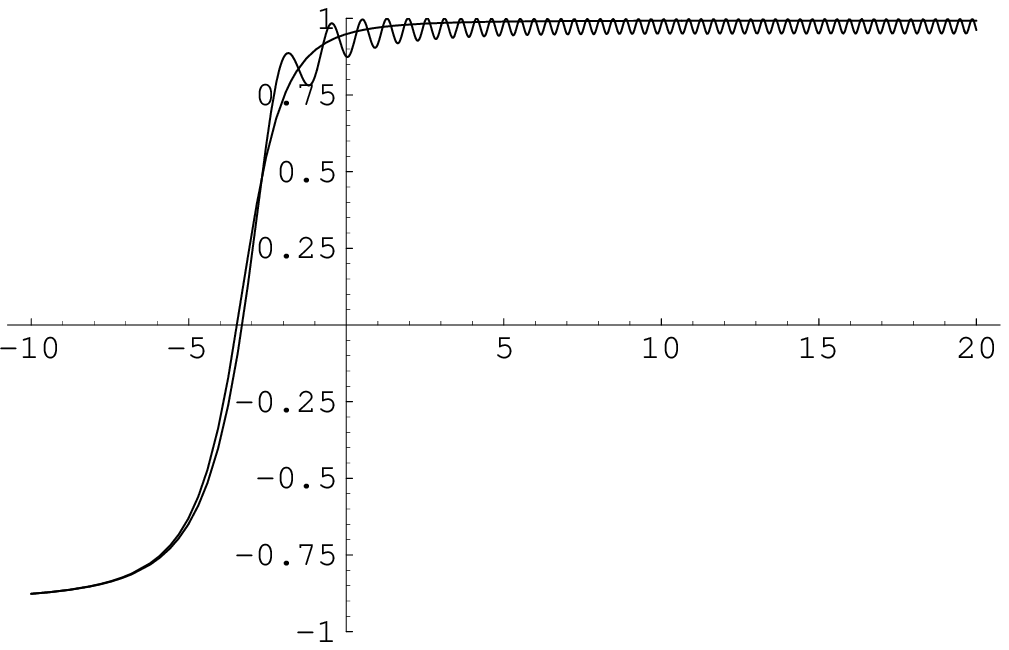,width=5.5cm,angle=0}
{\footnotesize \hs{1} {\bf (C).} $p_z^{{\rm in}} = -2$. }
\end{minipage}\hs{.5}\begin{minipage}[b]{5.5cm}
\epsfig{figure=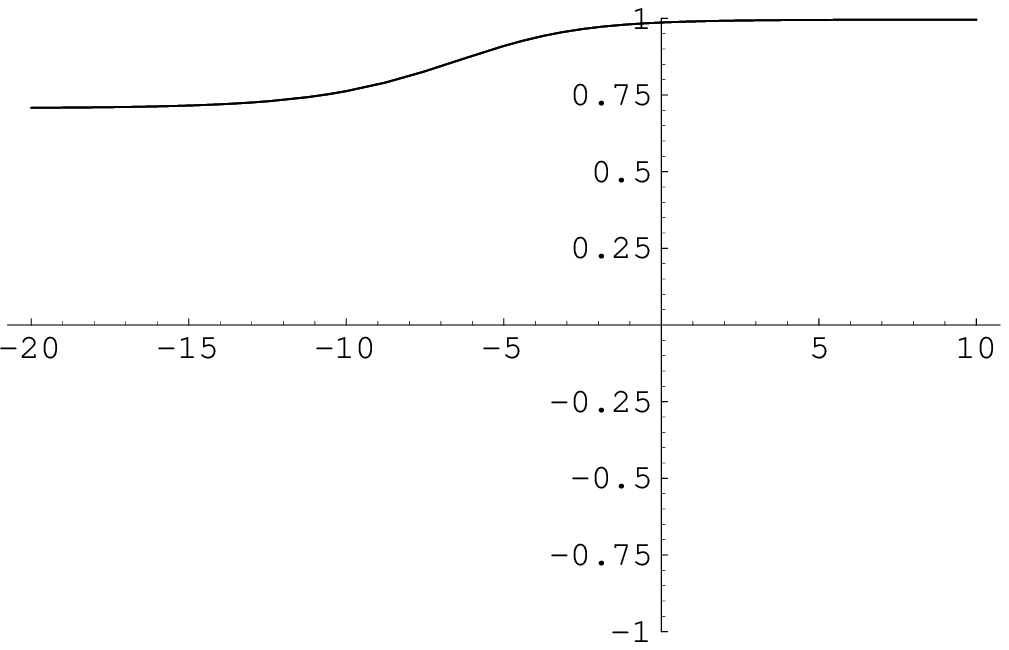,width=5.5cm,angle=0}
{\footnotesize \hs{1} {\bf (D).} $p_z^{{\rm in}} = 1$. }
\end{minipage}

{\footnotesize {\bf FIG. 16.} $J(u_{\bp_{{\rm in}},\lambda;t_0}(\bx,t_0 + \frac{\tau}{\sqrt{e E}}))$ as a 
function of $\tau$, for $a_0 = \frac{E}{E_c} = 1$, $\frac{|\bp_{\perp}|}{m} = 0$, $\rho = 5$ and $p_z^{{\rm in}} = -8$ (A), $-5$ (B), $-2$ (C) and $1$ (D). Also included is the 
current expected classically  for a particle accelerated from rest in this electric field.}
\end{figure}

\section{CONCLUSIONS}

	We have applied the particle definition proposed in \cite{Me1,mythesis} to 
the study of particle creation in spatially uniform electric fields. 
 By 
incorporating the `Bogoliubov coefficient' and `tunnelling' approaches into 
a single consistent, gauge-invariant definition, we have 
resolved several problems raised by Sriramkumar et al.~\cite{SPad,Sr1}.
We have demonstrated the utility of a time-dependent 
particle interpretation by presenting the time-development of the 
particle creation process, concentrating on the cases of a time-invariant
electric field and an `adiabatic' electric field. For a 
time-invariant
 electric field we found, as  expected for a first-order linear evolution equation, 
that the particle content after a finite amount of time $T$ is the same \cite{GaGi,Hal} 
as after evolution of a free `in' vacuum in the presence 
of a background that was switched on only for a time $T$. For the adiabatic 
case we have given a coherent account of the time-development of the particle 
production process, in which the particles are created with small momentum (in 
the frame of the electric field), and are then accelerated by the electric field to fill
 out the `bulge' of created particles predicted 
by asymptotic calculations~\cite{GaGi}. The current in the evolved vacuum is 
consistent with this picture, although the $\Re(\beta\alpha)$ term corresponds to 
oscillations around the $|\beta|^2$ term and contains the well-known 
logarithmic divergence, proportional to $\ddot{a}(t)$. 
We have also considered an initial state with one particle, and described how this state evolves 
as the sum of two contributions: the `sea of current' produced by the evolved vacuum, 
and the extra current arising from the initial particle state. 

	The repeated mention of `small momentum' throughout this work may cause concern
 about the Lorentz covariance of our procedure. In a constant electric field the class of observers 
 moving at constant velocity in the $z$-direction all measure the same electric field 
in their rest frame. Can they really all measure the particles to be created with low $p_z$ in their rest frames?
Of course they do not all measure this. The reason is hidden in the initial conditions. 
Insistence  that the state be vacuum at time $t=t_0$ effectively chooses a fixed reference frame $O$. 
An observer in another frame $O^\prime$ would find no time $t^\prime$  for which the state prepared as above contains no
particles. Consequently, no contradiction arises.

	We hope this work, along with the formalism presented in \cite{Me1,mythesis}, has shown the computational and
conceptual value of working with a concrete representation of the Dirac Sea. We strongly
support  Jackiw's claim~\cite{Jackiw} that ``physical consequences can be drawn from Dirac's construction''.

\begin{acknowledgments}

	 We thank Anton Garrett and 
Dr Emil Mottola for helpful discussions. Carl Dolby also thanks Trinity College and Cavendish 
Astrophysics for financial support, and thanks the Institute for Nuclear Theory, 
University of Washington and the Wellcome Trust Centre for Human Genetics, Oxford, for hospitality during his visits.

\end{acknowledgments}

\end{article}
\end{document}